\documentstyle[12pt,aaspp4]{article}

\newcommand{\bq}{\begin{equation}} 
\newcommand{\eq}{\end{equation}}

\newcommand{\beq}{\begin{equation}}
\newcommand{\eeq}{\end{equation}}
\newcommand{\beqa}{\begin{eqnarray}}
\newcommand{\eeqa}{\end{eqnarray}}
\newcommand{\om}{\Omega_m}

\begin{document} 
\def\refitem{\par\parskip 0pt\noindent\hangindent 20pt} 

\normalsize 

\title{New {\it Hubble Space Telescope} Discoveries of Type Ia Supernovae
at $z \geq 1$: Narrowing Constraints on the Early Behavior of 
Dark Energy\altaffilmark{1}}

\vspace*{0.3cm}

Adam G. Riess\altaffilmark{2,3}, 
Louis-Gregory Strolger\altaffilmark{4}, 
Stefano Casertano\altaffilmark{3}, 
Henry C. Ferguson\altaffilmark{3}, 
Bahram Mobasher\altaffilmark{3}, 
Ben Gold\altaffilmark{2},
Peter J. Challis\altaffilmark{5}, 
Alexei V. Filippenko\altaffilmark{6}, 
Saurabh Jha\altaffilmark{6}, 
Weidong Li\altaffilmark{6}, 
John Tonry\altaffilmark{7}, 
Ryan Foley\altaffilmark{6}, 
Robert P. Kirshner\altaffilmark{5}, 
Mark Dickinson\altaffilmark{8}, 
Emily MacDonald\altaffilmark{8}, 
Daniel Eisenstein\altaffilmark{9}, 
Mario Livio\altaffilmark{3}, 
Josh Younger\altaffilmark{5},
Chun Xu\altaffilmark{3},
Tomas Dahl\'{e}n\altaffilmark{3},
and
Daniel Stern\altaffilmark{10}

\altaffiltext{1}{Based on observations with the NASA/ESA {\it Hubble Space 
Telescope}, obtained at the Space Telescope Science Institute, which is 
operated by AURA, Inc., under NASA contract NAS 5-26555.} \altaffiltext{2}{Department of Physics and Astronomy, Johns Hopkins University, 
Baltimore, MD  21218.}
\altaffiltext{3}{Space Telescope Science Institute, 3700 San Martin 
Drive, Baltimore, MD 21218.} 
\altaffiltext{4}{Department of Physics and Astronomy, Western Kentucky University, 1906 College Heights Blvd., Bowling Green, KY 42101-1077}
\altaffiltext{5}{Harvard-Smithsonian Center for Astrophysics, 60 Garden St., 
Cambridge, MA 02138.} 
\altaffiltext{6}{Department of Astronomy, 601 Campbell Hall, University of 
California, Berkeley, CA  94720-3411.} 
\altaffiltext{7}{Institute for Astronomy, University of Hawaii, 
2680 Woodlawn Drive, Honolulu, HI 96822.} 
\altaffiltext{8}{NOAO, Tucson, AZ 85726-6732.}
\altaffiltext{9}{Steward Observatory, University of Arizona,
 Tucson, AZ 85721-0065.}
\altaffiltext{10}{Jet Propulsion Laboratory, California Institute of
  Technology, Pasadena, CA 91109}

\begin{abstract} 

We have discovered 21 new Type Ia supernovae (SNe~Ia) with the {\it
Hubble Space Telescope (HST)} and have used them to trace the
history of cosmic expansion over the last 10 billion years. 
These objects, which include 13
spectroscopically confirmed SNe~Ia at $z \geq 1$, were discovered
during 14 epochs of reimaging of the GOODS fields North and South over
two years with the Advanced Camera for Surveys on {\it HST}.  Together
with a recalibration of our previous {\it HST}-discovered SNe~Ia, the
full sample of 23 SNe~Ia at $z \geq 1$ provides the highest-redshift sample
known.
Combined with previous SN~Ia datasets, we measured $H(z)$ at discrete,
uncorrelated epochs, reducing the uncertainty of $H(z>1)$ from 50\% to
under 20\%, strengthening the evidence for a cosmic jerk--the transition from
deceleration in the past to acceleration in the present.  The unique leverage of the
{\it HST} high-redshift SNe~Ia provides the first meaningful
constraint on the dark energy equation-of-state parameter at $z \geq
1$.   The result remains consistent with a cosmological constant ($w(z)=-1$), and rules out rapidly evolving dark energy ($dw /dz >>1$).  The defining
property of dark energy, its negative pressure, appears to be present
at $z>1$, in the epoch preceding acceleration, with $\sim$ 98\% confidence in our primary fit. Moreover, the $z>1$
sample-averaged spectral energy distribution is consistent with that
of the typical SN~Ia over the last 10 Gyr, indicating that any
spectral evolution of the properties of SNe~Ia with redshift is still below our
detection threshold.

\end{abstract} 
subject headings: galaxies: distances and redshifts ---
cosmology: observations --- cosmology: distance scale --- 
supernovae: general

\section{Introduction} 

   The accelerating cosmic expansion first inferred from observations
of distant type Ia supernovae (SNe~Ia; Riess et al. 1998; Perlmutter
et al. 1999) indicates unexpected gravitational physics, frequently
attributed to the dominating presence of a ``dark energy'' with
negative pressure.  Increasingly incisive samples of SNe~Ia at $z<1$
have reinforced the significance of this result (Tonry et al. 2003;
Knop et al. 2003; Barris et al. 2004; Conley et al. 2006; Astier et
al. 2006).  Using the new Advanced Camera for Surveys (ACS) and
refurbished NICMOS camera on the {\it Hubble Space Telescope (HST)},
our collaboration secured observations of a sample of the most-distant
known SNe~Ia.  These half-dozen SNe Ia, all at $z>1.25$, helped
confirm the reality of cosmic acceleration by delineating the
transition from preceding cosmic deceleration during the
matter-dominated phase and by ruling out simple sources of
astrophysical dimming (Riess et al. 2004b, hereafter R04).  The expanded sample of 23 SNe Ia at $z \geq 1$ presented here are now used to begin characterizing the  early behavior of dark energy.

  Other studies independent of SNe~Ia now strongly favor something
like dark energy as the dominant component in the mass-energy budget
of the Universe. Perhaps most convincingly, observations of large-scale
structure and the cosmic microwave background radiation provide
indirect evidence for a dark-energy component (e.g., Spergel et
al. 2006).  Measurements of the integrated Sachs-Wolfe effect (e.g.,
Afshordi, Loh, \& Strauss 2004; Boughn \& Crittenden 2004; Fosalba et
al. 2003; Nolta et al. 2004; Scranton et al. 2005) more directly
suggest the presence of dark energy with a negative
pressure. Additional, albeit more tentative, evidence is provided by
observations of X-ray clusters (Allen et al. 2004) and baryon
oscillations (e.g., Eisenstein et al. 2005).

The unexplained existence of a dominant, dark-energy-like phenomenon
presents a stiff challenge to the standard model of cosmology and
particle physics.  The apparent acceleration may result from exotic
physics such as the repulsive gravity predicted for a medium with
negative pressure or from entirely new physics.  The explanation of
strongest pedigree is Einstein's famous ``cosmological constant''
$\Lambda$ (i.e., vacuum energy; Einstein 1917), followed by a decaying scalar field
similar to that already invoked for many inflation models (i.e.,
quintessence --Wetterich 1995, Caldwell, Dav\'e, \& Steinhardt 1998;
Peebles \& Ratra 2003).  Competitors include the Chaplygin gas
(Bento, Bertolami, \& Sen 2002), topological defects, and a massless scalar field at low
temperature.  Alternatively, alterations to
General Relativity may be required as
occurs from the higher-dimensional transport of gravitons in string
theory models (Deffayet et al. 2002) and braneworlds, or by finely-tuned,
long-range modifications (e.g., Cardassian type, Freese 2005; or
Carroll et al.  2004; see Szydlowski, Kurek, \& Krawiec 2006 for a
review).  Empirical clues are critical for testing
hypotheses and narrowing the allowed range of possible models.

SNe~Ia remain one of our best tools for unraveling the properties of
dark energy because their individual measurement precision is
unparalleled and they are readily attainable in sample sizes of order
$10^2$, statistically sufficient to measure dark-energy-induced changes to the
expansion rate of $\sim$1\%. Specifically, the equation-of-state
parameter of dark energy, ${w}$ (where $P = w\rho c^2$) determines
both the evolution of the density of dark energy,
\bq
\rho_{DE}=\rho_{DE,0} \exp \{ 3 \int_a^1 {da \over a} (1+w(a)) \}, \eq
and its gravitational effect on expansion, 
\bq \ddot a / a = (-4\pi G/3)[\rho_m + \rho_{DE}(1+3w(a))],\eq 
where $\rho_{DE,0}$ is the present dark-energy density.  Measuring
changes in the scale factor, $a$, with time from the distance and
redshift measurements of SNe~Ia,
\bq {d_l(z) \over c(1+z)}=\int_t^{t0} {dt' \over a(t')}=\int_0^z {dz'
\over H(z')}, \eq
constrains the behavior of $w(a)$ or $w(z)$ and is most easily
accomplished at $z<2$ during the epoch of dark-energy dominance.

Ideally, we seek to extract the function $w(z)$ for dark energy or its mean value at a wide range of epochs.  Alternatively, we might constrain its
recent value $w_0 \equiv w(z=0)$ and a derivative, $dw/dz \equiv w'$, which are
exactly specified for a cosmological constant to be ($-1$,0).  Most other models
make less precise predictions.  For example, the presence of a ``tracker''
dark-energy field whose evolution is coupled to the (decreasing) dark
matter or radiation density may be detected by a measured value of $w'
> 0$ or $w(z>1) \sim 0$.  In truth, we know almost nothing of what to expect for $w(z)$, so the safest approach is to assume nothing and
measure $w(z)$ across the redshift range of interest.  SN Ia at $z>1$ are crucial to constrain variations of $w$ with redshift.
These measurements can only be made from space, and we report here on that endeavor. We have discovered and measured 21 new SN Ia with HST and used them to constrain the properties
of the dark energy.  We present the follow-up spectroscopy and photometry of the new SNe~Ia in \S 2, light-curve analysis and cosmological constraints
in \S 3, a discussion in \S 4, and a summary in \S 5.

\section{Further Discoveries and Data Reprocessing}

\subsection{ACS Searches, 2003--2005}

  In {\it HST} Cycle 11 (July 2002 -- June 2003) we initiated the first
space-based program designed to find and monitor SNe (R04).  Our
search was conducted by imaging the two high-latitude fields of the
GOODS Treasury Program (the {\it Chandra} Deep Field South and the {\it Hubble}
Deep Field North) with 15 ACS pointings, 5 times each at 6 or 7-week
intervals (chosen to match the rise time of SNe~Ia at $z \approx 1$).
Multiple exposures in the $F850LP$ bandpass were differenced to find
transients.  Contemporaneous color measurements and host-galaxy
photometric redshifts were utilized to identify promising candidate
SNe~Ia at $z>1$ for target-of-opportunity (ToO) follow-up observations
(Riess et al. 2004a).

  In {\it HST} Cycles 12 and 13 (July 2003 -- June 2005) we continued
our past efforts, imaging the GOODS fields (Giavalisco et al. 2004, Strolger et al. 2004) 
14 more times and following
newly discovered SN~Ia candidates with {\it HST}.  The observational
methods we used in Cycles 12 and 13 were very similar to those of
Cycle 11 and are extensively described by R04.  Readers are directed
to that publication for the sake of brevity; here we describe only
{\it changes} to our observing and candidate-selection strategies.

To improve our search efficiency with finite observing time, we compressed our primary search exposures 
from 2 orbits to 1.  We omitted the short-wavelength filters, whose use in the GOODS program was 
primarily to constrain the properties of galaxy formation.  Our typical search sequence contained four dithered exposures of length 400~s in
$F850LP$ and one 400~s exposure in $F775W$ for a typical orbit instead of the previous four exposures of 500~s length in $F850LP$ (as
well as two such exposures in each $F606W$ and $F775W$).
  At
times when we had unusually long orbits, especially approaching the
continuous viewing zone for the HDFN, we included a $\sim$200~s
exposure in $F606W$ as well to help constrain SN type and redshift.

Although our net exposure time was reduced by 20\% from the Cycle 11 observations, the accumulation of additional template images for the host galaxies without the supernova during the GOODS program increased their total depth and more 
than compensated for the slightly decreased signal in this search.  

We benefited from two modest improvements in our ability to identify
high-redshift SNe~Ia before initiating a ToO over the strategy utilized by R04.  The first was
the availability of spectroscopic host-galaxy redshifts obtained
before the appearance of a SN candidate (e.g., from our Keck and other
spectra assembled for the GOODS catalog, and from ``Team Keck''; Wirth et al. 2004).  The second was the result of building up a 3 year baseline of variability for each galaxy.  This allowed us to 
distinguish between transient signals caused by supernovae and transient noise introduced by the variable 
emission from active galactic nuclei. 

The 21 new SNe~Ia are listed in Table 1 and their discovery images are
shown in Figure 1.  Internal names for the SNe are used in the following.
Color images of all of the SN Ia host galaxies (where filters $F850LP$, $F606W$, and $F435W$ correspond to red, green, and blue, respectively) are shown in
Figure 2, with the position of the SN indicated in each case.  

\begin{deluxetable}{lllll} 
\footnotesize
\tablecaption{SN Discovery Data}
\tablehead{\colhead{Name} & \colhead{Nickname}&\colhead{JD$-$2,400,000}&\colhead{SN
$\alpha$(J2000)}&\colhead{SN $\delta$(J2000)}}
\startdata
\hline
\hline
HST04Sas & Sasquatch	& 53148.7 & 12:36:54.11 &	+62:08:22.76 \nl
HST04Man & Manipogo	& 53146.3 & 12:36:34.81 &	+62:15:49.06 \nl
HST04Yow & Yowie	        & 53145.2 & 12:36:34.33	& +62:12:12.95 \nl
HST04Pat & Patuxent	& 53196.3 & 12:38:09.00	& +62:18:47.24 \nl
HST04Tha & Thames	& 53196.5 & 12:36:55.17	 & +62:13:04.05 \nl
HST04Cay & Cayanne$^*$	& 53245.2 & 12:37:27.11	& +62:12:07.68 \nl
HST04Eag & Eagle	& 53284.9 & 12:37:20.75	& +62:13:41.50 \nl
HST04Haw & Hawk	& 53284.0 & 12:35:41.16	 & +62:11:37.19 \nl
HST05Spo & Spock	& 53376.8 & 12:37:06.53	 & +62:15:11.70 \nl
HST04Mcg & McGuire	& 53265.0 & 03:32:10.02	& $-$27:49:49.98 \nl
HST04Gre & Greenberg	& 53265.0 & 03:32:21.49	& $-$27:46:58.30 \nl
HST04Omb & Ombo	& 53356.3 & 03:32:25.34 & $-$27:45:03.01 \nl
HST04Rak & Rakke	& 53356.3 & 03:32:18.15	 & $-$27:44:10.55 \nl
HST04Kur & Kurage	 & 53355.2 & 03:32:36.03	& $-$27:51:17.66 \nl
HST05Dic & Dickinson & 53472.5 & 12:35:49.61  & +62:10:11.96 \nl
HST05Fer & Ferguson & 53472.6 & 12:36:25.10 & +62:15:23.84 \nl
HST05Koe & Koekemoer  & 53472.7 & 12:36:22.92  & +62:18:23.20 \nl
HST05Str & Strolger & 53474.5 & 12:36:20.63 & +62:10:50.58 \nl
HST05Gab & Gabi    & 53474.5 & 12:36:13.83 & +62:12:07.56 \nl
HST05Red & Redford & 53426.6 & 12:37:01.70 & +62:12:23.98 \nl
HST05Lan & Lancaster & 53427.1 & 12:36:56.72 & +62:12:53.33 \nl
HS05Zwi & Zwicky & 53403.2 & 03:32:45.65 &$-$27:44:24.30 \nl
\enddata 
\tablenotetext{*}{not positively identified as SN Ia}
\end{deluxetable}

\subsection{Photometry}

   Our follow-up observations of candidate SNe~Ia were similar to those
previously obtained and described by R04.  After the search phase,
all images were reprocessed using up-to-date reference files and the
CALACS pipeline in the STSDAS package in IRAF.\footnote[8]{IRAF is
distributed by the National Optical Astronomy Observatories, which are
operated by the Association of Universities for Research in Astronomy,
Inc., under cooperative agreement with the National Science
Foundation.}  Because improvements have been made to the standard
CALACS pipeline (including an improved distortion map and more precise
flatfield images), we also reprocessed all preceding SN data from our
first year.

   To produce light curves, we developed an automated pipeline to
retrieve all images obtained by ACS and NICMOS in the GOODS fields
from the Multimission Archive (reprocessing all data from R04 as well
as all of the new frames).  The enhanced set of images available from
before and after each SN increased the depth of the constructed
template image (without SN light) for each SN, thereby increasing the subsequent
precision of the SN photometry.

   Two modest improvements to measurements of SN flux were made from
the methods described by R04. First, due to backside scattering from
the ACS WFC CCD mounting, the point-spread function (PSF) redward of
$\sim$ 8500~\AA\ displays a halo of scattered light which needs to be
quantified for precise photometry in the $F850LP$ bandpass.  R04 used
a red star with $i-z$ color similar to a SN Ia at peak at $z \approx
1.3$ ($i-z = 1.0$ mag, Vega) as a model PSF for SN photometry.  However,
variations in SN redshift and phase around that model make the match with a single PSF at a single color inexact resulting in
photometric errors ranging from $\pm$0.05 mag for our sample of phases
and redshifts.  Here, we adopted an individualized method for the PSF
modeling of high-redshift SNe~Ia.  As described in detail by Casertano
\& Riess (2007), we used 6 observed monochromatic PSFs from ACS
narrow-band filters between 8150~\AA\
and 10500~\AA\ to derive an empirical wavelength dependent correction to the
average PSF model.  At these monochromatic endpoints the difference
between (nearly) infinite-aperture photometry and that determined from
a fit to the previously used red-star PSF is $-0.14$ mag and +0.46
mag, respectively, with no difference occurring near 9300~\AA.  These
differences were evaluated for all monochromatic wavelengths.  At
every SN redshift and phase, the same representative model SN~Ia
spectral energy distribution (SED) used to derive K-corrections was
used to integrate the monochromatic PSFs within the $F850LP$ band
response and determined their photometric differences.  In Figure 3 we
show the required photometric correction as a function of SN~Ia age
and redshift.   Spatial variations in the ACS WFC PSF are small (Krist 2003)
and variations in the position of the SNe result in a PSF noise of 1-2\%, well below
the sky and read noise.

Second, we adopted an updated zeropoint for the {\it HST} ACS
``Vegamag'' photometric system from Sirianni et al. (2005) which uses
the spectrophotometry of Vega from Bohlin \& Gilliland (2004) and
improved quantum efficiency data for ACS to set Vega to 0.00 mag in
all passbands.  We then utilized the same magnitudes
of Vega assumed by Landolt (1992) 
 to calibrate the {\it HST} photometry on the same Vega system
as the nearby SN~Ia sample. Thus our data is calibrated 
on the Landolt system.  The resulting zeropoints corresponding to
1 electron s$^{-1}$ are $F850LP = 24.35$, $F775W = 25.28,$ and $F606W
= 26.43$ mag. These are fainter than those used in R04 by 0.02 mag.
For NICMOS we use the values of $F110W=22.92$ mag and $F160W=22.11$ mag given
by the STScI NICMOS Handbook.
  
An improvement in our ability to estimate photometric errors was
provided by the increased set of host images long before or after the
appearance of the SNe. For each such image we added and measured an
ideal PSF at a range of magnitudes at the {\it same} pixel position as the site of the SN
and derived statistics from the recovered magnitudes such as their
dispersion and their bias at a given magnitude.\footnote{We found
biases to exist only for the case of bright, sharp hosts which were
``softened'' by image interpolation in the registration of the template.
The worst case hosts such as 2003XX and 2002hp seen in Figures 1 and 2 without correction would bias the magnitudes too bright by as much as 0.2 mag at the end of the observed light curve.
We identified two valid solutions to this bias which gave consistent results: (1) equally soften the
SN image by reinterpolation, or (2) estimate the bias from the
simulated SNe~Ia.  This latter option is preferable as it is nearly
noise-free.}

The calibration of {\it HST} NICMOS NIC2 photometry has changed since
R04 due to the recent detection of an apparent non-linearity in all
NICMOS detectors.  This effect was initially discovered by Bohlin,
Lindler, \& Riess (2005) and has now been calibrated to good precision
by de Jong et al. (2006) using pairs of star-cluster images obtained
with the flatfield lamps switched on and off.  The result for NIC2 is
a reduction in the apparent flux by 0.06 and 0.03 mag per dex in
$F110W$ and $F160W$, respectively.\footnote{The correction is described
by de Jong et al. (2006) in the form: count rate $\propto$
flux$^\alpha$, where $\alpha=1.025 \pm 0.002$ and $\alpha=1.012 \pm
0.006$ for $F110W$ and $F160W$, respectively.}  

For faint sources such as high-redshift SNe~Ia, the nonlinearity in apparent flux
pertains to the flux difference between the calibration
stars (G191B2B and P330E, both $\sim$12 mag in $J$ and $H$) and the faint SN.  For
SN fluxes fainter than the sky level (all of those presented here) the nonlinearity pertains
to the flux difference between the calibration stars and the sky level, 
below which any additional nonlinearity of sources is effectively ``quenched''.  For SN~Ia plus host fluxes
near or below the sky (a typical sky level is 0.17 electron s$^{-1}$
in $F110W$ and 0.14 electron s$^{-1}$ in $F160W$), the correction we
calculate and apply is 0.220 mag brighter (than the uncorrected
zeropoints) in $F110W$ and 0.086 mag brighter in $F160W$.
Interestingly, the change in distance modulus from R04 due to these
corrections is mitigated by their compensating effect in distance and
reddening.\footnote{Because the change in the bluer band ($F110W$) is
larger, the net change in the distance moduli from R04 to first order
is approximately $\Delta F110W - R(\Delta F110W - \Delta F160W)$,
which for a reddening ratio $R \approx 3$ is 0.17 mag.  In practice, a
second-order change occurs as the apparent color impacts other
parameters in the fit as well as the individual K-corrections.  For
the highest-redshift SNe which rely heavily on NICMOS data, the
average change from R04 was found to be 5\% (0.10 mag) closer in
distance, but depends on the individual SN~Ia.}

SNe measured in our detection and follow-up observations are
generally of high enough signal-to-noise ratio (S/N $>$ 10) to support the use
of magnitudes (where a magnitude is zeropoint $-2.5\, {\rm log}_{10}
{\rm flux}$) without skewing the interpretation of photometric errors.
The infrequent observation (less than 5\% of observations) with (S/N $<$ 5)
on the post-maximum light curve of any SN has insignificant weight
when used in conjunction with the higher signal-to-noise ratio points.
However, our {\it pre-discovery} images require the use of flux measurement and
error estimates due to their greater leverage on the determination of the time of maximum.
  For these images we measured the sky-subtracted
flux (electrons per second) in a $0.17''$ radius aperture within which the zeropoint for a 1 electron per second source is 
is 24.68 mag for $F850LP$.

  Our final photometry for all SNe~Ia is listed in Table 2 and shown
compared to the individual multicolor light-curve shape (MLCS) fits
in Figure 4.

\subsection{Spectroscopy and SN Identification}

  The ACS grism spectroscopy we obtained for the SNe~Ia is listed in
Table 3 (they will also be made
 available at the University of Oklahoma supernova spectra database, SUSPECT).
  In general, our reduction and analysis methods were the same
as in R04 with the following exceptions.  A more realistic skyflat was
used to separate the contributions to each pixel from the sky and the
source as described in Pirzkal et al. (2005).  We also utilized a
Lanczos kernel in the drizzling procedure to decrease the effective
size of the pixel convolved with the images and improve the separation
for cases where the SN and host were very close together (e.g.,
SN HST04Sas).

  The spectra we obtained and used to classify the SNe are shown in
Figure 5.  As in R04, to classify the SNe, the detected SN spectra
were cross-correlated with template spectra (after removal of the
continuum) to identify their type and redshift using the ``SNID''
algorithm (Tonry et al. 2003, Blondin et al. 2006).  For the cases listed in Table 3 for
which narrow-line host emission was identified, the redshift was
constrained to the value determined from the host emission before
cross-correlation, improving the significance of the cross-correlation
peak.  For all 12 spectra shown in Figure 5, SNID provided a
significant classification for each as type Ia. Although the
diagnostic used by the SNID algorithm relies on the whole spectrum,
the majority of these SNe can also be classified as type Ia from the
presence of Si II absorption at 4130~\AA\ (Coil et
al. 2000). Specifically, evidence of Si II absorption was seen in the
two highest-redshift spectra presented here.  Broad Ca~II absorption
near 3750~\AA\ is visible in all the spectra as well, but this feature
is less secure than Si~II for SN~Ia classification due to its
appearance in the spectra of SNe~Ic (Filippenko 1997).  For the highest redshift spectrum shown, HST04Sas, $F110W$, $F160W$, and $F205W$ band NIC2 imaging of the host was obtained to constrain the phot-z of the red, elliptical host to $z=1.4 \pm 0.15$ which compares well with the SN value of $z=1.39 \pm 0.01$.

For SN HST04Cay cross-correlation peaks exist, but none
with high significance and the redshift of the host is not known.  For HST05Red, the spectral match to a SN Ia at the host $z=1.189$ in Table 3 is fair, but not secure.
  SN 2003XX, identified by R04 as an SN~Ia on
the basis of its elliptical host, is now spectroscopically classified
as an SN Ia due to the acquisition of an ACS grism spectroscopic
galaxy-only template on 2005-03-07 to subtract from the grism spectrum
of the SN obtained on 2003-04-16 and heavily contaminated by galaxy light.
  We also analyzed
the ACS grism spectra of two more apparent high-redshift SNe observed
by another collaboration (GO-9729; P.I. Perlmutter) during the course of Cycle 12 in
the GOODS fields (SN150G $\alpha$ = 12:37:09.456, $\delta$ =
+62:22:15.59; SN150I at $\alpha$ = 12:37:51.533, $\delta$ =
+62:17:08.24) but failed to find a significant peak in the
cross-correlation so we cannot determine their type or redshift.

   Because of the large range in the quality and breadth of the
photometric record of individual SNe~Ia, R04 developed a two-tiered
approach to the confidence of our SN~Ia identifications.   By distinguishing or selecting data based on objective criteria of their good quality we can mitigate systematic errors caused by undersampling of light curves and misidentifications without introducing cosmological biases.

   To summarize this approach: we classify as ``high-confidence'' 
   (hereafter as ``Gold'')
SNe~Ia  those with a compelling classification
and whose photometric record is sufficient to yield a robust distance
estimate easily characterized by its measurement uncertainty.  ``Likely but 
uncertain SNe~Ia'' (hereafter ``Silver'') are those with an aspect of
the spectral or photometric record which is absent or suspect and
whose distance error is described with a caveat rather than a
quantitative uncertainty.  As in R04, the three primary reasons for
rejecting a SN~Ia from the Gold set are that (1) the classification,
though plausible, was not compelling, (2) the
photometric record is too sparse to yield a robust distance (i.e., the
number of model parameters is approximately equal to the effective
number of samplings of the light and color curves, and (3) the
extinction is so large as to be uncertain due to our ignorance of
extragalactic extinction laws.  Although R04 set this extinction
threshold to 1 mag of visual extinction, here we adopt the more
conservative threshold of 0.5 mag utilized elsewhere (Miknaitis et
al. 2007; Tonry et al. 2003; Riess et al. 2005).    SNe with two liens against its
confidence (i.e., those rejected from the ``Silver'' set) are not included in the
remainder of this paper (e.g., HST05Cay).

The measured SN~Ia distances for the full Gold and Silver set of {\it HST}-discovered SNe are given in Table 6, including the revised distance
measurements of the HST objects first presented by R04.    These distances
use the same distance scale as in R04.  
To this sample we add the same ground-discovered sample of SNe Ia employed in R04 for the following analyses\footnote{
As in R04, past SN data such as light curves and spectra from Perlmutter et
al. (1999) and spectra from the sample of Knop et al. (2003) remain unavailable.
Thus we resort to the same reliance as in R04 on their published distances normalized to a consistent distance
scale using SNe in common and classification confidences from these
authors. In \S 4 we also consider the impact of rejecting these and other older data.}

An important and recent addition to the ground-discovered sample is 
the first-year SNLS dataset from Astier et al. (2006) containing 73
new SNe~Ia.  We sought to define a similarly high-confidence subsample from this
set using the information available from Astier et al. (2006) and Howell et al. (2005).    

We removed from the
full-sample 2 events discarded by the SNLS (SNLS-03D4au and
SNLS-03D4bc), as well as the 15 objects classified by Howell et
al. (2005) and Astier et al. (2006) as ``SN Ia*,'' meaning they
are``probable SN~Ia'' but among the ``least secure identification,''
and another type (such as a SN~Ic) is not excluded.  
 As discussed by Astier et al. (2006), most of
the objects at $z>0.8$ suffer from imprecise color measurements which
dominate the distance error and would lead to a distance bias after
application of a low-extinction cut.  However, the low-extinction cut
is valuable to reduce the sensitivity to unexpected or evolving
extinction laws.  Our solution is to first eliminate objects with
highly uncertain color measurements, $\sigma_{color} > 0.15$ mag,
which removes the following 6 objects: SNLS-03D1ew, SNLS-03D4cn,
SNLS-03D4cy, SNLS-04D3cp, SNLS-04D3dd, and SNLS-04D3ny.  The distance
uncertainty of these objects is typically 2 to 3 times that of the
rest of the objects and thus the value of these 6 is very low (roughly
the equivalent of losing a single well-measured SN~Ia).  Lastly, the
same color cut of $A_V < 0.5$ mag used for all Gold sample SNe removes 3 objects: SNLS-03D1gt,
SNLS-03D3ba, and SNLS-04D2gc.  Thus, 47 high-confidence SNLS objects
remain classified as ``Gold''.  The SNLS SNe were fit with our present MLCS2k2 algorithm to estimate their distances.
  Alternatively, we found that
after the addition of 0.19 mag to distances measured to the SNLS SNe by Astier et al. (2006) (to account for the arbitrary choice of distance scale
and determined from low-redshift SNe in common), the agreement between our MLCS2k2 measured distances on the distance scale
used in R04 and here and those from Astier et al. (2006)
were consistent in the mean to better than 0.01 mag.  Thus, either set of distances measured to the SNLS SNe provide 
a comparable and suitable addition to our cosmological sample.
We verified (next section) that either yields the same inferences for $w(z)$ and provide results fitting the SNLS SNe with either method in Table .  \footnote{We made use of the Astier et al. (2006) distances for our primary fit as their light curves were not initially published by Astier et al.
and only recently made available during the preparation of this work.}

We have not made use of any of the seven high-redshift SNe~Ia
from Krisciunas et al. (2005) due to the apparently biased selection
of the sample, as discovered by Krisciunas et al. but not yet fully
modeled and corrected.

Thus, a simple description of the full sample used here for the cosmological analyses consists of the addition of Table 5 from R04 to Table 6 provided here with
the revised distances to the SNe in common (i.e., the leading 20 SNe in Table 6) superceding those given in R04 Table 5,
 and the SNLS objects from Astier et al. (2006) identified in \S 2.3.  Either the Astier et al. (2006) provided distances (with the aforementioned addition
of 0.19 mag) or our own fits are comparable and suitable.  The sample can also be found at  http://braeburn.pha.jhu.edu/$\sim$ariess/R06 or upon request to ariess@stsci.edu.

Upcoming revisions to the ground-discovered samples and improvements to the distance-fitting algorithms are expected and will change the membership and distance measures in the full cosmological sample
and these should be considered before construction of a cosmological sample of SNe Ia.  In response to such improvements we will attempt to provide distance estimates
to the HST-discovered sample with updated fitting tools or distance scaling as warranted at
http://braeburn.pha.jhu.edu/$\sim$ariess/R06 or upon request to ariess@stsci.edu.

\begin{deluxetable}{llll} 
\footnotesize
\tablecaption{SN~Ia Imaging}
\tablehead{\colhead{Date$^a$}&\colhead{Vega Mag}&\colhead{Epoch(rest)}&\colhead{K-Corr}}
\startdata
\hline
\multicolumn{4}{c}{HST04Pat ($z=0.97$)} \nl
\hline
\multicolumn{1}{c}{ } & \multicolumn{1}{l}{$F775W$} &
\multicolumn{2}{r}{$F775W\rightarrow U$} \nl
53196.3&23.86(0.04)& -1.1&-0.63(0.10)\nl
53205.1&23.80(0.10)&  3.3&-0.73(0.08)\nl
\multicolumn{1}{c}{ } & \multicolumn{1}{l}{$F850LP$} &
\multicolumn{2}{r}{$F850LP\rightarrow B$} \nl
53196.3&23.77(0.04)& -1.1&-1.25(0.05)\nl
53205.1&23.81(0.04)&  3.3&-1.28(0.04)\nl
53205.4&23.77(0.04)&  3.4&-1.28(0.04)\nl
53225.3&24.45(0.04)& 13.5&-1.35(0.05)\nl
53377.7&27.01(0.40)& 90.9&-1.49(0.01)\nl
\hline
\multicolumn{4}{c}{HST04Mcg ($z=1.37$)} \nl
\hline
\multicolumn{1}{c}{ } & \multicolumn{1}{l}{$F850LP$} &
\multicolumn{2}{r}{$F850LP\rightarrow U$} \nl
53265.0&24.44(0.04)&  5.9&-1.01(0.03)\nl
53275.6&24.61(0.06)& 10.4&-1.06(0.04)\nl
53285.6&25.39(0.14)& 14.6&-1.11(0.02)\nl
53294.7&25.69(0.17)& 18.4&-1.12(0.01)\nl
53306.5&25.95(0.14)& 23.4&-1.15(0.02)\nl
53312.3&27.00(0.50)& 25.9&-1.16(0.02)\nl
\multicolumn{1}{c}{ } & \multicolumn{1}{l}{$F110W$} &
\multicolumn{2}{r}{$F110W\rightarrow B$} \nl
53277.6&24.22(0.05)& 11.2&-1.71(0.07)\nl
53286.6&24.59(0.09)& 15.0&-1.77(0.12)\nl
\multicolumn{1}{c}{ } & \multicolumn{1}{l}{$F160W$} &
\multicolumn{2}{r}{$F160W\rightarrow R$} \nl
53277.7&23.99(0.05)& 11.3&-1.93(0.05)\nl
53285.6&24.37(0.12)& 14.6&-1.90(0.02)\nl
53217.9& flux:0.180(0.06)&-13.9&------(---)\nl
\hline
\multicolumn{4}{c}{HST05Fer ($z=1.02$)} \nl
\hline
\multicolumn{1}{c}{ } & \multicolumn{1}{l}{$F775W$} &
\multicolumn{2}{r}{$F775W\rightarrow U$} \nl
53486.8&24.55(0.10)& 10.1&-0.82(0.04)\nl
\multicolumn{1}{c}{ } & \multicolumn{1}{l}{$F850LP$} &
\multicolumn{2}{r}{$F850LP\rightarrow B$} \nl
53472.6&23.58(0.03)&  3.1&-1.39(0.02)\nl
53486.8&24.14(0.04)& 10.1&-1.40(0.01)\nl
53493.8&24.28(0.06)& 13.6&-1.40(0.02)\nl
53504.3&24.99(0.08)& 18.8&-1.43(0.03)\nl
53513.2&25.24(0.08)& 23.2&-1.45(0.01)\nl
53527.3&26.08(0.17)& 30.2&-1.45(0.01)\nl
\multicolumn{1}{c}{ } & \multicolumn{1}{l}{$F110W$} &
\multicolumn{2}{r}{$F110W\rightarrow V$} \nl
53486.4&23.73(0.05)&  9.9&-1.40(0.05)\nl
53491.2&23.87(0.06)& 12.3&-1.37(0.05)\nl
53427.7& flux:-0.03(0.03)&-19.0&------(---)\nl
\hline
\multicolumn{4}{c}{HST05Koe ($z=1.23$)} \nl
\hline
\multicolumn{1}{c}{ } & \multicolumn{1}{l}{$F775W$} &
\multicolumn{2}{r}{$F775W\rightarrow U$} \nl
53472.7&25.18(0.05)&  8.5&-0.15(0.05)\nl
53486.8&26.11(0.10)& 14.8&-0.07(0.03)\nl
\multicolumn{1}{c}{ } & \multicolumn{1}{l}{$F850LP$} &
\multicolumn{2}{r}{$F850LP\rightarrow B$} \nl
53472.7&24.34(0.06)&  8.5&-1.43(0.01)\nl
53486.8&25.11(0.07)& 14.8&-1.40(0.04)\nl
53493.8&25.34(0.11)& 18.0&-1.34(0.08)\nl
53504.3&25.90(0.15)& 22.7&-1.24(0.04)\nl
53513.2&26.10(0.16)& 26.7&-1.19(0.02)\nl
53527.3&27.10(0.40)& 33.0&-1.18(0.03)\nl
\multicolumn{1}{c}{ } & \multicolumn{1}{l}{$F110W$} &
\multicolumn{2}{r}{$F110W\rightarrow V$} \nl
53485.3&24.48(0.07)& 14.1&-1.46(0.08)\nl
53491.3&24.66(0.09)& 16.8&-1.42(0.09)\nl
\hline
\multicolumn{4}{c}{HST05Dic ($z=0.638$)} \nl
\hline
\multicolumn{1}{c}{ } & \multicolumn{1}{l}{$F606W$} &
\multicolumn{2}{r}{$F606W\rightarrow U$} \nl
53472.5&23.42(0.02)&  4.3&0.160(0.03)\nl
\multicolumn{1}{c}{ } & \multicolumn{1}{l}{$F775W$} &
\multicolumn{2}{r}{$F775W\rightarrow B$} \nl
53472.5&22.61(0.02)&  4.3&-0.96(0.03)\nl
\multicolumn{1}{c}{ } & \multicolumn{1}{l}{$F850LP$} &
\multicolumn{2}{r}{$F850LP\rightarrow V$} \nl
53472.5&22.44(0.02)&  4.3&-1.06(0.02)\nl
53530.8&23.97(0.02)& 39.9&-1.09(0.01)\nl
\hline
\multicolumn{4}{c}{HST04Gre ($z=1.14$)} \nl
\hline
\multicolumn{1}{c}{ } & \multicolumn{1}{l}{$F775W$} &
\multicolumn{2}{r}{$F775W\rightarrow U$} \nl
53265.1&23.61(0.04)&  1.0&-0.46(0.01)\nl
53288.5&24.58(0.05)& 12.0&-0.46(0.01)\nl
53299.5&25.22(0.07)& 17.1&-0.46(0.01)\nl
53313.2&25.57(0.10)& 23.5&-0.47(0.01)\nl
\multicolumn{1}{c}{ } & \multicolumn{1}{l}{$F850LP$} &
\multicolumn{2}{r}{$F850LP\rightarrow B$} \nl
53265.1&23.25(0.02)&  1.0&-1.51(0.02)\nl
53275.6&23.38(0.02)&  6.0&-1.49(0.02)\nl
53288.5&24.00(0.05)& 12.0&-1.47(0.03)\nl
53299.5&24.68(0.10)& 17.1&-1.43(0.05)\nl
53313.2&24.98(0.07)& 23.5&-1.36(0.02)\nl
53356.3&26.26(0.31)& 43.7&-1.32(0.03)\nl
53378.4&26.31(0.38)& 54.0&-1.37(0.02)\nl
53388.6&25.94(0.26)& 58.8&-1.36(0.01)\nl
53398.1&26.79(0.32)& 63.2&-1.37(0.01)\nl
53404.2&26.10(0.23)& 66.0&-1.37(0.01)\nl
\multicolumn{1}{c}{ } & \multicolumn{1}{l}{$F110W$} &
\multicolumn{2}{r}{$F110W\rightarrow V$} \nl
53290.5&23.62(0.05)& 12.9&-1.49(0.06)\nl
53301.2&23.90(0.05)& 17.9&-1.42(0.05)\nl
53217.2& flux:-0.05(0.04)&-21.2&------(---)\nl
\hline
\multicolumn{4}{c}{HST04Omb ($z=0.975$)} \nl
\hline
\multicolumn{1}{c}{ } & \multicolumn{1}{l}{$F775W$} &
\multicolumn{2}{r}{$F775W\rightarrow U$} \nl
53356.3&23.82(0.05)& -5.5&-0.58(0.05)\nl
53373.7&23.81(0.05)&  3.2&-0.74(0.07)\nl
53378.5&24.06(0.06)&  5.7&-0.79(0.05)\nl
53388.6&24.46(0.06)& 10.8&-0.86(0.05)\nl
53398.1&24.77(0.07)& 15.6&-0.93(0.02)\nl
53404.1&25.09(0.08)& 18.6&-0.94(0.01)\nl
\multicolumn{1}{c}{ } & \multicolumn{1}{l}{$F850LP$} &
\multicolumn{2}{r}{$F850LP\rightarrow B$} \nl
53356.3&23.78(0.04)& -5.5&-1.26(0.02)\nl
53373.7&23.58(0.03)&  3.2&-1.30(0.03)\nl
53378.4&23.61(0.03)&  5.6&-1.32(0.02)\nl
53388.6&23.99(0.05)& 10.8&-1.34(0.03)\nl
53398.1&24.31(0.06)& 15.6&-1.39(0.05)\nl
53404.2&24.40(0.06)& 18.7&-1.44(0.06)\nl
53407.9&24.50(0.07)& 20.6&-1.47(0.05)\nl
\multicolumn{1}{c}{ } & \multicolumn{1}{l}{$F110W$} &
\multicolumn{2}{r}{$F110W\rightarrow V$} \nl
53368.9&23.31(0.04)&  0.8&-1.43(0.02)\nl
53377.8&23.54(0.04)&  5.3&-1.41(0.06)\nl
\hline
\multicolumn{4}{c}{HST05Red ($z=1.19$)} \nl
\hline
\multicolumn{1}{c}{ } & \multicolumn{1}{l}{$F775W$} &
\multicolumn{2}{r}{$F775W\rightarrow U$} \nl
53426.6&25.52(0.05)& 12.0&-0.21(0.03)\nl
53436.6&25.92(0.07)& 16.6&-0.19(0.01)\nl
\multicolumn{1}{c}{ } & \multicolumn{1}{l}{$F850LP$} &
\multicolumn{2}{r}{$F850LP\rightarrow B$} \nl
53426.6&24.42(0.04)& 12.0&-1.38(0.04)\nl
53436.6&24.74(0.06)& 16.6&-1.34(0.07)\nl
53439.6&25.10(0.11)& 18.0&-1.30(0.07)\nl
53445.5&25.47(0.14)& 20.7&-1.26(0.07)\nl
53454.8&26.22(0.21)& 24.9&-1.20(0.02)\nl
53463.3&26.58(0.25)& 28.8&-1.19(0.03)\nl
53471.7&26.75(0.34)& 32.6&-1.21(0.02)\nl
53472.9&26.65(0.20)& 33.2&-1.21(0.03)\nl
\multicolumn{1}{c}{ } & \multicolumn{1}{l}{$F110W$} &
\multicolumn{2}{r}{$F110W\rightarrow V$} \nl
53440.4&24.31(0.05)& 18.3&-1.30(0.05)\nl
\multicolumn{1}{c}{ } & \multicolumn{1}{l}{$F160W$} &
\multicolumn{2}{r}{$F160W\rightarrow R$} \nl
53440.5&23.51(0.08)& 18.4&-1.69(0.13)\nl
\hline
\multicolumn{4}{c}{HST05Lan ($z=1.23$)} \nl
\hline
\multicolumn{1}{c}{ } & \multicolumn{1}{l}{$F775W$} &
\multicolumn{2}{r}{$F775W\rightarrow U$} \nl
53427.1&25.50(0.05)& -2.4&-0.24(0.05)\nl
53436.6&25.75(0.05)&  1.8&-0.19(0.05)\nl
53436.7&25.88(0.10)&  1.8&-0.19(0.05)\nl
\multicolumn{1}{c}{ } & \multicolumn{1}{l}{$F850LP$} &
\multicolumn{2}{r}{$F850LP\rightarrow B$} \nl
53427.2&24.75(0.10)& -2.4&-1.40(0.03)\nl
53436.6&24.56(0.08)&  1.8&-1.37(0.03)\nl
53439.6&24.72(0.09)&  3.1&-1.36(0.04)\nl
53445.5&24.96(0.06)&  5.8&-1.34(0.02)\nl
53454.8&25.29(0.08)&  9.9&-1.33(0.02)\nl
53463.3&25.69(0.18)& 13.7&-1.31(0.04)\nl
53471.7&25.96(0.26)& 17.5&-1.26(0.08)\nl
53473.6&26.78(0.42)& 18.4&-1.24(0.08)\nl
\multicolumn{1}{c}{ } & \multicolumn{1}{l}{$F110W$} &
\multicolumn{2}{r}{$F110W\rightarrow V$} \nl
53439.4&24.40(0.06)&  3.0&-1.56(0.05)\nl
53445.4&24.43(0.06)&  5.7&-1.53(0.06)\nl
\multicolumn{1}{c}{ } & \multicolumn{1}{l}{$F160W$} &
\multicolumn{2}{r}{$F160W\rightarrow R$} \nl
53438.4&24.36(0.14)&  2.6&-1.80(0.07)\nl
53444.4&24.43(0.14)&  5.3&-1.73(0.06)\nl
\hline
\multicolumn{4}{c}{HST04Tha ($z=0.954$)} \nl
\hline
\multicolumn{1}{c}{ } & \multicolumn{1}{l}{$F775W$} &
\multicolumn{2}{r}{$F775W\rightarrow U$} \nl
53196.5&24.19(0.05)&  9.8&-0.91(0.05)\nl
53207.7&25.16(0.10)& 15.5&-0.99(0.02)\nl
\multicolumn{1}{c}{ } & \multicolumn{1}{l}{$F850LP$} &
\multicolumn{2}{r}{$F850LP\rightarrow B$} \nl
53196.5&23.87(0.03)&  9.8&-1.37(0.03)\nl
53207.7&24.37(0.06)& 15.5&-1.43(0.06)\nl
53216.3&24.72(0.11)& 19.9&-1.51(0.06)\nl
53221.7&25.18(0.10)& 22.6&-1.56(0.03)\nl
53231.0&25.36(0.11)& 27.4&-1.58(0.01)\nl
53244.4&25.91(0.14)& 34.3&-1.58(0.02)\nl
53284.6&26.54(0.51)& 54.8&-1.51(0.02)\nl
\multicolumn{1}{c}{ } & \multicolumn{1}{l}{$F110W$} &
\multicolumn{2}{r}{$F110W\rightarrow V$} \nl
53208.2&24.17(0.07)& 15.7&-1.25(0.03)\nl
53146.4& flux:-0.01(0.05)&-15.8&------(---)\nl
\hline
\multicolumn{4}{c}{HST04Rak ($z=0.74$)} \nl
\hline
\multicolumn{1}{c}{ } & \multicolumn{1}{l}{$F606W$} &
\multicolumn{2}{r}{$F606W\rightarrow U$} \nl
53356.3&24.15(0.05)&  2.5&0.362(0.01)\nl
\multicolumn{1}{c}{ } & \multicolumn{1}{l}{$F775W$} &
\multicolumn{2}{r}{$F775W\rightarrow B$} \nl
53356.3&22.88(0.02)&  2.5&-1.11(0.01)\nl
53373.7&23.70(0.03)& 12.5&-1.10(0.01)\nl
53388.6&24.56(0.05)& 21.1&-1.12(0.01)\nl
53398.1&25.11(0.07)& 26.5&-1.12(0.01)\nl
53404.1&25.57(0.10)& 30.0&-1.11(0.01)\nl
53408.0&25.64(0.10)& 32.2&-1.11(0.02)\nl
\multicolumn{1}{c}{ } & \multicolumn{1}{l}{$F850LP$} &
\multicolumn{2}{r}{$F850LP\rightarrow V$} \nl
53356.3&22.96(0.02)&  2.5&-1.10(0.03)\nl
53373.7&23.47(0.02)& 12.5&-1.04(0.03)\nl
53378.4&23.61(0.04)& 15.2&-1.02(0.05)\nl
53388.6&23.92(0.03)& 21.1&-0.95(0.04)\nl
53398.1&24.28(0.05)& 26.5&-0.90(0.03)\nl
53404.2&24.63(0.06)& 30.0&-0.87(0.02)\nl
53404.2&24.42(0.07)& 30.0&-0.87(0.02)\nl
53407.9&24.80(0.08)& 32.2&-0.86(0.01)\nl
53313.0& flux:0.070(0.06)&-22.3&------(---)\nl
\hline
\multicolumn{4}{c}{HST05Zwi ($z=0.521$)} \nl
\hline
\multicolumn{1}{c}{ } & \multicolumn{1}{l}{$F606W$} &
\multicolumn{2}{r}{$F606W\rightarrow B$} \nl
53415.9&22.94(0.02)&  5.3&-0.27(0.03)\nl
53435.0&24.13(0.02)& 17.9&-0.14(0.06)\nl
53443.1&24.52(0.10)& 23.2&-0.04(0.03)\nl
\multicolumn{1}{c}{ } & \multicolumn{1}{l}{$F775W$} &
\multicolumn{2}{r}{$F775W\rightarrow V$} \nl
53403.2&22.40(0.02)& -2.9&-0.86(0.02)\nl
53415.9&22.19(0.10)&  5.3&-0.81(0.04)\nl
53435.1&23.04(0.02)& 17.9&-0.71(0.07)\nl
53443.2&23.30(0.02)& 23.3&-0.61(0.07)\nl
\multicolumn{1}{c}{ } & \multicolumn{1}{l}{$F850LP$} &
\multicolumn{2}{r}{$F850LP\rightarrow R$} \nl
53403.2&22.04(0.01)& -2.9&-0.84(0.04)\nl
53415.9&22.16(0.02)&  5.3&-0.90(0.04)\nl
53435.1&22.95(0.02)& 17.9&-0.97(0.04)\nl
53443.2&22.91(0.02)& 23.3&-0.90(0.06)\nl
\hline
\multicolumn{4}{c}{HST04Haw ($z=0.490$)} \nl
\hline
\multicolumn{1}{c}{ } & \multicolumn{1}{l}{$F606W$} &
\multicolumn{2}{r}{$F606W\rightarrow B$} \nl
53375.7&25.68(0.04)& 49.4&-0.16(0.01)\nl
53427.5&26.16(0.04)& 84.2&-0.17(0.01)\nl
53472.5&26.82(0.04)&114.4&-0.17(0.01)\nl
\multicolumn{1}{c}{ } & \multicolumn{1}{l}{$F775W$} &
\multicolumn{2}{r}{$F775W\rightarrow V$} \nl
53284.0&23.97(0.03)&-12.0&-0.80(0.01)\nl
53332.9&23.15(0.02)& 20.7&-0.71(0.03)\nl
53375.7&24.45(0.02)& 49.4&-0.62(0.02)\nl
53427.5&25.08(0.05)& 84.2&-0.70(0.01)\nl
\multicolumn{1}{c}{ } & \multicolumn{1}{l}{$F850LP$} &
\multicolumn{2}{r}{$F850LP\rightarrow R$} \nl
53284.0&23.81(0.03)&-12.0&-0.84(0.05)\nl
53333.0&22.90(0.01)& 20.8&-0.91(0.04)\nl
53375.7&23.83(0.05)& 49.4&-0.82(0.01)\nl
53427.6&24.94(0.12)& 84.3&-0.84(0.01)\nl
53472.5&25.82(0.21)&114.4&-0.84(0.01)\nl
53243.9& flux:0.020(0.08)&-38.9&------(---)\nl
\hline
\multicolumn{4}{c}{HST04Kur ($z=0.359$)} \nl
\hline
\multicolumn{1}{c}{ } & \multicolumn{1}{l}{$F775W$} &
\multicolumn{2}{r}{$F775W\rightarrow V$} \nl
53355.2&23.60(0.02)& -5.3&-0.70(0.02)\nl
53366.1&23.28(0.05)&  2.6&-0.71(0.02)\nl
53375.9&23.74(0.05)&  9.9&-0.73(0.01)\nl
53404.0&24.77(0.04)& 30.5&-0.85(0.01)\nl
53415.4&25.05(0.05)& 38.9&-0.84(0.01)\nl
53425.9&25.03(0.05)& 46.6&-0.85(0.01)\nl
\multicolumn{1}{c}{ } & \multicolumn{1}{l}{$F850LP$} &
\multicolumn{2}{r}{$F850LP\rightarrow R$} \nl
53355.3&23.04(0.02)& -5.2&-0.70(0.02)\nl
53366.1&22.79(0.02)&  2.6&-0.68(0.02)\nl
53375.9&23.12(0.02)&  9.9&-0.65(0.01)\nl
53404.0&24.09(0.05)& 30.5&-0.72(0.01)\nl
53415.5&24.18(0.06)& 39.0&-0.70(0.01)\nl
53425.9&24.26(0.06)& 46.6&-0.69(0.02)\nl
\hline
\multicolumn{4}{c}{HST04Yow ($z=0.46$)} \nl
\hline
\multicolumn{1}{c}{ } & \multicolumn{1}{l}{$F775W$} &
\multicolumn{2}{r}{$F775W\rightarrow V$} \nl
53145.1&22.73(0.01)&  9.4&-0.75(0.02)\nl
53157.0&23.18(0.01)& 17.6&-0.72(0.03)\nl
53169.4&23.70(0.01)& 26.1&-0.67(0.02)\nl
53195.2&24.42(0.30)& 43.7&-0.66(0.02)\nl
\multicolumn{1}{c}{ } & \multicolumn{1}{l}{$F850LP$} &
\multicolumn{2}{r}{$F850LP\rightarrow R$} \nl
53145.2&22.53(0.01)&  9.5&-0.85(0.01)\nl
53157.1&23.04(0.01)& 17.7&-0.86(0.04)\nl
53169.4&23.31(0.03)& 26.1&-0.76(0.02)\nl
53195.2&24.06(0.05)& 43.7&-0.78(0.02)\nl
53216.3&24.80(0.09)& 58.2&-0.80(0.02)\nl
53221.7&24.90(0.06)& 61.9&-0.81(0.01)\nl
53284.4&25.49(0.17)&104.8&-0.82(0.01)\nl
53332.7&25.79(0.20)&137.9&-0.82(0.01)\nl
53097.2& flux:0.050(0.06)&-23.3&------(---)\nl
\hline
\multicolumn{4}{c}{HST04Man ($z=0.854$)} \nl
\hline
\multicolumn{1}{c}{ } & \multicolumn{1}{l}{$F775W$} &
\multicolumn{2}{r}{$F775W\rightarrow B$} \nl
53146.3&23.38(0.04)& -1.0&-1.17(0.03)\nl
53157.0&23.55(0.04)&  4.7&-1.13(0.02)\nl
53169.4&24.12(0.05)& 11.4&-1.11(0.03)\nl
53195.6&25.92(0.10)& 25.5&-0.91(0.01)\nl
\multicolumn{1}{c}{ } & \multicolumn{1}{l}{$F850LP$} &
\multicolumn{2}{r}{$F850LP\rightarrow V$} \nl
53146.3&23.35(0.05)& -1.0&-1.31(0.01)\nl
53157.1&23.45(0.02)&  4.8&-1.28(0.07)\nl
53169.4&23.69(0.04)& 11.4&-1.16(0.07)\nl
53195.6&25.04(0.10)& 25.5&-0.75(0.06)\nl
53207.7&25.94(0.21)& 32.0&-0.68(0.02)\nl
53098.0& flux:0.030(0.04)&-27.0&------(---)\nl
\hline
\multicolumn{4}{c}{HST05Spo ($z=0.839$)} \nl
\hline
\multicolumn{1}{c}{ } & \multicolumn{1}{l}{$F775W$} &
\multicolumn{2}{r}{$F775W\rightarrow B$} \nl
53427.1&24.06(0.04)& 14.6&-1.12(0.04)\nl
53427.3&23.95(0.04)& 14.7&-1.12(0.04)\nl
53436.6&24.67(0.05)& 19.8&-1.04(0.06)\nl
53436.6&24.76(0.10)& 19.8&-1.04(0.06)\nl
53473.6&26.30(0.30)& 39.9&-0.98(0.01)\nl
\multicolumn{1}{c}{ } & \multicolumn{1}{l}{$F850LP$} &
\multicolumn{2}{r}{$F850LP\rightarrow V$} \nl
53376.8&24.78(0.09)&-12.6&-1.17(0.04)\nl
53427.2&23.66(0.03)& 14.7&-1.09(0.10)\nl
53427.4&23.61(0.04)& 14.8&-1.09(0.10)\nl
53436.6&24.00(0.05)& 19.8&-0.93(0.10)\nl
53439.6&24.17(0.07)& 21.4&-0.88(0.08)\nl
53445.5&24.35(0.06)& 24.6&-0.80(0.08)\nl
53454.8&24.91(0.10)& 29.7&-0.71(0.03)\nl
53463.3&25.11(0.09)& 34.3&-0.70(0.04)\nl
53471.7&25.37(0.12)& 38.9&-0.73(0.02)\nl
53473.6&25.31(0.21)& 39.9&-0.73(0.03)\nl
\hline
\multicolumn{4}{c}{HST04Eag ($z=1.02$)} \nl
\hline
\multicolumn{1}{c}{ } & \multicolumn{1}{l}{$F775W$} &
\multicolumn{2}{r}{$F775W\rightarrow U$} \nl
53284.9&23.95(0.05)& -4.4&-0.63(0.03)\nl
53332.1&26.59(0.60)& 18.9&-0.84(0.01)\nl
\multicolumn{1}{c}{ } & \multicolumn{1}{l}{$F850LP$} &
\multicolumn{2}{r}{$F850LP\rightarrow B$} \nl
53284.9&23.76(0.05)& -4.4&-1.34(0.02)\nl
53296.9&23.60(0.02)&  1.5&-1.36(0.02)\nl
53305.6&23.70(0.03)&  5.8&-1.37(0.01)\nl
53316.5&24.19(0.04)& 11.2&-1.38(0.02)\nl
53326.2&24.62(0.07)& 16.0&-1.40(0.03)\nl
53332.2&24.93(0.11)& 19.0&-1.42(0.03)\nl
53334.9&24.96(0.07)& 20.3&-1.42(0.03)\nl
\multicolumn{1}{c}{ } & \multicolumn{1}{l}{$F110W$} &
\multicolumn{2}{r}{$F110W\rightarrow V$} \nl
53297.5&23.57(0.07)&  1.8&-1.49(0.02)\nl
53306.3&23.60(0.07)&  6.1&-1.45(0.06)\nl
\hline
\multicolumn{4}{c}{HST05Gab ($z=1.12$)} \nl
\hline
\multicolumn{1}{c}{ } & \multicolumn{1}{l}{$F775W$} &
\multicolumn{2}{r}{$F775W\rightarrow U$} \nl
53474.5&24.52(0.05)& -6.0&-0.50(0.01)\nl
53485.7&24.16(0.10)& -0.7&-0.50(0.02)\nl
\multicolumn{1}{c}{ } & \multicolumn{1}{l}{$F850LP$} &
\multicolumn{2}{r}{$F850LP\rightarrow B$} \nl
53474.5&23.83(0.04)& -6.0&-1.48(0.01)\nl
53485.7&23.58(0.03)& -0.7&-1.48(0.02)\nl
53493.9&23.66(0.03)&  3.0&-1.48(0.02)\nl
53504.2&23.88(0.04)&  7.9&-1.47(0.02)\nl
53513.3&24.24(0.04)& 12.2&-1.44(0.03)\nl
53530.8&25.38(0.13)& 20.5&-1.38(0.03)\nl
\multicolumn{1}{c}{ } & \multicolumn{1}{l}{$F110W$} &
\multicolumn{2}{r}{$F110W\rightarrow V$} \nl
53486.4&23.68(0.07)& -0.4&-1.57(0.02)\nl
53492.3&23.60(0.07)&  2.3&-1.57(0.03)\nl
\hline
\multicolumn{4}{c}{HST05Str ($z=1.01$)} \nl
\hline
\multicolumn{1}{c}{ } & \multicolumn{1}{l}{$F775W$} &
\multicolumn{2}{r}{$F775W\rightarrow U$} \nl
53474.5&23.93(0.04)& -0.0&-0.67(0.08)\nl
53485.7&24.25(0.10)&  5.4&-0.76(0.03)\nl
\multicolumn{1}{c}{ } & \multicolumn{1}{l}{$F850LP$} &
\multicolumn{2}{r}{$F850LP\rightarrow B$} \nl
53474.5&23.67(0.03)& -0.0&-1.33(0.04)\nl
53485.7&23.90(0.04)&  5.4&-1.36(0.01)\nl
53493.9&24.14(0.05)&  9.5&-1.36(0.01)\nl
53504.2&24.51(0.08)& 14.7&-1.38(0.03)\nl
53513.3&25.08(0.08)& 19.2&-1.42(0.03)\nl
\multicolumn{1}{c}{ } & \multicolumn{1}{l}{$F110W$} &
\multicolumn{2}{r}{$F110W\rightarrow V$} \nl
53485.4&23.85(0.12)&  5.3&-1.45(0.06)\nl
53492.2&23.99(0.12)&  8.7&-1.41(0.04)\nl
53426.8& flux:0.043(0.03)&-23.8&------(---)\nl
\hline
\multicolumn{4}{c}{HST04Sas ($z=1.39$)} \nl
\hline
\multicolumn{1}{c}{ } & \multicolumn{1}{l}{$F775W$} &
\multicolumn{2}{r}{$F775W\rightarrow U$} \nl
148.670&26.22(0.25)& -2.4&0.091(0.14)\nl
\multicolumn{2}{r}{$F850lp\rightarrow U$} \nl
148.700&24.75(0.07)& -2.4&-0.96(0.06)\nl
156.200&25.01(0.10)&  0.7&-0.99(0.07)\nl
163.600&25.13(0.08)&  3.8&-1.04(0.05)\nl
169.300&25.74(0.15)&  6.2&-1.07(0.03)\nl
176.100&25.54(0.14)&  9.0&-1.08(0.03)\nl
183.500&25.98(0.20)& 12.1&-1.10(0.03)\nl
190.400&26.29(0.25)& 15.0&-1.12(0.02)\nl
\multicolumn{1}{c}{ } & \multicolumn{1}{l}{$F110W$} &
\multicolumn{2}{r}{$F110W\rightarrow B$} \nl
164.050&24.50(0.07)&  4.0&-1.68(0.05)\nl
168.060&24.48(0.07)&  5.6&-1.69(0.05)\nl
\multicolumn{1}{c}{ } & \multicolumn{1}{l}{$F160W$} &
\multicolumn{2}{r}{$F160W\rightarrow R$} \nl
164.180&24.06(0.10)&  4.0&-2.01(0.04)\nl
169.050&24.05(0.10)&  6.0&-2.00(0.03)\nl
98.6000& flux:-0.06(0.04)&-23.3&------(---)\nl
\hline
\multicolumn{4}{c}{2003aj ($z=1.307$)} \nl
\hline
\multicolumn{1}{c}{ } & \multicolumn{1}{l}{$F775W$} &
\multicolumn{2}{r}{$F775W\rightarrow U$} \nl
52673.1&26.62(0.10)&  1.7&-0.05(0.08)\nl
\multicolumn{1}{c}{ } & \multicolumn{1}{l}{$F850LP$} &
\multicolumn{2}{r}{$F850LP\rightarrow B$} \nl
52673.2&25.31(0.08)&  1.7&-1.25(0.03)\nl
52680.2&25.53(0.06)&  4.8&-1.21(0.03)\nl
52694.5&26.38(0.16)& 11.0&-1.16(0.05)\nl
\multicolumn{1}{c}{ } & \multicolumn{1}{l}{$F110W$} &
\multicolumn{2}{r}{$F110W\rightarrow V$} \nl
52684.2&25.13(0.11)&  6.5&-1.49(0.06)\nl
\multicolumn{1}{c}{ } & \multicolumn{1}{l}{$F160W$} &
\multicolumn{2}{r}{$F160W\rightarrow R$} \nl
52685.1&24.69(0.12)&  6.9&-1.83(0.04)\nl
52627.4& flux:0.062(0.03)&-18.0&------(---)\nl
\hline
\multicolumn{4}{c}{2002fx ($z=1.400$)} \nl
\hline
\multicolumn{1}{c}{ } & \multicolumn{1}{l}{$F775W$} &
\multicolumn{2}{r}{$F775W\rightarrow U$} \nl
52495.0&28.02(1.00)&-14.2&0.008(0.07)\nl
52537.8&27.12(0.25)&  3.6&0.300(0.10)\nl
52580.0&29.02(1.00)& 21.2&0.812(0.05)\nl
\multicolumn{1}{c}{ } & \multicolumn{1}{l}{$F850LP$} &
\multicolumn{2}{r}{$F850LP\rightarrow B$} \nl
52537.8&25.21(0.07)&  3.6&-1.20(0.09)\nl
52580.5&27.10(0.27)& 21.4&-0.82(0.08)\nl
52490.3& flux:0.016(0.03)&-16.1&------(---)\nl
\hline
\multicolumn{4}{c}{2003eq ($z=0.84$)} \nl
\hline
\multicolumn{1}{c}{ } & \multicolumn{1}{l}{$F606W$} &
\multicolumn{2}{r}{$F606W\rightarrow U$} \nl
52783.7&24.58(0.01)&  0.8&0.707(0.05)\nl
\multicolumn{1}{c}{ } & \multicolumn{1}{l}{$F775W$} &
\multicolumn{2}{r}{$F775W\rightarrow B$} \nl
52783.7&23.22(0.01)&  0.8&-1.18(0.03)\nl
52799.1&23.67(0.02)&  9.2&-1.14(0.02)\nl
52807.3&24.15(0.04)& 13.6&-1.12(0.04)\nl
52819.8&25.05(0.07)& 20.4&-1.02(0.06)\nl
52838.3&26.22(0.30)& 30.5&-0.97(0.02)\nl
\multicolumn{1}{c}{ } & \multicolumn{1}{l}{$F850LP$} &
\multicolumn{2}{r}{$F850LP\rightarrow V$} \nl
52783.7&23.12(0.01)&  0.8&-1.27(0.02)\nl
52792.1&23.28(0.03)&  5.4&-1.22(0.07)\nl
52799.1&23.45(0.02)&  9.2&-1.16(0.05)\nl
52807.3&23.64(0.07)& 13.6&-1.10(0.09)\nl
52819.8&24.30(0.10)& 20.4&-0.89(0.08)\nl
52838.3&25.34(0.12)& 30.5&-0.69(0.02)\nl
\multicolumn{1}{c}{ } & \multicolumn{1}{l}{$F110W$} &
\multicolumn{2}{r}{$F110W\rightarrow R$} \nl
52792.9&23.20(0.10)&  5.8&-1.22(0.01)\nl
52735.5& flux:0.087(0.10)&-25.3&------(---)\nl
\hline
\multicolumn{4}{c}{2003es ($z=0.954$)} \nl
\hline
\multicolumn{1}{c}{ } & \multicolumn{1}{l}{$F775W$} &
\multicolumn{2}{r}{$F775W\rightarrow U$} \nl
52784.5&24.12(0.03)&  8.0&-0.87(0.05)\nl
\multicolumn{1}{c}{ } & \multicolumn{1}{l}{$F850LP$} &
\multicolumn{2}{r}{$F850LP\rightarrow B$} \nl
52784.5&23.73(0.02)&  8.0&-1.33(0.03)\nl
52792.4&24.09(0.06)& 12.0&-1.36(0.06)\nl
52801.3&24.68(0.06)& 16.6&-1.42(0.07)\nl
52807.9&24.98(0.09)& 19.9&-1.49(0.06)\nl
52821.0&25.63(0.14)& 26.6&-1.56(0.01)\nl
52838.1&25.96(0.24)& 35.4&-1.58(0.02)\nl
\multicolumn{1}{c}{ } & \multicolumn{1}{l}{$F110W$} &
\multicolumn{2}{r}{$F110W\rightarrow V$} \nl
52792.8&24.20(0.08)& 12.2&-1.28(0.05)\nl
52734.6& flux:0.005(0.04)&-17.5&------(---)\nl
\hline
\multicolumn{4}{c}{2003az ($z=1.265$)} \nl
\hline
\multicolumn{1}{c}{ } & \multicolumn{1}{l}{$F775W$} &
\multicolumn{2}{r}{$F775W\rightarrow U$} \nl
52690.9&25.10(0.05)&  3.0&-0.15(0.06)\nl
52701.2&25.46(0.05)&  7.5&-0.09(0.05)\nl
\multicolumn{1}{c}{ } & \multicolumn{1}{l}{$F850LP$} &
\multicolumn{2}{r}{$F850LP\rightarrow B$} \nl
52690.9&24.36(0.04)&  3.0&-1.40(0.03)\nl
52701.2&24.48(0.04)&  7.5&-1.37(0.02)\nl
52709.1&24.64(0.05)& 11.0&-1.35(0.03)\nl
52716.9&25.09(0.06)& 14.4&-1.32(0.05)\nl
52726.5&25.52(0.08)& 18.7&-1.25(0.08)\nl
52733.2&25.75(0.09)& 21.6&-1.17(0.07)\nl
\multicolumn{1}{c}{ } & \multicolumn{1}{l}{$F110W$} &
\multicolumn{2}{r}{$F110W\rightarrow V$} \nl
52703.6&24.10(0.06)&  8.6&-1.51(0.06)\nl
52710.6&24.25(0.06)& 11.7&-1.46(0.07)\nl
52642.2& flux:0.054(0.03)&-18.4&------(---)\nl
\hline
\multicolumn{4}{c}{2002kc ($z=0.216$)} \nl
\hline
\multicolumn{1}{c}{ } & \multicolumn{1}{l}{$F606W$} &
\multicolumn{2}{r}{$F606W\rightarrow V$} \nl
52629.6&22.30(0.01)& -7.4&-0.23(0.05)\nl
52672.3&23.28(0.01)& 27.6&0.222(0.02)\nl
\multicolumn{1}{c}{ } & \multicolumn{1}{l}{$F775W$} &
\multicolumn{2}{r}{$F775W\rightarrow R$} \nl
52629.6&21.81(0.02)& -7.4&-0.43(0.01)\nl
52642.5&21.35(0.10)&  3.1&-0.47(0.04)\nl
52672.3&22.11(0.01)& 27.6&-0.43(0.01)\nl
\multicolumn{1}{c}{ } & \multicolumn{1}{l}{$F850LP$} &
\multicolumn{2}{r}{$F850LP\rightarrow I$} \nl
52629.7&21.70(0.01)& -7.4&-0.45(0.01)\nl
52672.3&21.87(0.01)& 27.6&-0.19(0.07)\nl
\hline
\multicolumn{4}{c}{2003eb ($z=0.90$)} \nl
\hline
\multicolumn{1}{c}{ } & \multicolumn{1}{l}{$F606W$} &
\multicolumn{2}{r}{$F606W\rightarrow U$} \nl
52734.6&24.26(0.02)& -1.1&0.824(0.07)\nl
52783.5&27.07(0.15)& 24.5&1.233(0.02)\nl
\multicolumn{1}{c}{ } & \multicolumn{1}{l}{$F775W$} &
\multicolumn{2}{r}{$F775W\rightarrow B$} \nl
52734.6&23.05(0.02)& -1.1&-1.14(0.05)\nl
52745.6&23.15(0.02)&  4.5&-1.10(0.03)\nl
52783.5&25.22(0.06)& 24.5&-0.83(0.02)\nl
52799.1&25.92(0.10)& 32.7&-0.80(0.03)\nl
\multicolumn{1}{c}{ } & \multicolumn{1}{l}{$F850LP$} &
\multicolumn{2}{r}{$F850LP\rightarrow V$} \nl
52734.7&22.83(0.01)& -1.1&-1.32(0.01)\nl
52745.7&22.84(0.01)&  4.6&-1.27(0.09)\nl
52751.2&22.98(0.01)&  7.5&-1.20(0.06)\nl
52763.6&23.47(0.01)& 14.0&-1.09(0.14)\nl
52774.0&24.02(0.04)& 19.5&-0.92(0.13)\nl
52783.6&24.39(0.04)& 24.5&-0.75(0.10)\nl
52792.1&24.77(0.11)& 29.0&-0.60(0.03)\nl
52792.4&24.63(0.08)& 29.2&-0.60(0.03)\nl
52799.1&25.06(0.15)& 32.7&-0.55(0.02)\nl
52801.3&24.84(0.06)& 33.9&-0.56(0.04)\nl
52807.9&25.03(0.10)& 37.3&-0.61(0.02)\nl
52821.0&25.40(0.09)& 44.2&-0.63(0.02)\nl
52838.1&25.36(0.10)& 53.2&-0.69(0.05)\nl
52692.5& flux:0.058(0.05)&-23.3&------(---)\nl
\hline
\multicolumn{4}{c}{2003XX ($z=0.935$)} \nl
\hline
\multicolumn{1}{c}{ } & \multicolumn{1}{l}{$F606W$} &
\multicolumn{2}{r}{$F606W\rightarrow U$} \nl
52733.6&25.17(0.05)&  5.7&1.042(0.09)\nl
\multicolumn{1}{c}{ } & \multicolumn{1}{l}{$F775W$} &
\multicolumn{2}{r}{$F775W\rightarrow B$} \nl
52733.6&23.57(0.05)&  5.7&-1.08(0.02)\nl
52745.6&24.12(0.05)& 11.9&-1.05(0.04)\nl
52783.5&26.40(0.25)& 31.5&-0.77(0.02)\nl
\multicolumn{1}{c}{ } & \multicolumn{1}{l}{$F850LP$} &
\multicolumn{2}{r}{$F850LP\rightarrow V$} \nl
52692.5&26.20(0.13)&-15.4&-1.12(0.06)\nl
52733.7&23.26(0.01)&  5.7&-1.27(0.10)\nl
52745.7&23.63(0.01)& 11.9&-1.14(0.10)\nl
52751.2&23.88(0.02)& 14.8&-1.07(0.15)\nl
52763.6&24.56(0.04)& 21.2&-0.77(0.11)\nl
52774.0&24.87(0.05)& 26.6&-0.58(0.05)\nl
52783.6&25.20(0.06)& 31.5&-0.48(0.02)\nl
52792.1&25.36(0.13)& 35.9&-0.50(0.05)\nl
52799.1&26.04(0.20)& 39.5&-0.53(0.04)\nl
52807.3&26.45(0.19)& 43.8&-0.56(0.02)\nl
\hline
\multicolumn{4}{c}{2002hr ($z=0.526$)} \nl
\hline
\multicolumn{1}{c}{ } & \multicolumn{1}{l}{$F606W$} &
\multicolumn{2}{r}{$F606W\rightarrow U$} \nl
52579.6&24.04(0.03)& -6.4&0.103(0.08)\nl
52629.5&25.82(0.15)& 26.2&-0.36(0.03)\nl
52674.1&27.43(0.30)& 55.5&-0.35(0.07)\nl
\multicolumn{1}{c}{ } & \multicolumn{1}{l}{$F775W$} &
\multicolumn{2}{r}{$F775W\rightarrow B$} \nl
52579.6&23.56(0.03)& -6.4&-0.84(0.01)\nl
52589.7&23.20(0.12)&  0.1&-0.87(0.05)\nl
52590.0&23.28(0.11)&  0.3&-0.87(0.05)\nl
52596.6&23.44(0.14)&  4.7&-0.93(0.09)\nl
52614.5&23.91(0.08)& 16.4&-1.27(0.18)\nl
52629.5&24.42(0.05)& 26.2&-1.66(0.03)\nl
52674.1&25.67(0.10)& 55.5&-1.58(0.02)\nl
\multicolumn{1}{c}{ } & \multicolumn{1}{l}{$F850LP$} &
\multicolumn{2}{r}{$F850LP\rightarrow V$} \nl
52579.7&23.38(0.02)& -6.3&-0.96(0.05)\nl
52629.5&23.76(0.03)& 26.2&-1.18(0.04)\nl
52674.1&25.04(0.10)& 55.5&-1.14(0.02)\nl
\hline
\multicolumn{4}{c}{2003bd ($z=0.67$)} \nl
\hline
\multicolumn{1}{c}{ } & \multicolumn{1}{l}{$F606W$} &
\multicolumn{2}{r}{$F606W\rightarrow U$} \nl
52691.9&24.57(0.04)&  8.1&0.162(0.02)\nl
52735.4&27.50(0.12)& 34.1&0.170(0.02)\nl
\multicolumn{1}{c}{ } & \multicolumn{1}{l}{$F775W$} &
\multicolumn{2}{r}{$F775W\rightarrow B$} \nl
52691.9&23.41(0.03)&  8.1&-1.05(0.02)\nl
52735.4&25.77(0.10)& 34.1&-1.24(0.02)\nl
52745.6&26.05(0.10)& 40.2&-1.25(0.01)\nl
\multicolumn{1}{c}{ } & \multicolumn{1}{l}{$F850LP$} &
\multicolumn{2}{r}{$F850LP\rightarrow V$} \nl
52692.0&23.18(0.01)&  8.1&-1.07(0.01)\nl
52735.5&24.82(0.08)& 34.2&-1.02(0.01)\nl
52745.7&25.14(0.08)& 40.3&-1.02(0.01)\nl
52751.2&25.21(0.09)& 43.6&-1.02(0.01)\nl
52763.6&25.47(0.14)& 51.0&-1.03(0.01)\nl
52774.0&25.56(0.19)& 57.2&-1.04(0.01)\nl
52792.1&25.96(0.21)& 68.1&-1.04(0.01)\nl
52807.3&25.85(0.26)& 77.2&-1.04(0.01)\nl
52819.8&25.88(0.27)& 84.6&-1.04(0.01)\nl
52838.3&26.57(0.31)& 95.7&-1.04(0.01)\nl
52642.6& flux:0.057(0.03)&-21.4&------(---)\nl
\hline
\multicolumn{4}{c}{2002kd ($z=0.735$)} \nl
\hline
\multicolumn{1}{c}{ } & \multicolumn{1}{l}{$F606W$} &
\multicolumn{2}{r}{$F606W\rightarrow U$} \nl
52629.4&24.96(0.03)& -8.9&0.335(0.01)\nl
52673.5&25.81(0.05)& 16.4&0.335(0.03)\nl
\multicolumn{1}{c}{ } & \multicolumn{1}{l}{$F775W$} &
\multicolumn{2}{r}{$F775W\rightarrow B$} \nl
52629.4&23.78(0.02)& -8.9&-1.06(0.04)\nl
52644.5&22.96(0.17)& -0.2&-1.10(0.03)\nl
52645.5&22.76(0.10)&  0.3&-1.10(0.03)\nl
52673.5&24.18(0.05)& 16.4&-1.11(0.02)\nl
\multicolumn{1}{c}{ } & \multicolumn{1}{l}{$F850LP$} &
\multicolumn{2}{r}{$F850LP\rightarrow V$} \nl
52629.5&23.76(0.02)& -8.8&-1.10(0.01)\nl
52639.4&23.09(0.01)& -3.1&-1.12(0.03)\nl
52673.7&23.68(0.02)& 16.5&-1.00(0.05)\nl
\hline
\multicolumn{4}{c}{2003be ($z=0.640$)} \nl
\hline
\multicolumn{1}{c}{ } & \multicolumn{1}{l}{$F606W$} &
\multicolumn{2}{r}{$F606W\rightarrow U$} \nl
52692.0&24.47(0.03)& 12.1&0.096(0.03)\nl
52732.5&27.26(0.20)& 36.8&0.086(0.03)\nl
\multicolumn{1}{c}{ } & \multicolumn{1}{l}{$F775W$} &
\multicolumn{2}{r}{$F775W\rightarrow B$} \nl
52641.3&28.02(0.80)&-18.7&-0.68(0.12)\nl
52692.0&23.45(0.03)& 12.1&-1.04(0.06)\nl
52732.5&25.50(0.10)& 36.8&-1.29(0.02)\nl
52784.3&26.09(0.10)& 68.4&-1.22(0.02)\nl
\multicolumn{1}{c}{ } & \multicolumn{1}{l}{$F850LP$} &
\multicolumn{2}{r}{$F850LP\rightarrow V$} \nl
52692.1&23.04(0.01)& 12.2&-1.07(0.01)\nl
52732.5&24.45(0.05)& 36.8&-1.08(0.01)\nl
52784.4&25.31(0.13)& 68.4&-1.07(0.01)\nl
52641.3& flux:0.087(0.10)&-18.7&------(---)\nl
\hline
\multicolumn{4}{c}{2003dy ($z=1.34$)} \nl
\hline
\multicolumn{1}{c}{ } & \multicolumn{1}{l}{$F850LP$} &
\multicolumn{2}{r}{$F850LP\rightarrow U$} \nl
52733.7&24.45(0.04)& -0.1&-0.96(0.09)\nl
52745.7&24.67(0.06)&  4.9&-1.05(0.05)\nl
52751.2&24.77(0.08)&  7.3&-1.09(0.04)\nl
52763.6&25.29(0.11)& 12.6&-1.16(0.04)\nl
52774.0&25.57(0.15)& 17.0&-1.21(0.01)\nl
52783.6&26.62(0.45)& 21.1&-1.22(0.02)\nl
52801.3&26.65(0.26)& 28.7&-1.27(0.03)\nl
\multicolumn{1}{c}{ } & \multicolumn{1}{l}{$F110W$} &
\multicolumn{2}{r}{$F110W\rightarrow B$} \nl
52751.6&24.28(0.07)&  7.4&-1.73(0.05)\nl
52754.6&24.38(0.08)&  8.7&-1.74(0.05)\nl
\multicolumn{1}{c}{ } & \multicolumn{1}{l}{$F160W$} &
\multicolumn{2}{r}{$F160W\rightarrow R$} \nl
52751.7&23.81(0.08)&  7.5&-1.89(0.01)\nl
52692.4& flux:0.037(0.04)&-17.8&------(---)\nl
\hline
\multicolumn{4}{c}{2002ki ($z=1.14$)} \nl
\hline
\multicolumn{1}{c}{ } & \multicolumn{1}{l}{$F775W$} &
\multicolumn{2}{r}{$F775W\rightarrow U$} \nl
52600.8&24.79(0.10)& -0.6&-0.46(0.01)\nl
52643.6&26.82(0.17)& 19.3&-0.46(0.01)\nl
\multicolumn{1}{c}{ } & \multicolumn{1}{l}{$F850LP$} &
\multicolumn{2}{r}{$F850LP\rightarrow B$} \nl
52600.8&23.89(0.04)& -0.6&-1.48(0.02)\nl
52643.6&25.80(0.10)& 19.3&-1.36(0.05)\nl
52652.2&25.99(0.10)& 23.3&-1.32(0.02)\nl
52663.7&26.89(0.20)& 28.7&-1.32(0.02)\nl
\multicolumn{1}{c}{ } & \multicolumn{1}{l}{$F160W$} &
\multicolumn{2}{r}{$F160W\rightarrow I$} \nl
52652.4&24.68(0.25)& 23.4&-1.65(0.03)\nl
52664.5&24.87(0.25)& 29.1&-1.55(0.07)\nl
\hline
\multicolumn{4}{c}{2003ak ($z=1.551$)} \nl
\hline
\multicolumn{1}{c}{ } & \multicolumn{1}{l}{$F850LP$} &
\multicolumn{2}{r}{$F850LP\rightarrow U$} \nl
52673.2&25.68(0.14)&  5.0&-0.68(0.03)\nl
52680.2&25.95(0.10)&  7.7&-0.67(0.03)\nl
52694.5&26.62(0.21)& 13.3&-0.63(0.03)\nl
\multicolumn{1}{c}{ } & \multicolumn{1}{l}{$F110W$} &
\multicolumn{2}{r}{$F110W\rightarrow B$} \nl
52681.1&25.03(0.10)&  8.1&-1.71(0.03)\nl
52693.0&25.03(0.10)& 12.8&-1.74(0.05)\nl
52708.5&25.38(0.15)& 18.8&-1.82(0.07)\nl
52715.4&25.80(0.15)& 21.5&-1.87(0.04)\nl
\multicolumn{1}{c}{ } & \multicolumn{1}{l}{$F160W$} &
\multicolumn{2}{r}{$F160W\rightarrow V$} \nl
52681.3&24.01(0.05)&  8.2&-2.29(0.03)\nl
52693.4&24.21(0.07)& 12.9&-2.25(0.03)\nl
52701.6&24.61(0.08)& 16.1&-2.26(0.11)\nl
52627.4& flux:0.050(0.03)&-12.9&------(---)\nl
\hline
\multicolumn{4}{c}{2002hp ($z=1.305$)} \nl
\hline
\multicolumn{1}{c}{ } & \multicolumn{1}{l}{$F775W$} &
\multicolumn{2}{r}{$F775W\rightarrow U$} \nl
52579.5&25.52(0.10)&  3.3&-0.03(0.07)\nl
\multicolumn{1}{c}{ } & \multicolumn{1}{l}{$F850LP$} &
\multicolumn{2}{r}{$F850LP\rightarrow B$} \nl
52537.1&26.83(0.39)&-15.0&-1.47(0.06)\nl
52579.5&24.32(0.04)&  3.3&-1.27(0.05)\nl
52589.1&24.89(0.06)&  7.4&-1.23(0.03)\nl
52595.4&25.06(0.08)& 10.1&-1.22(0.03)\nl
52603.8&25.67(0.08)& 13.8&-1.19(0.05)\nl
52613.8&26.09(0.14)& 18.1&-1.10(0.09)\nl
52629.3&26.60(0.17)& 24.9&-1.00(0.03)\nl
52639.4&27.30(0.35)& 29.2&-0.99(0.04)\nl
\multicolumn{1}{c}{ } & \multicolumn{1}{l}{$F110W$} &
\multicolumn{2}{r}{$F110W\rightarrow V$} \nl
52589.2&24.28(0.07)&  7.5&-1.52(0.06)\nl
52595.5&24.33(0.07)& 10.2&-1.46(0.06)\nl
\hline
\multicolumn{4}{c}{2002fw ($z=1.30$)} \nl
\hline
\multicolumn{1}{c}{ } & \multicolumn{1}{l}{$F775W$} &
\multicolumn{2}{r}{$F775W\rightarrow U$} \nl
52536.8&25.37(0.05)& -5.8&-0.22(0.04)\nl
52548.3&25.01(0.08)& -0.8&-0.14(0.07)\nl
52578.4&26.45(0.10)& 12.2&0.111(0.08)\nl
\multicolumn{1}{c}{ } & \multicolumn{1}{l}{$F850LP$} &
\multicolumn{2}{r}{$F850LP\rightarrow B$} \nl
52536.9&24.49(0.05)& -5.8&-1.40(0.02)\nl
52548.3&24.17(0.04)& -0.8&-1.40(0.07)\nl
52552.5&24.44(0.06)&  0.9&-1.40(0.07)\nl
52557.4&24.36(0.05)&  3.1&-1.36(0.05)\nl
52567.9&24.48(0.06)&  7.6&-1.33(0.02)\nl
52577.5&24.85(0.07)& 11.8&-1.31(0.04)\nl
52578.5&25.07(0.07)& 12.2&-1.30(0.05)\nl
52595.5&25.88(0.17)& 19.6&-1.17(0.08)\nl
52603.8&26.68(0.25)& 23.2&-1.07(0.03)\nl
\multicolumn{1}{c}{ } & \multicolumn{1}{l}{$F110W$} &
\multicolumn{2}{r}{$F110W\rightarrow V$} \nl
52549.5&23.96(0.08)& -0.3&-1.67(0.04)\nl
52557.6&24.11(0.08)&  3.1&-1.64(0.07)\nl
\multicolumn{1}{c}{ } & \multicolumn{1}{l}{$F160W$} &
\multicolumn{2}{r}{$F160W\rightarrow R$} \nl
52549.8&23.83(0.08)& -0.2&-1.92(0.09)\nl
52557.8&23.76(0.09)&  3.2&-1.85(0.07)\nl
\hline
\multicolumn{4}{c}{2002dc ($z=0.475$)} \nl
\hline
\multicolumn{1}{c}{ } & \multicolumn{1}{l}{$F775W$} &
\multicolumn{2}{r}{$F775W\rightarrow B$} \nl
52405.4&22.41(0.02)& -0.0&-0.78(0.05)\nl
52438.0&23.57(0.03)& 22.0&-1.66(0.13)\nl
52449.3&24.07(0.05)& 29.7&-1.78(0.02)\nl
\multicolumn{1}{c}{ } & \multicolumn{1}{l}{$F850LP$} &
\multicolumn{2}{r}{$F850LP\rightarrow V$} \nl
52415.2&22.38(0.02)&  6.6&-0.89(0.06)\nl
52438.1&23.23(0.03)& 22.1&-1.02(0.08)\nl
52449.4&23.51(0.06)& 29.7&-1.14(0.04)\nl
52455.2&23.85(0.05)& 33.7&-1.15(0.05)\nl
52600.0&27.54(0.65)&131.8&-0.94(0.01)\nl
\hline
\multicolumn{4}{c}{2002dd ($z=0.95$)} \nl
\hline
\multicolumn{1}{c}{ } & \multicolumn{1}{l}{$F775W$} &
\multicolumn{2}{r}{$F775W\rightarrow U$} \nl
52405.4&23.66(0.02)&  0.8&-0.66(0.12)\nl
52438.0&24.99(0.06)& 17.5&-0.98(0.01)\nl
52449.3&25.51(0.09)& 23.3&-1.01(0.03)\nl
\multicolumn{1}{c}{ } & \multicolumn{1}{l}{$F850LP$} &
\multicolumn{2}{r}{$F850LP\rightarrow B$} \nl
52405.4&23.24(0.05)&  0.8&-1.27(0.05)\nl
52415.2&23.45(0.02)&  5.8&-1.31(0.02)\nl
52438.1&24.34(0.08)& 17.6&-1.44(0.08)\nl
52449.4&24.85(0.12)& 23.4&-1.56(0.02)\nl
52455.2&24.96(0.09)& 26.3&-1.58(0.01)\nl
\multicolumn{1}{c}{ } & \multicolumn{1}{l}{$F110W$} &
\multicolumn{2}{r}{$F110W\rightarrow V$} \nl
52438.0&23.98(0.05)& 17.5&-1.27(0.03)\nl

\enddata 
\tablenotetext{a}{JD$-$2,400,000. \\ 
Uncertainties in magnitudes are listed in parentheses.}
\end{deluxetable}

\begin{deluxetable}{llll} 
\footnotesize
\tablecaption{Spectroscopic Data}
\tablehead{\colhead{SN}&\colhead{MJD(age)}&\colhead{Instrument}&\colhead{$z$}}
\startdata
\hline
\hline
HST04Pat     & 53205.1(+3)& {\it HST} ACS & $0.97^{a,1}$ \nl
HST04Cay      & 53254.9 & {\it HST} ACS & $? ^2$ \nl
HST04Mcg      & 53275.6(+10) & {\it HST} ACS & $ 1.357 ^{c,1} $ \nl
HST05Fer      & 53486.8(+10) & {\it HST} ACS & $1.02^{a,1}$ \nl
HST05Koe     & 53486.8(+15) & {\it HST} ACS &  $1.23^ {a,1} $ \nl
HST05Dic     & --- & --- & $0.638^{b,3}$ \nl
HST04Gre    & 53275.6(+6) & {\it HST} ACS &  $1.14 ^{a,1}$ \nl
HST04Omb    & 53373.7(+3) & {\it HST} ACS & $0.975^{a,d,1} $ \nl
HST05Red    &  53436.6 & {\it HST} ACS & $1.189^{b,2} $ \nl 
HST05Lan    &  53436.6(+2) & {\it HST} ACS & $1.235^{c,1}$ \nl
HST04Tha    & 53207.7(+15) & {\it HST} ACS & $0.954^{a,d,1}$ \nl
HST04Rak     & 53373.7(+12) & {\it HST} ACS & $0.739^{c,1} $ \nl
HST05Zwi     & 53415.9(+5) & {\it HST} ACS & $0.521^ {a,d,1}$ \nl
HST04Haw     & --- & --- & $0.490^{d,3} $ \nl
HST04Kur     & 53366.1(+3) & {\it HST} ACS & $0.359^{c,1}$ \nl
HST04Yow     & 53157.0(+17) & {\it HST} ACS & $0.457^ {a,d,1} $ \nl     
HST04Man  & 53157.0(+5) & {\it HST} ACS &   $ 0.854^{a,d,1} $ \nl
HST05Spo     & 53436.6(+20) & {\it HST} ACS & $0.839^{a,d,1}$ \nl
HST04Eag      & 53296.9(+2) & {\it HST} ACS & $1.019^{c,1}$ \nl
HST05Gab      & 53485.7(-1) & {\it HST} ACS & $1.12^{a,1}$ \nl 
HST05Stro      & 53485.7(+5) & {\it HST} ACS & $1.027^{a,d,1}$ \nl
HST04Sas   & 53156.2(+1) & {\it HST} ACS &   $1.39^{a,1}$ \nl
\enddata 
\tablenotetext{a}{From cross-correlation with broad SN features.}
\tablenotetext{b}{From narrow features in our own additional Keck 
host-galaxy spectrum.}
\tablenotetext{c}{From both (a) and (b).}
\tablenotetext{d}{From Team Keck.}
\tablenotetext{1}{Classified as SN~Ia with high confidence from
spectrum.}
\tablenotetext{2}{Photometric properties indicate likely (but not certain) SN~Ia.}
\end{deluxetable}

\section{Distance Fitting and Cosmology Constraints}

 Distance estimates from SN~Ia light curves are derived from the luminosity
distance, 
\bq d_{L} = \left(\frac{{\cal L}}{4 \pi {\cal
F}}\right)^{\frac{1}{2}}, \eq 
where ${\cal L}$ and ${\cal F}$ are the intrinsic
luminosity and the absorption-free flux of the SN within a given passband,
respectively.  Equivalently, logarithmic measures of the flux
(apparent magnitude, $m$), luminosity (absolute magnitude, $M$), and
colors (to quantify the selective absorption) were used to derive
extinction-corrected distance moduli, $\mu_0=m-M=5\log d_L +25 $
($d_L$ in units of megaparsecs).  In this context, the luminosity is a
``nuisance parameter'' whose fiducial value vanishes from a reconstruction of the expansion history (which makes use of differences in distance with redshift).  We have adopted the MLCS2k2 method (Riess,
Press, \& Kirshner 1996; Jha 2002) used by R04 and the data described in
\S 2 to derive accurate and individual {\it relative} distance moduli
for the sets of SNe described in \S 3 and given in Table 6\footnote{Although a new version
of the MLCS2k2 algorithm is currently under development with an
expanded training set (Jha, Riess, \& Kirshner 2006), we utilized the
same version used by R04 to maintain, as far as possible, a
consistent and tested process to measure all available SN Ia data.
However, tests comparing the R04 and Jha et al. (2006) versions of the
algorithm show very good agreement when applied to high quality data (Lampeitl 2006).  
In addition we have found and corrected a minor numerical error in our calculation
of $K$-corrections which affect a few SNe Ia at the few percent level.
In the
future, it will be valuable to reanalyze all available data
consistently with continually improving algorithms and we will attempt to provide updated distance esimates
using updated versions of data and algorithms at http://braeburn.pha.jhu.edu/$\sim$ariess/R06 or upon request to ariess@stsci.edu.}.  As in Riess, Press, \& Kirshner (1996),
we require for robust fits of light curves that SN photometry 
 commence no less than 10 days after maximum, although in practice degeneracies
in light curve fits for late-commencing SN Ia photometry are also alleviated by flux limits from the
preceding epoch (typically 3 weeks earlier than the discovery point in the restframe).

In Figure 6 we show the Hubble diagram of distance moduli and
redshifts for all of the {\it HST}-discovered SNe~Ia in the Gold and
Silver sets from our program.  The new SNe~Ia span a wide range of
redshift ($0.21 < z < 1.55$), but their most valuable contribution to
the SN~Ia Hubble diagram remains in the highest-redshift region where
they now well delineate the range at $z \geq 1$ with 23 SNe~Ia, 13 new
objects since R04.  This territory remains uniquely accessible to {\it
HST}, which has discovered the dozen highest-redshift SNe~Ia known,
and its exploration is the focus of the rest of this paper.

In the inset to Figure 6 we show the residual Hubble diagram (from an
empty Universe) with the Gold data uniformly binned.  Here and elsewhere, we will utilize uniform, unbiased binning
achieved with a fixed value of $n\Delta z$, where $\Delta z$ is the
bin width in redshift and $n$ is the number of SNe in the
bin.\footnote{The last bin ends abruptly with the highest-redshift SN;
thus, its $n\Delta z \leq$ value is smaller than the rest.}  In
Figure 6 we use $n\Delta z=6$ which yields seven bins for our sample.  Although binning
is for illustrative purposes in the Hubble diagram, there are some
specific advantages of binning such as the removal of
lensing-induced asymmetrical residuals by flux averaging (Wang 2005) and
the ease of accounting for systematic uncertainties introduced by zeropoint errors in sets of photometric 
passbands used at similar redshifts. 

The distance-redshift relation of SNe~Ia is one of few powerful
tools available in observational cosmology.  A number of different
hypotheses and models can be tested with it, including kinematic
descriptions of the expansion history of the Universe, the existence
of mass-energy terms on the right-hand side of the Friedman equation,
and the presence of astrophysical sources of contamination.  Testing
all interesting hypotheses is well beyond the scope of this paper and
is best left for future work. Instead, we now undertake a few
narrowly posed investigations.

For the following analyses we limit the low-redshift boundary of our sample to $cz > 7000$ km s$^{-1}$ to avoid the influence of a possible local, ''Hubble Bubble'' (Jha 2002, Zehavi et al. 1998).    The Gold sample above this velocity boundary consists of 182 supernovae.

\subsection{Kinematics and Model-Independent Observables}

   The distance-redshift relation of SNe~Ia contains a cumulative
record of changes in the cosmic expansion rate, $H(z)$, over the last 10 Gyr,
described in flat space as given in equation 3.  
Although the interesting kinematic information, $H(z)$, appears to be
degraded by the physical integration across its temporal variations,
good sampling of SNe Ia over a wide range of redshifts allows us to
recover its value at discrete, uncorrelated epochs, independent of the
cosmological model.  Such information may be more general and
of longer-lasting value than constraints on any single, specific model 
of dark energy.

Following Wang \& Tegmark (2005; see also Daly \& Djorgovski 2004), we
transform the Gold sample of luminosity distances to comoving
coordinate distances, $r(z)$, as
\bq r(z)={1 \over 2997.9(1+z)}10^{\mu_0/5-5}. \eq  
We assume spatial flatness (as motivated by theoretical considerations, i.e., that most inflation models
predict $\Omega_K < 10^{-5}$ or by similar resolutions to the ``flatness problem'') for ease of calculations, but the
following approach can be generalized to allow for nontrivial spatial
curvature.  After sorting the SNe Ia by redshift, we define
the quantity 
\bq x_i={r_{i+1}-r_i \over z_{i+1}-z_i}, \eq 
where the mean value of $x_i$ gives an unbiased estimate of the
inverse of $H(z)$ at the redshift, $z_i$.  As in Wang \& Tegmark
(2005), we flux average the data first ($n\Delta z \approx 1$) to remove
possible lensing bias.  We then calculate the minimum-variance values
of $H(z)$ in 3, 4, or 5 even-sized bins across the sample, with
$n\Delta z$ chosen to be 40, 20, or 15, respectively, to achieve the
desired number of bins.

  In the upper panel of Figure 7 we show sets of 3, 4, or 5 samplings
of $H(z)$ versus redshift from the Gold sample.  As seen, $H(z)$
remains well constrained until $z \approx 1.3$, beyond which the SN
sample is too sparse to usefully determine $H(z)$.  For comparison, we
show the dynamical model of $H(z)$ derived from
$H(z)^2=H_0^2(\Omega_M(1+z)^3+\Omega_\Lambda)$ with ``concordance'' values of $\Omega_M=0.29$ and
$\Omega_\Lambda=0.71$.

  In the lower panel of Figure 7 we show the kinematic quantity 
${\dot a} = H(z)/(1+z)$ versus redshift.  In the uncorrelated ${\dot a}$
versus redshift space, it is very easy to evaluate the sign of the
change in expansion rate independent of the cosmological model.  For
comparison we show three simple kinematic models: purely
accelerating, decelerating, and coasting, with $q(z) \equiv (-\ddot a
/a)/H^2(z) = d(H^{-1}(z))/dt - 1 = 0.5$, $-0.5$, and 0.0,
respectively. We also show a model with recent acceleration
($q_0 = -0.6$) and previous deceleration ${dq/dz}=1.2$, where
$q(z)=q_0+z{dq/dz}$, which is a good fit to the data.

In Figure 8 we demonstrate the improvement in the measure of $H(z)$ at
$z>1$ realized from the addition of the new SNe~Ia, presented here, to
the sample from R04: we have reduced the uncertainty of $H(z)$ at
$z>1$ from just over 50\% to just under 20\%.\footnote{Monte Carlo
simulations of the determination of uncorrelated components of $H(z)$ show that the increase in
precision proceeds as approximately $n^{2/3}$, significantly faster
than $n^{1/2}$, where $n$ is the number of SNe due to the rate of increase in unique pairs of SNe.}

 We also repeated the analysis of R04 in which the deceleration
parameter, $q(z) \equiv (-\ddot a /a)/H^2(z) = d(H^{-1}(z))/dt
- 1$, is parameterized by $q(z) = q_0 + z dq/dz$ and determined from
the data and Eq. (5).  As in R04, we find that the Gold set strongly
favors a Universe with recent acceleration ($q_0 < 0$) and previous
deceleration ($dq/dz > 0$) with 99.96\% and 99.999\% likelihood
(summed within this quadrant), respectively.  Summing the probability
density in the $q_0$ vs. ${dq/dz}$ plane along lines of constant
transition redshift, $z_t = -q_0/(dq/dz)$, yields the transition
redshift of $z_t = 0.43 \pm 0.07$.

However, as shown by Shapiro \& Turner (2005), different
parameterizations of $q(z)$ can lead to different redshifts for the
transition and to different confidence levels for the epochs of
acceleration and deceleration (though all appear sufficiently high to
yield a robust conclusion similar to our own; see also Daly \&
Djorgovski 2004). Thus, our uncertainty is only statistical within a
linear form for $q(z)$, and is not useful for comparing to the
transition expected for a cosmological constant ($z_t \approx 0.7$) due
to the different functional forms used for evaluating the redshift
at which ${\ddot a}=0$.

\subsection{Alternatives to Dark Energy}

     After the detection of the apparent acceleration of cosmic
expansion (and dark energy) by Riess et al. (1998) and Perlmutter et
al. (1999), alternative hypotheses for the apparent faintness of
high-redshift SNe~Ia were posed.  These included extragalactic gray
dust with negligible tell-tale reddening or additional dispersion
(Aguirre 1999a,b; Rana 1979, 1980), and pure luminosity evolution
(Drell, Loredo, \& Wasserman 2000).

As reported elsewhere, there is considerable evidence against these
possibilities and little evidence in favor of either of them (Riess et
al. 1998; Perlmutter et al. 1999; Riess et al. 2001; Leibundgut 2001;
Sullivan et al. 2003; Knop et al. 2003; Filippenko 2004, 2005).
However, it is important to remain vigilant for such possibilities.
The redshift range $z>1$, where deceleration dominates over
acceleration for the simplest cosmological models, is crucial for
breaking degeneracies between astrophysical effects and cosmological
effects (e.g., Riess et al. 2001).

In R04 we found that the first significant sample of SNe~Ia at $z>1$
from {\it HST} rejected with high confidence the simplest model of
gray dust by Goobar, Bergstrom, \& M\"{o}rtsell (2002), in which a smooth
background of dust is present (presumably ejected from galaxies) at a
redshift greater than the SN sample (i.e., $z > 2$) and diluted as the
Universe expands.   This model and its opacity was invented to match  the 1998 evidence for dimming of supernovae at $z~0.5$ without invoking 
dark energy in a universe with $\Omega_m = 1$. 
This model is
shown in the inset of Figure 6.  The present Gold sample (at the best fitting value of $H_0$) rejects this
model at even higher confidence ($\Delta \chi^2=194$, i.e., 14 $\sigma$, see Table 4),
 beyond a level worthy of further consideration.  

However, a more pernicious kind of dust was also suggested by Goobar,
Bergstrom, \& M\"{o}rtsell (2002), a ``replenishing dust'' in which a
constant density of gray dust is continually replenished at just the
rate it is diluted by the expanding Universe.  This latter dust is
virtually indistinguishable from an $\Omega_\Lambda$ model (see Table 4) via the
magnitude-redshift relation because the dimming is directly
proportional to distance traveled and thus mathematically similar to
the effects of a cosmological constant.  Dust of this sort with the
required opacity, replenishing rate, and ejection velocity from
galaxies ($> 1000$ km s$^{-1}$ for it to fill space uniformly without
adding detectable dispersion) may always be virtually undetectable in
the Hubble diagram of SNe~Ia, but its degree of fine tuning makes it
unattractive as a simple alternative to a cosmological constant.  

More recently, \"{O}stman \& M\"{o}rtsell (2005) used 11,694 quasars
from the Sloan Digital Sky Survey (SDSS), at $0.1 < z < 2$, to limit
the dimming of SNe~Ia at $z = 1$ by Milky Way dust to less than 0.03
mag, and by very gray dust ($R_V = 12$) to less than 0.1 mag, for
either the high-$z$ dust or replenishing dust models.  Petric et al. (2006) 
has ruled out the presence of intergalactic gray dust at a level of $\Omega_{dust} \leq 10^{-6}$
by the lack of an X-ray scattering halo around a quasar at $z=4.3$, limiting dimming 
due to micron-sized grains to less than a few percent (less than a percent to $z\sim0.5$)
 and this dust scenario
has also been limited by the resolution of the far infrared background (Aguirre \& Haiman 2000) 

     We also find, as in R04 but with even greater confidence, that
the data are inconsistent ($\Delta \chi^2 = 116$) with the simplest SN
luminosity evolution (proportional to redshift) in lieu of a
cosmological constant; see inset of Figure 6 and Table 4.

\begin{deluxetable}{ll} 
\tablecaption{$\chi^2$ Comparison of Gold Set Data to Models, $cz> 7000$ km s$^{-1}$ }
\tablehead{\colhead{Model}&\colhead{$\chi^2$(for 184 SNe Ia)}}
\startdata
$\Omega_M=0.29, \Omega_\Lambda=0.71$ & 150$^{a}$ \nl
$\Omega_M=1.00, \Omega_\Lambda=0.00$ & 285$^{a}$  \nl
$\Omega_M=0.00, \Omega_\Lambda=0.00$ & 164$^{a}$  \nl
High-redshift gray dust (with $\Omega_M=1.00, \Omega_\Lambda=0.00$) & 344$^{b}$  \nl
Replenishing dust (with $\Omega_M=1.00, \Omega_\Lambda=0.00$) & 150$^{b}$  \nl
Dimming $\propto z$ (with $\Omega_M=1.00, \Omega_\Lambda=0.00$)  & 266$^{b}$  \nl 
\enddata 
\tablenotetext{a}{best $\chi^2$ after marginalizing over $H_0$}
\tablenotetext{a}{best $\chi^2$ for best $H_0$}
\end{deluxetable}

Less direct but significant failures to detect SN evolution come from
detailed comparisons of the composite state of the abundance,
temperature, and outward velocity of the SN photosphere as recorded
through the spectra (Sullivan 2005, Blondin et al. 2006, Hook et al. 2005, Balland et al. 2006).  Greater
leverage for this test can be gained by extending the redshift range
of the observed SED using the present sample.  In Figure 16 we show
the average, composite spectrum of the 13 best-observed SNe~Ia at $z
\geq 1$.  For the spectra listed in Table 3 (and from R04), we
transformed to the rest frame and calculated a $3\sigma$-clipped
average at each wavelength point.  Shown are the mean and dispersion
about the mean. For comparison, we used the template spectra from
Nugent et al. (2002) (with colors matched to the MLCS empirical model)
averaged over the week following maximum light.  As shown, the
high-redshift composite, with a mean redshift of 1.2, bears a striking
resemblance to the model of a post-maximum SN~Ia.  Two specific
indications of SNe~Ia are seen in the Si feature at 4130~\AA\ and the
width of the Ca~II and Si~II blended feature at 3750~\AA\ which
indicates the presence of both ions.  We put less reliance on the region redward of 4300 \AA\
corresponding to redward of observed 9500 \AA\ because 1) the quantum efficiency
of the detector is rapidly decaying and thus harder to calibrate and 2) modest color
differences may exist between the spectral template and the mean of these objects (due
to intrinsic variation or reddening).

We find no evidence for a
difference in the mean SED of SNe~Ia across 10 Gyr of look-back time
or 1.2 units of redshift.

SNe Ia hosted by red, early-type hosts are expected to be particularly robust against
 absorption by dust (where the dust content is low) and may be expected to arise from the
 earliest forming progenitors (as compared to late-type hosts).    
Thus differences in distances measured to SNe Ia hosted by early and late-type hosts form
 an important probe of systematics (Sullivan et al. 2003).   We find that the 6 SNe Ia we
 discovered at $z \geq 1$ with ACS residing in red, elliptical hosts (HST05Lan, 2002hp,
 HST04Sas, 2003es, HST04Tha, and 2003XX) all have low measured extinction ($A_V < 0.25$ mag)
 and the same dispersion and mean (closer for their redshift by 0.07 $\pm 0.10$ mag) 
as the full sample,
 indicating that at $z>1$ no differences are apparent for distances measured in early type hosts.

\subsection{Dark Energy}

Strong evidence suggests that high-redshift SNe~Ia provide accurate distance
measurements and that the source of the apparent acceleration they reveal
lies in the negative pressure of a ``dark energy'' component.  Proceding from this conclusion, our
hard-earned sample of SNe~Ia at $z>1.0$ can provide unique constraints
on its properties.  Strong motivation for this investigation comes
from thorough studies of high-redshift and low-redshift SNe~Ia,
yielding a consensus that there is no evidence for evolution or
intergalactic gray dust at or below the current statistical
constraints on the average high-redshift apparent brightness of SN~Ia
(see Filippenko 2004, 2005 for recent reviews).  We summarize the key
findings here.  (1) Empirically, analyses of 
SN~Ia distances (after accounting for the light curve shape-luminosity relation) versus host stellar age, morphology or dust content (Riess et al. 1998, Sullivan et al. 2003, Jha, Riess, \& Kirshner 2006), or metallicity or star formation rate (Gallagher et al. 2005, Riess et al. 1999) indicate that SN~Ia distances are relatively
indifferent to the evolution of the Universe.  (2)
Detailed examinations of the distance-independent properties of SNe~Ia
(including the far-UV flux, e.g., as presented in the last section)
provide strong evidence for uniformity across redshift and no
indication (thus far) of redshift-dependent differences (e.g., Sullivan et
al. 2005; Howell et al. 2005; Blondin et al. 2006).  (3) SNe~Ia are uniquely qualified as standard candles
because a well understood, physical limit (the Chandrasekhar limit) 
provides the source of their homogeneity.  Based on
these studies, we adopt a limit on redshift-dependent systematics is
to be 5\% per $\Delta z=1$ at $z>0.1$ and make quantitative use of this
in \S 4.1.

   Many have studied the constraint placed by the redshift-magnitude
relation of SNe~Ia on the parameter combination $\Omega_M$-$w$, where
$w$ (assumed to be constant) is the dark energy equation-of-state
parameter.  There are few models for dark energy that predict an equation of state that is {\it constant}, different from the 
cosmological constant, and not already ruled out by the data.
On the other hand, a prominent
class of models does exist whose defining feature is a time-dependent
dark energy (i.e., quintessence).     While the rejection of
$w=-1$ for an assumed constant value of $w$ would invalidate a
cosmological constant, it is also possible that apparent consistency
with $w=-1$ in such an analysis would incorrectly imply a cosmological
constant.  For example, if $w(z)$ is rising, declining, or even
sinusoidal, a measured derivative could be inconsistent with zero
while the average value remains near $-1$.  Therefore, when using
$w(z)$ to discriminate between dark-energy models, it is important to
allow for time-varying behavior, or else valuable information may be
lost.  Here, we seek to constrain the value of $w(z>1)$ and bound its
derivative across the range $0.2 < z < 1.3$.  This is unique information
afforded by the {\it HST}-discovered SN~Ia sample.

Unfortunately, present dynamical dark-energy models in the literature
(see Szydlowski, Kurek, \& Krawiec 2006 for a review) do not suggest a
universal or fundamental parametric form for $w(z)$.  Instead, most
models contain embedded, free-form functions (e.g., the shape of a
scalar potential).  Thus, we proceed with simple, empirical studies of
variations in the equation-of-state parameter of dark energy.

The luminosity distance to SNe Ia, through the solution to the
Freidman equation, is sensitive to $w(z)$:
\bq
d_l=c H_0^{-1}(1+z) \int_0^z {dz' \over E(zÍ)}, 
\eq
Here, $E(z) \equiv \{ \om(1+z')^3 +(1-\om) \times
exp[+3\int_0^{\ln(1+z)} d\ln(1+z') (1+w(z'))] \}^{1/2}, $
\noindent
where $\om$ is the dimensionless matter density $8\pi\rho_m/(3H_0^2)$, 
$H_0$ is the Hubble constant (the present value of the Hubble 
parameter), and $z$ is the redshift of any SN~Ia.

We can determine the likelihood for parameters of $w(z)$ from a
$\chi^2$ statistic,
\bq \chi^2(H_0,\Omega_M,{\bf w_i})=\sum_i 
{ (\mu_{p,i}(z_i; H_0, \Omega_M,{\bf w_i})-\mu_{0,i})^2 \over
\sigma_{\mu_{0,i}}^2+\sigma_v^2}, \eq 
where $\sigma_v$ is the dispersion in supernova redshift (transformed
to units of distance moduli) due to peculiar velocities $\sigma_{\mu_{0,i}}$ is the uncertainty in the
individual distance moduli, and ${\bf w_i}$ is a set of dark-energy
parameters describing $w(z)$.  Due
to the extreme redshift of our distant sample and the abundance of
objects in the nearby sample, our analysis is insensitive to the value
we assume for $\sigma_v$ within its likely range of 200 km s$^{-1}$
$\leq \sigma_v \leq$ 500 km s$^{-1}$.  For our analysis we adopt
$\sigma_v=400$ km s$^{-1}$.  For high-redshift SNe~Ia whose redshifts
were determined from the broad features in the SN spectrum, we assume
an uncertainty of $\sigma_v = 2500$ km s$^{-1}$ in quadrature to
the peculiar velocity term.  Marginalizing our likelihood functions
over the nuisance parameter, $H_0$ (by integrating the probability
density $P \propto e^{- \chi^2 / 2}$ for all values of $H_0$), and the
use of the described independent priors, yields the confidence
intervals considered below.

Strong degeneracies exist in the effect of $w(z)$ and $\Omega_M$ on
the expansion history, requiring independent constraints to make
significant progress.  Here we consider the use of one or more of the
following constraints.

\noindent (1) {\it The SDSS luminous red galaxy, baryon acoustic
oscillation (BAO) distance parameter to $z = 0.35$}: $ A \equiv
\Omega_M^{1 \over 2} E(z)^{-1/3}[{1 \over z} \int_0^z {dzÍ \over
E(zÍ)}]^{2 \over 3},$ where $z = 0.35$ and $A = 0.469(n/0.98)^{-0.35}
\pm 0.017$ from Eisenstein et al. (2005). The 3-year Wilkinson
Microwave Anisotropy Probe (WMAP) results yield $n = 0.95$ (Spergel et
al. 2006).

\noindent (2) {\it The present local mass density}: $\Omega_M=0.28 \pm
0.04$, a consensus value when combining large-scale structure (LSS)
measurements from the 2dF and SDSS of $\Omega_Mh$ (Tegmark et
al. 2004; Cole et al. 2005) and the value of $H_0$ from {\it HST}
Cepheids (Freedman et al. 2001; Riess et al. 2005).

\noindent (3) {\it The distance to last scattering, $z = 1089$}: In the
$H_0$-independent form (Bond, Efstathiou, \& Tegmark 1997), $R_{CMB}
\equiv \Omega_M^{1 \over 2} \int_0^{1089} {dz' \over E(zÍ)}$ from the
3-year integrated WMAP analysis (Spergel et al. 2006), updated by Wang
\& Mukherjee (2006) to be $1.70 \pm 0.03$ independent of the dark-energy 
model.  

\noindent (4) {\it The distance ratio from $z = 0.35$ to $z = 1089$}:
Measured by the SDSS BAOs from Eisenstein et al. (2005),
$R_{0.35}={0.35 \over E(0.35)}^{(1/3)} [\int_0^{0.35}
(dz/E(0.35))]^{(2/3)} / \int_0^{1089} dz/E(1089) = 0.0979 \pm 0.0036$.

Additional cosmological constraints exist, but in general provide less
leverage and may be less robust than the above.

Unfortunately, constraints anchored at $z=1089$ such as $R_{CMB}$ and
$R_{0.35}$ in (3) and (4) above require careful consideration when
using continuous descriptions of $w(z)$.  Specifically, we must
consider how to evaluate integrals depending on $w(z)$ in a region
where we have no data and little intuition, i.e., between the
highest-redshift SN~Ia at $z \approx 1.8$ and the surface of last
scattering at $z=1089$.  This span contributes significantly to
$R_{CMB}$ (and even more to $R_{0.35}$), accounting for approximately
two-thirds of the total distance to last scattering.  Constraints
resulting from a naive leap across this present ``cosmological
desert'' are likely to be unjustifiably strong, and extremely
sensitive to our assumptions about the early behavior of dark energy
or on our chosen, parametric form for $w(z)$ (as we will see).

To evaluate $w(z)$ as well as its sensitivity to assumptions about its
high-redshift behavior, we will make use of three different priors.
Our ``weak'' prior (the most conservative) will utilize only
$\Omega_M$ and $A$ which are determined at low redshifts, making no
attempt to guess the behavior of dark energy in the span $1.8 < z <
1089$.  Our ``strong'' constraint integrates the expressions
containing $w(z)$ in $R_{CMB}$ and $R_{0.35}$ between $z=0$ and the
highest-redshift SN ($z \approx 1.8$) and beyond; it assumes that the
influence of dark energy on $H(z)$ is minimal by relaxing the
equation-of-state parameter to $w=-1$ at $z>1.8$.  This ``strong''
prior strikes a balance between making maximal use of all information,
but only broadly guesses at the importance of dark energy where there
are no discrete data (i.e., by assuming that the high-redshift Universe
is fully dark-matter dominated).  Our ``strongest'' prior is the naive
extension of $w(z)$ from $z=1.8$ to $z=1089$ strictly along its
empirical or parametric description.

The minimum complexity required to detect time variation in dark
energy is to add a second parameter to measure a change in the
equation-of-state parameter with redshift.  The expansion
$w(z)=w_0+w'z$, where $w' \equiv {dw \over dz}|_{z=0}$ was proposed by Cooray \& Huterer (1999), 000first used
by Di Pietro \& Claeskens (2003) and later by R04.  However, in this
form, $w(z)$ diverges, making it unsuitable at high redshift or
requiring guesses as to the form of its graceful exit (Wang \&
Mukherjee 2006).  Chevalier \& Polarski (2001) and Linder (2003) suggest $w(z)=w_0+w_a z/(1+z)$,
which solves the divergence problem but at a cost of demanding stiffer
behavior for $w(z)$ than we may presume {\it a priori}.  As we shall
see, invoking the $(w_0,w_a)$ parameterization can act like an
additional, strong prior itself .

In Figure 9 we show the constraints on $w_0-w_a$ plane for the three priors
for the Gold sample.  Here we focus on what is learned from the
inclusion of the high-redshift SNe~Ia from {\it HST}, both the new
objects and those with improved calibration from R04.

The addition of SNe~Ia at $z>1$ with the weak prior provides valuable
leverage in the $w_0-w_a$ space, well in excess of their fractional
contribution to the sample.  Citing a popular metric, the area of the
95\% confidence interval (Huterer \& Turner 2001, Kolb et al. 2006; Albrecht \& Bernstein 2006), we note a
reduction by 40\% or an increase in the figure of merit (inverse area)
by a factor of
1.7 by the inclusion of the Gold-quality SNe~Ia from {\it HST} with
the weak prior.  The reason is readily apparent; because the prior
contains no information regarding the evolution of dark energy, what
little is gleaned comes exclusively from the discrete data.  The
highest-redshift SNe~Ia are critical to breaking the degeneracy
between $w_0$ and $w_a$ affecting their lower-redshift brethren.
Constraints measured using the parameterization, $w(z)=w_0+w'z$ yield
similar improvements, 50\% in area, when the prior is weak.  The best
fit is consistent with a cosmological constant at $w_0=-1$ and
$w_a=0$.  However, the overall level of empirical knowledge about
dark-energy evolution remains very modest.  We may conclude that if
dark energy evolves, the evolution is not very rapid (though it
remains difficult to predict the natural level of expected evolution;
see Caldwell \& Linder 2005).

As our prior becomes more ambitious (maintaining the flatness prior),
 the confidence intervals rapidly
shrink.  For the strong and strongest prior, the 95\% confidence
interval is 1.5 and 4.4 times smaller than the weak prior,
respectively.  For the strongest prior, even modest values of $w_a$
(and even more so for $w'$) are strongly excluded because of their
implication that dark energy would gradually and unhaltingly grow in
importance with redshift to last scattering.\footnote{For comparison,
a constant $w$ model requires $w<-0.7$ if dark energy does not evolve
to last scattering.  From this we can see that even modest evolution,
if unwavering, will become incompatible with a high-redshift
constraint.  This result is very similar to that of Wang \& Mukerjee
2006.}  For these priors, a cosmological constant (i.e., $w_0=-1$,
$w_a=0$) is separated from the best fit along the direction of the
major axis of the error ellipse, lying within the boundary of the {\it
joint} $1\sigma$ to $2\sigma$ confidence level.  However, it appears that the
gains from the stronger priors come at a cost of reliance on the merit
of our prior at high redshifts and with diminished regard for the
data sampling $w(z)$ at lower redshifts.

We now explore the sensitivity of the constraints on $w(z)$ to the
assumption that it can be described by a simple, parametric form.  To
do so we compare the constraints on $w(z)$ derived from the previous
$w(z)=w_0+ w_a z/(1+z)$ to a higher-order polynomial expansion in
powers of ${\rm ln} (1+z)$ which assumes little about $w(z)$.  We will
construct $w(z)=\sum_i^4 w_i ({\rm ln} (1+z))^i $, using Eq. (3) and a
similar $\chi^2$ statistic to determine the likelihood for the terms
$w_i$ to $i=4$ (4th order).  The seven-dimensional likelihood manifold
for $(w_i,\Omega_M,H_0)$ is computed using a Monte Carlo Markov Chain
(MCMC).  The MCMC procedure
involves randomly choosing values for $w_i$, $\Omega_M$, and $H_0$, computing the luminosity distance, evaluating the $\chi^2$ of the fit to supernova data, and then determining whether to accept or reject the set
of parameters based on whether $\chi^2$ is improved (using a Metropolis-Hastings algorithm).
A set of parameters which is accepted to the chain forms a new starting point for the next guess
and the process is repeated for a sufficient number of steps 
(convergence can be checked via the method of Dunkley et al. 2005).
Ultimately, the list (aka chain) of the randomly chosen and accepted parameters
 forms a good approximation of the likelihood distribution (Knox, Christensen, \& Skordis 2001).

The best solution and its
uncertainty for $w(z)$ is shown in Figure 10 for the quartic polynomial and
low-order descriptions of $w(z)$.  As shown, the low-order fit implies
much greater precision concerning dark energy (with a 95\% confidence
region which is 7.2 times smaller).

The higher-order fit suffers in comparison, tolerating non-monotonic
and even oscillating solutions for $w(z)$ as well as recent changes.
The only (natural) limitation to the high-order fit comes from direct
confrontation with data.  Inclusion of the {\it HST}-discovered SNe~Ia
alone reduces the confidence intervals by a factor of 1.6.  Yet,
despite the extra freedom, the recent equation-of-state parameter
remains well constrained.  In Figure 11 we show the uncertainty as a
function of redshift for $w(z)$ for the low-order and high-order fits
with and without the high-redshift {\it HST} data.  The difference in
implied precision on $w(z)$ from the two parameterizations is very
similar to the difference seen using a weak or strong prior at high
redshift.  Both a simple parameterization and a strong, high-redshift
prior greatly restrict the allowed wandering of $w(z)$, though neither
can be well-justified.  In this way (and others discussed in the next
section) we can see that a simple dark-energy parameterization {\it is
equivalent} to a strong and unjustified prior on the nature of dark
energy.  The conclusions we draw either from the polynomial form with a strong prior or from the simple form with a 
weak prior are powerfully shaped by the data and are unlikely to be completely mislead by a bad assumption.

  An alternative approach to parameterizing $w(z)$ and constraining
the parameters is to extract discrete, uncorrelated estimates of
$w(z)$ as a function of redshift, analogous to the uncorrelated
estimates of $H(z)$ derived in the last section.  Following a method
established by Huterer \& Cooray (2005), we can extract the
uncorrelated and model-independent band power estimates of $w(z)$
similar to a principal component analysis.

  The evolution of $w$ may be usefully resolved across several redshift bins or steps.
Let $w_i$ be a constant value of $w(z)$ in bin $i$.  For a given experiment,
the $w_i$ will generally be correlated with each other because
measurements of $d_l$ constrain redshift integrals of $w(z)$; see
Eq. (8).  We can use a MCMC to obtain the likelihood surface for the
variables in the vector $\vec w = [w_1, w_2, \ldots w_n]$.

Assume for simplicity that the mean value of $\vec w$ from the likelihood surface has already been subtracted so that now $\langle w_i \rangle = 0$.  In general, the variables will be correlated, with a correlation matrix
\[
	\mathbf{C} = \langle \vec w \vec w^T \rangle = \left(
	\begin{array}{ccc} \sigma_{w_1}^2 & \cdots & \sigma_{w_1 w_n} \\ \vdots &  & \vdots \\
	\sigma_{w_n w_1} & \cdots & \sigma_{w_n}^2 \end{array}
	\right)
\]

What we're interested in is finding a linear transformation from $\vec w$ into some new vector $\vec {\mathcal{W}}$ such that the correlation matrix for the new vector is diagonal (in other words, we want $\langle {\mathcal{W}}_i {\mathcal{W}}_j \rangle = 0$ for $i \ne j$).  Since there are an infinite number of matrices that diagonalize $\mathbf{C}$ there are an arbitrary number of ways to do this.

The simplest is to decompose $\mathbf{C}$ into eigenvalues and eigenvectors, and use the eigenvectors to multiply $\vec w$.  If
\[
	\mathbf{C} = \mathbf{V}^T \mathbf{\Lambda V},
\]
where $\mathbf{\Lambda}$ is diagonal, then in turn
\[
	\mathbf{V C V}^T = \mathbf{\Lambda}.
\]
Therefore if we choose the vector $\vec {\mathcal{W}}$ to be equal to $\mathbf{V} \vec w$, then 
\begin{eqnarray*}
	&\langle \vec{\mathcal{W}} \vec{\mathcal{W}^T}  \rangle 
	&= \langle \left( \mathbf{V} \vec w \right) \left( \mathbf{V} \vec w \right)^T \rangle \\
	&&= \mathbf{V} \langle \vec w \vec w^T \rangle \mathbf{V}^T
	= \mathbf{V C V}^T = \mathbf{\Lambda}.
\end{eqnarray*}

So in this case $\vec {\mathcal{W}}$ has a diagonal correlation matrix.  However, transforming from $\vec w$ to $\vec {\mathcal{W}}$ in this way will often involve both adding and subtracting individual entries of $\vec w$ to get a particular entry of $\vec {\mathcal{W}}$ so physically interpreting what $\vec {\mathcal{W}}$ means can be difficult.
\
It's actually possible to use a different transformation to get a $\vec {\mathcal{W}}$ that is in some sense ``closer'' to $\vec w$.  A slightly more complicated transformation is to define a matrix $\mathbf{T}$ such that 
\[
	\vec{\mathcal{W}} = \mathbf{T} \vec w = \mathbf{V}^T \mathbf{\Lambda}^{-\frac12} \mathbf{V} \vec w,
\]
where $\mathbf{\Lambda}^{-\frac12}$ is just the diagonal matrix of the reciprocal of the square root of each eigenvalue.  With this transformation the covariance matrix for $\vec {\mathcal{W}}$ is

\begin{eqnarray*}
	&\langle \vec{\mathcal{W}} \vec{\mathcal{W}^T} \rangle 
	&= \langle \left( \mathbf{V}^T \mathbf{\Lambda}^{-\frac12} \mathbf{V} \vec w \right) 
		\left( \mathbf{V}^T \mathbf{\Lambda}^{-\frac12} \mathbf{V}  \vec w \right)^T \rangle \\
	&&= \mathbf{V}^T \mathbf{\Lambda}^{-\frac12} \mathbf{V} \langle \vec w \vec w^T \rangle  
		\mathbf{V}^T \mathbf{\Lambda}^{-\frac12}\mathbf{V} \\
	&&= \mathbf{V}^T \mathbf{\Lambda}^{-\frac12} \mathbf{V C V}^T \mathbf{\Lambda}^{-\frac12}\mathbf{V} \\
	&& =\mathbf{V}^T \mathbf{\Lambda}^{-\frac12} \mathbf{\Lambda} \mathbf{\Lambda}^{-\frac12}\mathbf{V}
	= \mathbf{I} .
\end{eqnarray*}

So with this transformation the correlation matrix for $\vec {\mathcal{W}}$ is just the identity matrix.  Further, with a matrix of the form $\mathbf{V}^T \mathbf{\Lambda}^{-\frac12} \mathbf{V}$, the entries of $\mathbf{T}$ are much more likely to be positive (the eigenvalues are positive, so any negative entries in the eigenvector matrix $\mathbf{V}$ will tend to multiply together and become positive).

This is the transformation we used, with one further modification.  Explicitly writing out the transformation for a single entry in $\vec {\mathcal{W}}$
\[
	{\mathcal{W}}_i = \sum_j T_{ij} w_j.
\]
For a given $i$, $T_{ij}$ can be thought of as weights for each $w_j$ in the transformation from $\vec w$ to $\mathcal{W_i}$.  We are free to rescale each $\mathcal{W_i}$ without changing the diagonality of the correlation matrix, so we then multiply both sides of the equation above by an amount such that the sum of the weights $\sum_j T_{ij}$ is equal to one. This allows for easy interpretation of the weights as a kind of discretized window function.  This is, in fact, what we plot as the window function for the decorrelated parameters in the paper.

  The resultant values of ${\mathcal{W}}_i$ are thus uncorrelated (in
that there is no covariance between the errors of any pair) and free
to vary independently, subject only to the constraints of the data.
Each uncorrelated measure of ${\mathcal{W}}_i$ can be tested against
the cosmological constant expectation of ${\mathcal{W}}_i=-1$,
independently as well as collectively.  Again, we use the weak,
strong, and strongest priors, the last two including $R_{CMB}$ and
$R_{0.35}$ with $w(z>1.8)$ equal to $-1$ or the value in the
highest-redshift bin, respectively.  We also used the same three
redshift bins defined by $n\Delta z=40$, resulting in three
independent measures of $w(z)$ anchored at approximately z=0.25, 0.70,
and 1.35 (with boundaries of $z=0,0.45,0.935$, and $1.8$) for the terms ${\mathcal{W}}_{0.25}$, ${\mathcal{W}}_{0.70}$,
and ${\mathcal{W}}_{1.35}$, respectively.  Thus there are five free parameters in the MCMC chain
corresponding to the 3 values for ${\mathcal{W}}_i$, $\Omega_M$, and $H_0$.  The likelihoods and
confidence intervals for these terms are shown in Figure 12, 13, and
14 (respectively) and given in Table 5.  The rows of $\mathbf{T}$ are
represented as the window functions in Figures 12, 13, and 14.  As
expected, the lowest--redshift measure of $w(z)$,
${\mathcal{W}}_{0.25}$, is derived primarily from the lowest-redshift
bin which provides over 90\% of the weight.  The highest-redshift
measure, ${\mathcal{W}}_{1.35}$, is mostly derived from the
highest-redshift bin which contributes two-thirds of the weight of the
measurement, but also relies on the lower-redshift bins to decouple
high-redshift measurements of $d_l$ from the low-redshift behavior of
$w(z)$.

For the two lowest-redshift bins, the likelihood distributions are
close to Gaussian, but for ${\mathcal{W}}_{1.35}$ the distribution can
be quite skewed, requiring an explanation of the definition of our
confidence intervals and reported values of ${\mathcal{W}}_{z}$ .  In Table 5 and Figures 12, 13, and 14, we define
a $+1\sigma$ and $-1\sigma$ region to be the boundaries at which the likelihood
falls to 0.6065 of the peak on either side (i.e., as for a Gaussian),
a definition which neglects the non-Gaussian tails to provide the
frequently sought error bar.  Likewise, we define a $2 \sigma$ region
as the boundaries of equal likelihood which contain 95\% of the likelihood.
We also report the values of the peak of the likelihood for ${\mathcal{W}}_{z}$ in the table.

\begin{deluxetable}{cccc}
\tablecaption{Likelihood Regions For ${\mathcal{W}}_{z}$}
\tablehead{\colhead{\it $w_z$}&\colhead{peak}& \colhead{1 $\sigma$} & \colhead{2 $\sigma$ }}
\startdata
\hline
\multicolumn{4}{c}{Prior=Weak, Sample=All Gold} \nl
${\mathcal{W}}_{0.25}$& -1.05 & -1.15 to -0.95 & -1.26 to -0.85 \nl
${\mathcal{W}}_{0.70}$& -0.45 & -0.86 to -0.06 & -1.49 to 0.32 \nl
${\mathcal{W}}_{1.35}$& 0.59 & -2.62 to 3.03 & -16.6 to 6.15 \nl
\hline
\multicolumn{4}{c}{Prior=Weak, Sample=Gold minus HST} \nl
${\mathcal{W}}_{0.25}$& -1.06 & -1.16 to -0.95 & -1.27 to -0.86 \nl
${\mathcal{W}}_{0.70}$& 0.11 & -0.43 to 0.61 & -1.17 to 1.17 \nl
${\mathcal{W}}_{1.35}$& 10.77 & 1.86 to 18.55 & -20.1 to 27.92 \nl
\hline
\multicolumn{4}{c}{Prior=Strong, Sample=All Gold} \nl
${\mathcal{W}}_{0.25}$& -1.02 & -1.12 to -0.93 & -1.23 to -0.84 \nl
${\mathcal{W}}_{0.70}$& -0.15 & -0.57 to 0.131 & -1.05 to 0.46 \nl
${\mathcal{W}}_{1.35}$& -0.76 & -1.78 to -0.16 & -15.8 to 0.51 \nl
\hline
\multicolumn{4}{c}{Prior=Strong, Sample=Gold minus HST} \nl
${\mathcal{W}}_{0.25}$& -1.03 & -1.14 to -0.94 & -1.25 to -0.85 \nl
${\mathcal{W}}_{0.70}$& 0.151 & -0.26 to 0.61 & -0.80 to 1.00 \nl
${\mathcal{W}}_{1.35}$& -1.95 & -5.89 to -0.70 & -17.8 to 0.35 \nl
\hline
\multicolumn{4}{c}{Prior=Strongest, Sample=All Gold} \nl
${\mathcal{W}}_{0.25}$& -1.02 & -1.11 to -0.92 & -1.21 to -0.83 \nl
${\mathcal{W}}_{0.70}$& -0.13 & -0.47 to 0.17 & -0.88 to 0.48 \nl
${\mathcal{W}}_{1.35}$& -0.85 & -1.81 to -0.46 & -17.0 to -0.30 \nl
\hline
\multicolumn{4}{c}{Prior=Strongest, Sample=Gold minus HST} \nl
${\mathcal{W}}_{0.25}$& -1.03 & -1.13 to -0.94 & -1.24 to -0.85 \nl
${\mathcal{W}}_{0.70}$& 0.24 & -0.17 to 0.64 & -0.70 to 1.06 \nl
${\mathcal{W}}_{1.35}$& -1.89 & -5.50 to -0.80 & -18.0 to -0.34 \nl
\hline
\multicolumn{4}{c}{Prior=Strong, Sample=All Gold with MLCS2k2 Fits to SNLS SNe} \nl
${\mathcal{W}}_{0.25}$& -1.05 & -1.14 to -0.94 & -1.26 to -0.84 \nl
${\mathcal{W}}_{0.70}$& -0.09 & -0.45 to 0.23 & -0.91 to 0.56 \nl
${\mathcal{W}}_{1.35}$& -1.01 & -2.23 to -0.26 & -15.8 to 0.37 \nl
\hline
\enddata
\end{deluxetable}

For the weak prior, the three measures (${\mathcal{W}}_{0.25}$,
${\mathcal{W}}_{0.70}$, and ${\mathcal{W}}_{1.35}$) are all consistent
with $w=-1$ at or near the 68\% confidence interval.  The
lowest-redshift bin remains well constrained with
${\mathcal{W}}_{0.25}$ $=-1.06 \pm 0.10$ despite the additional
freedom at higher redshifts.  The next bin is modestly well
constrained at ${\mathcal{W}}_{0.70}$ $=-0.46 \pm 0.46$, though the
likelihood in the highest-redshift bin has only begun to appear at all
localized.  Although the maximum likelihood values of $w(z)$ are
monotonic in redshift, any trend is not significant.  The constraints
are weakest at high redshift where the data are sparse and discrete
values of $w(z)$ become harder to isolate from integrated constraints.
Figure 15 shows the impact of adding SNe~Ia to the sample at the
highest redshifts from {\it HST}.  The first published sampled from
R04 began the process of localizing the highest redshift bin and
markedly improved the intermediate bin, with no impact on the nearest
bin.  For the weakest prior which lacks any complementary constraints
at high redshift, the increased sample has provided only modest gains.

For the strong prior, the nearby bin is unchanged, the uncertainty in
the intermediate bin tightens by 25\%, and the highest-redshift bin
becomes significantly more peaked with a factor of two reduction in
uncertainty.  A modest tension exists with $w=-1$ in the middle bin
(less than 2$\sigma$) (with no apparent trend in redshift).  
However, most of the data in this intermediate bin come from previous and on-going ground-based surveys
and is rapidly growing in size.  Thus we can expect this bin to soon come into better focus.  Interestingly, 
 the full sample considered here will soon be significantly augmented by the addition of 60 new SNe Ia with high-quality data from 
the first ESSENCE dataset of Wood-Vasey et al. (2007) which taken together with the past data, we note, results in a reduction of 
the difference with $w=-1$ at these redshifts to $\sim$ 1 $\sigma$.

For the strong
prior, the addition of the high-redshift {\it HST} data shown in
Figure 15 provides strong gains in the precision of
${\mathcal{W}}_{1.35}$, especially with the sample published here for
which we find ${\mathcal{W}}_{1.35}$ $=-0.8^{+0.6}_{-1.0}$, marking
significant inroads into the measurement of the early-time behavior of
dark energy.  Indeed, this measurement represents the most distinctive
contribution of the {\it HST}-discovered SNe~Ia. Combined with the
low-redshift measurement, ${\mathcal{W}}_{0.25}$, the change
${\mathcal{W}}_{0.25}$ $-{\mathcal{W}}_{1.35}$ $=-0.3^{+0.6}_{-1.0} $
over one unit of redshift, disfavoring rapid changes, i.e.,
$|dw/dz|>>1$.  Indeed, without the {\it HST} data, any measurement of the same
${\mathcal{W}}_{1.35}$ is not meaningful
because the sample would contain no SNe at $z \sim 1.35$.
The gain from the weak to the strong prior results from
the complementary nature of the contributions of the SN data, the BAO
measurement, and the WMAP measurement moderated by the high-redshift
integration.

For the strongest prior, the tension within the highest-redshift bin
has grown between the SNe contained within and the extrapolation of
the implied behavior of dark energy continuously to $z=1089$.  Any
growing mode or non-accelerating type behavior for dark energy ($w >
-1/3$) is curtailed.  While this plot looks convincing, it is worth noting that non-accelerating modes were mildly favored by the 
weak prior.  The embedded assumption of the way dark energy changes with redshift dominates over the 
information added by considering the data on $w(z)$.  Because this prior relies on the
correctness of guesses, we conclude that the apparent gain in
information is at best risky and at worst, illusionary.  Thus, we
discourage the use of other parameterizations of dynamic dark energy
in which the behavior of dark energy between $z=1.8$ and $z=1089$
remains important and a determining factor.

Not surprisingly, the addition of the SNLS sample to the full sample
provides the most improvement in the lowest-redshift bin, about 10\%
in $\sigma_w$, but little gain in the intermediate bin (3\%) and none
in the highest redshift bin.

The likelihood distribution for ${\mathcal{W}}_{1.35}$ also
empirically characterizes the most basic property of dark energy at
$z>1$, the way it gravitates.  Using the strong prior, the likelihood
that high-redshift dark energy has negative pressure, i.e.,
${\mathcal{W}}_{1.35} < 0$, is 97.6\% (or 99\% for the MLCS2k2 fits to the Astier et al. 
light curves).  (We note that the likelihood contained beyond a single boundary
such a $w<0$ 
cannot be directly inferred from our non-gaussian confidence regions because they have been defined using two boundaries
and the requirement that the boundary likelihoods match.)
\footnote{The strongest prior
would appear to provide even greater confidence for this statement.
However, as we discussed, this prior seems unjustifiably strong and
its use in this context is somewhat circular (i.e., guessing, and then
inferring, the nature of dark energy at high redshift).  In contrast,
the strong prior brings the most independent information to bear on
the early-time behavior of dark energy.}  {\it Thus, the defining property
of dark energy appears to be intact even during the epoch of matter
domination.}  We can say with greater confidence (99.99\%) that if a
decaying scalar field is responsible for this high-redshift dark
energy, its energy is not primarily kinetic ($w=1$).  With less
confidence, 93.4\%, we can say that early dark energy provided repulsive gravity ($w < - 1/3$).  The nature of dark energy at high redshift
($z>1$) may be of particular interest because classes of tracker
models predict rapid changes in $w(z)$ near epochs when one energy
density (e.g., radiation, dark matter, or dark energy) becomes
dominant over another.  For example, Albrecht and Skordis (2000) describe
a form of potential which gravitates as the dominant mass-energy component
at a given epoch.  Simply put, this model would predict that the equation of state
of the dark energy would mimic matter ($w=0$) during the decelerating, matter-dominated phase
(with a brief transition ``ripple'' during which $w>0$) and the aforementioned
likelihood for ${\mathcal{W}}_{1.35} \geq 0$ would appear inconsistent with this model.  
However, future observational and theoretical work should better define if variants of this
idea remain viable.

Finally, we may consider whether three additional parameters to
describe $w(z)$ are actually needed to improve upon a flat,
$\Lambda$-cold-dark-matter ($\Lambda$CDM) model fit to the data (i.e., the ``concordance''
value of $\Omega_M=0.29$, $\Omega_\Lambda=0.71$).  The residuals from this fit are shown in Figure 17 and have a dispersion for all data of 0.21 mag.  To
determine the need for complex forms of $w(z)$ we can calculate the improvement to the fit,
\begin{equation} \chi^2_{eff} \equiv -\Delta (2 {\rm ln} {\mathcal{L}}) =2 {\rm ln}
{\mathcal{L}}(w=-1)-2 {\rm ln}{\mathcal{L}}(w_i={\mathcal{W}}_i),
\end{equation} 
with $i$ additional free parameters.  For the weak, strong, and
strongest priors we find an improvement of $\chi^2_{eff}=$ 4, 5.5, and
5.5, respectively, for the three additional degrees of freedom, in no
case requiring the additional complexity in dark energy (improvements of $>14$ would be noteworthy).  Likewise,
there is no improvement at all for the Akaike Information Criterion (i.e.,
$\Delta$AIC=$\Delta \chi^2 - 2i$; Liddle et al. 2004) with changes of
$-2$, $-0.5$, and $-0.5$, respectively which fail to overcome the penalty of increased complexity in the model.

\section{Discussion}

   We have presented a sample of SNe~Ia at $z \geq 1.0$, found and followed
with {\it HST}, more than double the size of those in R04.  We have
also performed a set of limited cosmological analyses focusing on
tests and models we expect to be most sensitive to the additional
data.  Given the set of independent experimental constraints and the
panoply of dark-energy models, we expect that the most interesting
tests and constraints will be discovered by future work. We now
discuss caveats and tests of this dataset.

\subsection{Systematics}

Past research has sought to quantify and characterize possible
systematic errors in supernova-based cosmological inference (e.g.,
Riess et al. 1998; Perlmutter et al. 1999; Knop et al. 2003; Tonry et
al. 2003; Riess et al. 2004; Filippenko 2005; Astier et al. 2006).
Here we consider systematics which are more sensitive to the uniquely
high redshifts of our sample or otherwise merit new consideration.

Strong or weak lensing may skew the distribution of magnitudes but
will not bias the mean (Holz 1998).  Extensive analyses of the
expected lensing along the line-of-sight for the SNe found in the
GOODS fields in R04 by J\"{o}nsson et al. (2006) show that the
distribution of magnifications matches an unbiased sample having a
mean consistent with unity and that attempting to correct for this
effect has negligible impact on the derived cosmological parameters.
This is not surprising, as the dispersion of lensing-induced
magnitudes is less than 0.05 mag, far less than the intrinsic
dispersion.  The same was concluded in R04 using the methods of
Ben\'\i tez et al. (2002), though with less accuracy than in
J\"{o}nsson et al. (2006).

As high-redshift SN Ia surveys continue, sample sizes increase,
and data collection methods improve, it may become sensible to discard
older samples in an effort to reduce systematic errors in the cosmological sample
associated with inferior calibration (e.g., data from Riess, et al. 1998, 
Perlmutter et al. 1999, Tonry et al. 2003, and Knop et al. 2003).  At present we do not believe
this is warranted, especially since it is not yet possible to 
similarly replace the low-redshift sample whose data was collected with similar 
techniques as the older high-redshift data.  
Nevertheless, it may be instructive to determine the impact on the
determination of the nature of dark energy using at intermediate redshifts only
 the more modern and homogeneous
data from Astier et al. (2006) and with the same HST data at high redshift and
low-redshift sample found here.  From this more limited sample we find no significant change for the lowest redshift measure of the equation of state, ${\mathcal{W}}_{0.25}$,
 and a 30\% decrease in its precision.  For the intermediate redshift bin,
 ${\mathcal{W}}_{0.70}$, we find values {\it more} consistent
with a cosmological constant with $-1.11_{-0.82}^{+0.59}$ and 
$-0.76_{-0.56}^{+0.42}$ , for the weak and strong prior, respectively, 
representing a 50\% to 100\% decrease in the precision.  For the highest redshift
 measure, ${\mathcal{W}}_{1.35}$, we find values {\it less} consistent from 
a cosmological constant, with $3.39_{-2.91}^{+4.21}$ and $-0.23_{-0.72}^{+0.75}$ ,
 for the weak and strong prior, respectively,
representing little change in the mean precision.  The two most discrepant measurements
 from $w(z)=-1$ are with the full dataset and the strong prior
for ${\mathcal{W}}_{0.70}$=-0.15 (-1$\sigma$=-0.41, -2 $\sigma$ =-0.89) and for
 the limited dataset and the weak prior for which 
${\mathcal{W}}_{1.35}$=3.39 (-1 $\sigma$=-2.91, -2 $\sigma$=-23.1).
Both measurements are inconsistent at more than the 1 sigma level and less than the 2 
sigma level with a cosmological constant, an insignificant difference for 3 independent measurements of $w(z)$.
We also note that an increased sample of 60 new SNe Ia at $0.3 < z < 0.7$ from ESSENCE (Wood-Vasey et al. 2007)
further reduces the apparent difference from a cosmological constant from the full-sample for the intermediate redshift bin. 

Evolution of SN explosions remains the most pernicious systematic
source of uncertainty and is challenging to quantify (see, however, Riess \& Livio 2006 for a suggestion for future observations).
  The two most
direct lines of evidence that evolution with redshift, is limited to
less than $\sim$10\% in luminosity comes from studies of SN Ia
distances across host-galaxy (characteristic) ages (Sullivan et
al. 2003; Riess et al. 1998; Jha 2002; Jha, Riess, \& Kirshner 2006, Gallagher et al. 2005) and the consistency of SN~Ia distances
with the expected deceleration at $z>1$ (R04).

An alternative to evolution proportional to redshift would be
evolution that is proportional to look-back time (Wright 2002).
However, this seems unlikely, as the evolution of cosmic properties to
which SNe~Ia may respond is more closely correlated with redshift than
with time.  Two natural examples are the evolution of cosmic
metallicity as seen in damped Ly$\alpha$ systems (Kulkarni et
al. 2005) and the minimum stellar mass of a star which can turn off
the main sequence and donate the SN~Ia white dwarf.  Combined with the
necessary delay required to produce SNe~Ia ($>$ 1 Gyr), the gradient
in these physical parameters lies primarily with redshift, not time
(Riess \& Livio 2006).

Recent work has questioned the validity of the mean Galactic
reddening-law parameter, $R_V = A_V/E_{B-V} = 3.1$, for use in
determining the extinction in extragalactic SNe.  Estimates of the
typical value of $R_V$ in distant galaxies hosting SNe~Ia tend to
prefer values somewhat less than 3.1 (Riess, Press, \& Kirshner 1995;
Jha 2002; Jha, Riess, \& Kirshner 2006; Phillips et al. 1999), with
values as low as $R_V=2.3$ by Wang et al. (2006).  An apparent
correlation exists between $R_V$ and $A_V$: SNe with greater
extinction prefer lower values of $R_V$, and both of the well-studied
SNe~Ia with $A_V > 2$ mag (SN 1999cl and SN 1996ai) yield $R_V < 2.0$
(Jha, Riess, \& Kirshner 2006).  Wang et al. (2006) have suggested, 
as a natural explanation for reduced values of $R_V$, the occasional
dominance of scattering over absorption caused by dust clouds in the
circumstellar environment.  Evidence does exist for the occasional and
significant presence of circumstellar dust from at least 3 SNe~Ia
exhibiting light echos (SN 1995E, Quinn et al. 2006; SN 1998bu,
Cappellaro et al. 2001; SN 1991T, Schmidt et al. 1994) and a few
possible SNe~Ia with circumstellar hydrogen seen at late times (e.g.,
SN 2002ic; Hamuy et al. 2003).  In the presence of local dust, Wang
et al. (2006) show that $1.8 < R_V < 3.1$ for normal, Milky-Way-type dust.

However, past determinations of the mean $R_V$ for distant SNe~Ia are
strongly biased toward the few objects with large extinction and lower
$R_V$, which have greater weight in these analyses.  For example, Jha,
Riess, \& Kirshner (2006) analyzed the 33 nearby SNe~Ia with the
highest extinction ($A_V > 0.5$ mag) and found a median $R_V=2.8$,
well in excess of the weighted mean value of 1.8 (with half of this
difference explained by the two SNe Ia with $A_V > 2$ mag).  Because
our cosmological analysis excludes (from the Gold sample) all SNe with
$A_V > 0.5$ mag, we would expect the characteristic $R_V$ in our
sample to be better represented by the median, or perhaps closer to
Galactic for low-extinction events.  Further analyses of large samples
of low-extinction events may yield greater insight on this issue.

Analyses which seek to correlate SN color with luminosity, such as
those by Tripp \& Branch (1999), Guy et al. (2005), and Astier et
al. (2006), combine both the reddening-law parameter $R_V$ and
intrinsic luminosity-color relation parameter into a single relation.
If these independent, physical relations have the same dependence
between color and brightness, or if the relative contributions of dust
and populations to the relation do not change with redshift, then
combining these effects into one should succeed.  However, if one of
these assumptions fails significantly (e.g., there is more or less dust
on average at different redshifts), important systematic errors in
the cosmological inference may result.  In any case, the effective
value of $R_V$ from these analyses should be an average of the value
from dust and the physics of SN explosions.  The MLCS2k2 templates
suggest that the intrinsic relation from normal SNe Ia would be
$M/(B-V) \approx 1.8$, reducing the effective $R_V$ below the typical
value for the Galaxy as found by Tripp \& Branch (1999), Guy et
al. (2005), and Astier et al. (2006).

To determine the sensitivity of our cosmological analyses to a reduced
value of $R_V$, we refit all of the distances using $R_V=2.5$ and 2.0
and compared the results to our preceding distances determined with
$R_V=3.1$.  We found our mean distance to increase by 0.026 mag and
0.042 mag for $R_V=2.5$ and $R_V=2.0$, respectively.
However, we found the mean variation of this change to be negligible
across redshift with less than a 0.01 mag difference between the bins
used in Figure 7.  The distance determinations to objects with large
extinction ($A_V > 1$ mag) depend more critically on $R_V$, with an
increase of 0.10 and 0.35 mag for SN 2002kc ($A_V=1.35$ mag for
$R_V=3.1$) for $R_V$=2.5 and 2.0, respectively.  Such sensitivity is
exactly the reason it is unwise to make use of well-reddened objects
in cosmological analyses and why we limited our Gold sample to $A_V <
0.5$ mag.

Wood-Vasey et al (2007) has shown that a selection bias exists in the distribution of
observed extinctions and luminosities of SNe Ia found in the ESSENCE Survey
for supernovae discovered near the redshift limit of the survey.  This bias can be 
corrected using Monte Carlo simulations of the survey selection.  The bias becomes significant
for SNe Ia which can only be found at a signal-to-noise ratio of less than 15 due to their apparent faintness
and the survey's limitations.  For our HST SN survey, this limit occurs for SNe Ia at $z>1.4$
and thus such selection bias has little affect on our sample, the great majority of which is at $z<1.4$.
The reason for the absence of SNe Ia at the redshift limit of our survey ($z \sim 1.7$) has been attributed to
an apparent 2 to 3 Gyr delay between the formation of stars and the production of SNe Ia (Strolger et al. 2004).

At present, none of the {\it known}, well-studied sources of
systematic error rivals the statistical errors presented here.

Now we consider systematic errors whose definition includes our
present inability to discover them and to correct for them.  Here
we will quantify the translation of such systematic errors to our data
analysis and their propagated effects on the cosmology.  A simple form
for general systematic errors in SN distances to take is a correlation with redshift.  Such a model provides
an adequate (if naive) description of zeropoint errors tied to fixed
passbands, errors in K-corrections, random errors from evolution (tied
to look-back time), etc.  This model is good at addressing zeropoint
errors between small ground-based telescopes used to collect the
$z<0.1$ sample, errors from large ground-based programs which
collected the sample at $0.1 < z < 0.8$, and from {\it HST} at $z>1$.
It is similar to the model used by the Supernova Acceleration Probe
(SNAP) collaboration (Aldering 2005). 

We define the covariance between two SNe to be $< \sigma_{\mu,1}
\sigma_{\mu,2} > = A^2 {\rm exp}[-| z_1-z_2 |/z_T]$, where $A$
provides the correlation (in mag) for two SNe at the same redshift and
$z_T$ provides the decay length of the correlation in redshift.  For
the expected precision of calibration of {\it HST} and past datasets,
we examine a model with $(A,Z_T)= (0.05,0.05)$, and $(0.05,0.10)$
against the error-free case (0.0,0.0).  We consider (0.05,0.10) to be
a rough estimate of the covariance of the SN dataset, in which SNe at
like redshifts suffer the same 0.05 mag errors related to calibration,
those transformed to adjacent rest-frame bandpasses ($z_T > 0.20$)
share 0.02 mag of systematic error, and the entire Hubble flow sample
($z_T=0.1$) shares an 0.03 mag error in common (and is not improved
beyond the inclusion of $\sim$ 100 SNe~Ia).  For the more optimistic
model, (0.05,0.05), the correlation length of these errors is halved
in redshift.

For the Gold sample of SNe~Ia presented here, we use the redshifts
and the distance errors (but not the distance measurements) to compute
the Fisher Information Matrix combined with a typical
prior on $\sigma_{\Omega_M} = 0.03$.  To study these systematic errors we
include a covariance matrix with off-diagonal terms given by our model
and parameters.  We then compute the expected 68\% confidence level
interval for the parameter space $w_0-w'$ for the 3 cases.  We find
that our best guess at covariance, (0.05,0.10) broadens the interval
by 27\% whereas the more optimistic (0.05,0.05) broadens the
interval by 17\%.  Thus we conclude that our present results remain
dominated by statistical errors.  However, as the sample grows, for
this balance to remain true, ongoing improvements in calibration must
be realized.

\section{Summary and Conclusions}

(1) We present 21 new {\it HST}-discovered SNe~Ia and an improved
calibration of the previous sample from R04.  Together this sample
contains 23 SNe Ia at $z \geq 1$, extending the Hubble diagram over 10
Gyr.

(2) We derive uncorrelated, model-independent estimates of $H(z)$
which well-delineate current acceleration and preceding deceleration.
The {\it HST}-discovered SNe~Ia measure $H(z>1)$ to slightly better
than 20\% precision.

(3) The full {\it HST}-discovered SN~Ia sample, presented here,
provides a factor of two improvement over our present ability to
constrain simple parameterizations of the equation-of-state parameter
of dark energy ($w$) and its evolution.

(4) Stronger priors and tighter constraints on the preferred
cosmological model can be extracted from independent measurements tied
to the surface of last scattering, but the use of these requires
assumptions about the behavior of dark energy across a wide range of
redshift ($1.8 < z < 1089$).  The strongest of these priors, like the
simplest dark energy parameterizations, appears unjustified in the
presence of our current ignorance about dark energy.  Assuming the
effect of dark energy at $z>1.8$ is minimal, we derive meaningful
constraints on the early properties of dark energy:
$w(z>1)=-0.8^{+0.6}_{-1.0}$ and $w(z>1)<0$, i.e., negative pressure,
at 98\% confidence.

(5) At present, we find that the use of additional parameters to
describe $w(z)$ does not provide a statistically significant
improvement to the fit of the redshift-magnitude relation over the use
of a simple cosmological constant.

(6) An analysis of the $z>1$ sample-averaged spectrum shows it to be
consistent with the mean spectrum of SNe~Ia over the last 10 Gyr, failing
to reveal direct evidence for SN~Ia evolution.

\bigskip 
\medskip 

 We are grateful to Dorothy Fraquelli, Sid Parsons, Al Holm, Tracy
Ellis, Richard Arquilla, and Mark Kochte for their help in assuring
rapid delivery of the {\it HST} data.   We also wish to thank Ryan Chornock, Anton Koekemoer, Ray Lucas, Max Mutchler, Sherie Holfeltz, Helene McLaughlin, 
Eddie Bergeron, and Matt McMaster for their help.  Financial support for this work
was provided by NASA through programs GO-9352, GO-9728, GO-10189, and
GO-10339 from the Space Telescope Science Institute, which is operated
by AURA, Inc., under NASA contract NAS 5-26555. Some of the data
presented herein were obtained with the W. M. Keck Observatory, which
is operated as a scientific partnership among the California Institute
of Technology, the University of California, and NASA; the Observatory
was made possible by the generous financial support of the W. M. Keck
Foundation.

\begin{deluxetable}{lllllll} 
\footnotesize
\tablecaption{HST-discovered Sample$^{*}$; Distance Scale of Riess et al. 2004}
\tablehead{\colhead{SN}&\colhead{$z$}&\colhead{$\mu_0^a$} & \colhead{$\sigma^b$} & \colhead{host $A_V$} & \colhead{$\Delta$} & \colhead{sample}}
\startdata
\hline
\hline
1997ff & 1.755 & 45.35 & 0.35 & 0.00  & --- & Gold \nl
2002dc & 0.475 & 42.24 & 0.20 & 0.21  & 0.19 & Gold \nl
 2002dd & 0.950 & 43.98 & 0.34 & 0.35  & -0.34 & Gold \nl
 2003aj & 1.307 & 44.99 & 0.31 & 0.29  & 0.09 & Silver \nl
 2002fx & 1.400 & 45.28 & 0.81 & 0.50  & -0.01 & Silver \nl
 2003eq & 0.840 & 43.67 & 0.21 & 0.22  & -0.04 & Gold \nl
 2003es & 0.954 & 44.30 & 0.27 & 0.10  & -0.08 & Gold \nl
 2003az & 1.265 & 44.64 & 0.25 & 0.73  & -0.4 & Silver \nl
 2002kc & 0.216 & 40.33 & 0.19 & 1.35  & -0.31 & Silver \nl
 2003eb & 0.900 & 43.64 & 0.25 & 0.28  & -0.4 & Gold \nl
 2003XX & 0.935 & 43.97 & 0.29 & 0.26  & -0.31 & Gold \nl
 2002hr & 0.526 & 43.08 & 0.27 & 0.70  & -0.4 & Silver \nl
 2003bd & 0.670 & 43.19 & 0.24 & 0.34  & 0.02 & Gold \nl
 2002kd & 0.735 & 43.14 & 0.19 & 0.21  & 0.12 & Gold \nl
 2003be & 0.640 & 43.01 & 0.25 & 0.42  & -0.22 Gold \nl
 2003dy & 1.340 & 44.92 & 0.31 & 0.43  & -0.4 & Gold \nl
 2002ki & 1.140 & 44.71 & 0.29 & 0.13  & 0.04 & Gold \nl
 2003ak & 1.551 & 45.07 & 0.32 & 0.75  & -0.4 & Silver \nl
 2002hp & 1.305 & 44.51 & 0.30 & 0.22  & 0.32 & Gold \nl
 2002fw & 1.300 & 45.06 & 0.20 & 0.25  & -0.20 & Gold \nl
 HST04Pat & 0.970 & 44.67 & 0.36 & 0.19  & -0.4 & Gold \nl
 HST04Mcg & 1.370 & 45.23 & 0.25 & 0.14  & -0.4 & Gold \nl
 HST05Fer & 1.020 & 43.99 & 0.27 & 0.45  & -0.13 & Gold \nl
 HST05Koe & 1.230 & 45.17 & 0.23 & 0.13  & -0.4 & Gold \nl
 HST05Dic & 0.638 & 42.89 & 0.18 & 0.42  & -0.39 & Silver \nl
 HST04Gre & 1.140 & 44.44 & 0.31 & 0.11  & -0.4 & Gold \nl
 HST04Omb & 0.975 & 44.21 & 0.26 & 0.39  & -0.39 & Gold \nl
 HST05Red & 1.190 & 43.64 & 0.39 & 0.53  & 0.08 & Silver \nl
 HST05Lan & 1.230 & 44.97 & 0.20 & 0.23  & 0.26 & Gold \nl
 HST04Tha & 0.954 & 43.85 & 0.27 & 0.19  & 0.06 & Gold \nl
 HST04Rak & 0.740 & 43.38 & 0.22 & 0.20  & -0.10 & Gold \nl
 HST05Zwi & 0.521 & 42.05 & 0.37 & 0.56  & -0.18 & Silver \nl
 HST04Hawk & 0.490 & 42.54 & 0.24 & 0.18  & -0.40 & Silver \nl
 HST04Kur & 0.359 & 41.23 & 0.39 & 2.49  & -0.34 & Silver \nl
 HST04Yow & 0.460 & 42.23 & 0.32 & 0.43  & -0.04 & Gold \nl
 HST04Man & 0.854 & 43.96 & 0.29 & 0.13  & -0.01 & Gold \nl
 HST05Spo & 0.839 & 43.45 & 0.20 & 0.22  & -0.07 & Gold \nl
 HST04Eag & 1.020 & 44.52 & 0.19 & 0.18  & -0.27 & Gold \nl
 HST05Gab & 1.120 & 44.67 & 0.18 & 0.11  & -0.20 & Gold \nl
 HST05Str & 1.010 & 44.77 & 0.19 & 0.12  & -0.29 & Gold \nl
 HST04Sas & 1.390 & 44.90 & 0.19 & 0.26  & 0.35 & Gold \nl

\enddata 
\tablenotetext{*}{The full sample used here for the cosmological analyses consists of the union of Table 5 from Riess et al. 2004, this table (with
the new distances to SNe from 2002 to 2003 replacing those in Riess et al. 2004), and the SNLS sample from Astier et al. (2006) listed in \S 2.3 of this paper.
The full sample is also available at  http://braeburn.pha.jhu.edu/$\sim$ariess/R06 or upon request to ariess@stsci.edu. }
\tablenotetext{b}{Redshift and velocity error and intrinsic SN Ia dispersion of 0.08 mag already included.} 
\tablenotetext{a}{Distance normalization is arbitrary; see Appendix.}
\end{deluxetable}

\vfill \eject

\begin{figure}[h]
\vspace*{120mm}
\includegraphics{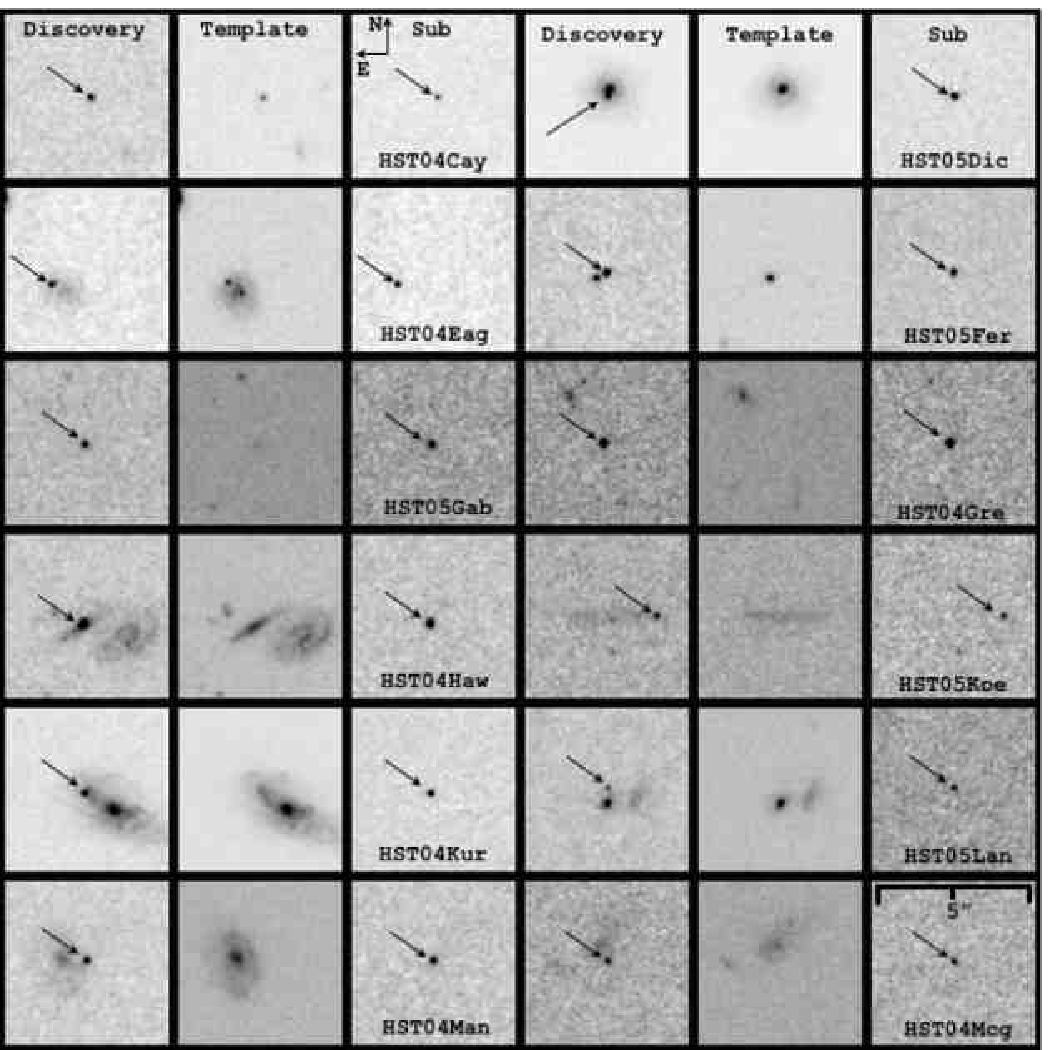}
\end{figure}

\vfill \eject

\begin{figure}[h]
\vspace*{120mm}
\includegraphics{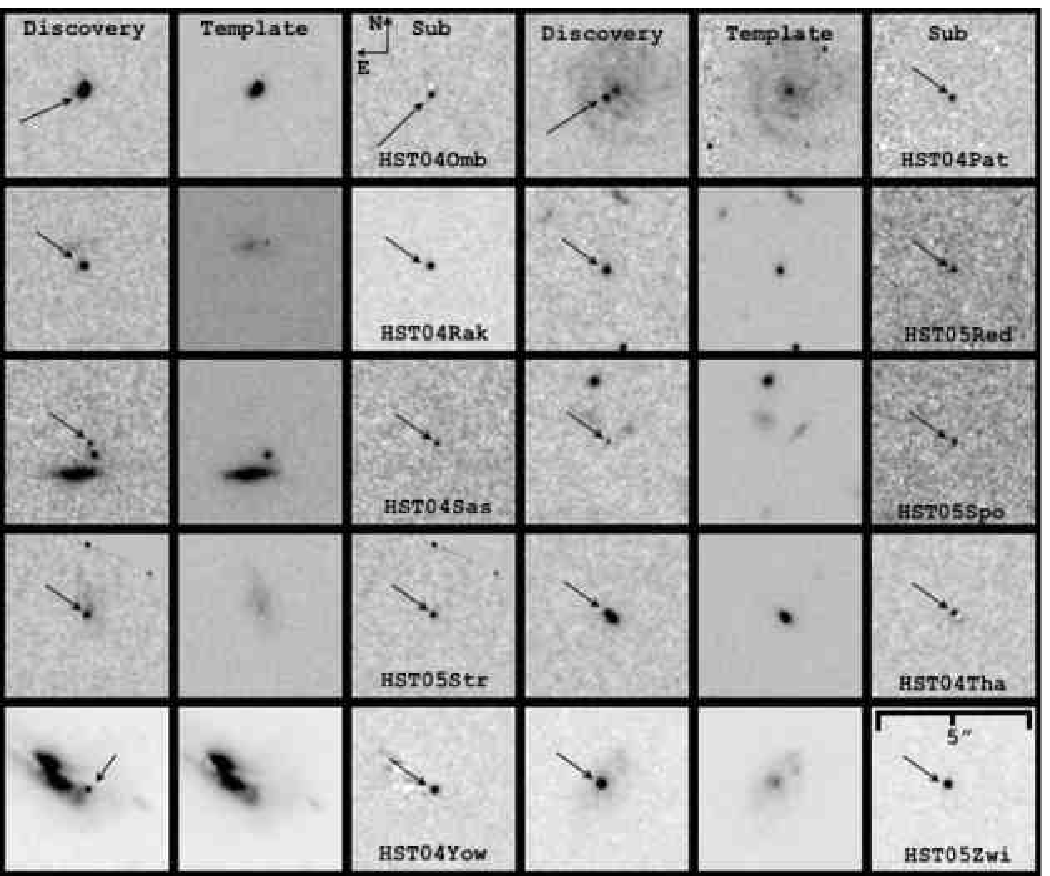}
\end{figure}

\vfill \eject

\begin{figure}[h]
\vspace*{120mm}
\includegraphics{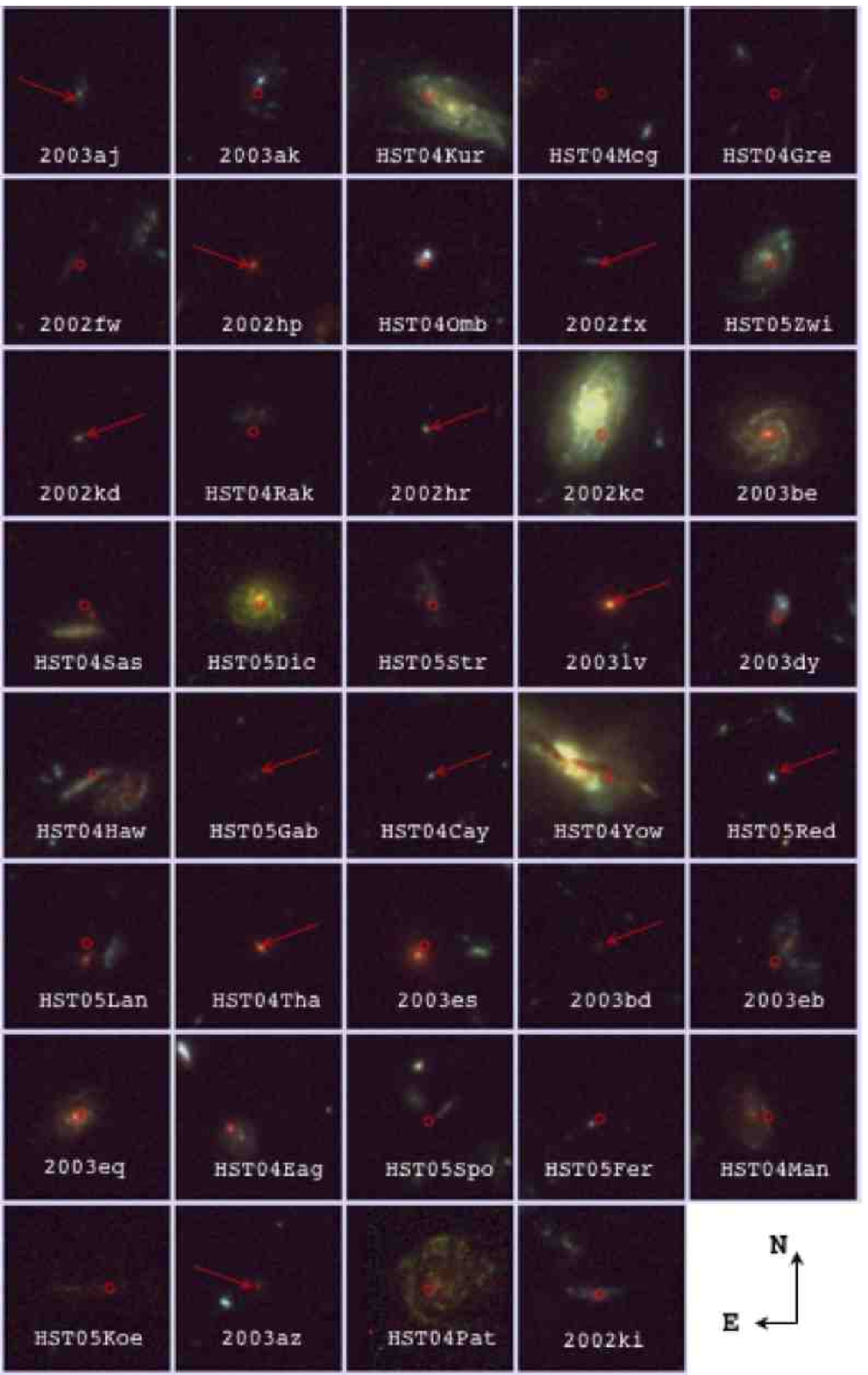}
\end{figure}

\vfill \eject

\begin{figure}[h]
\vspace*{120mm}
\includegraphics{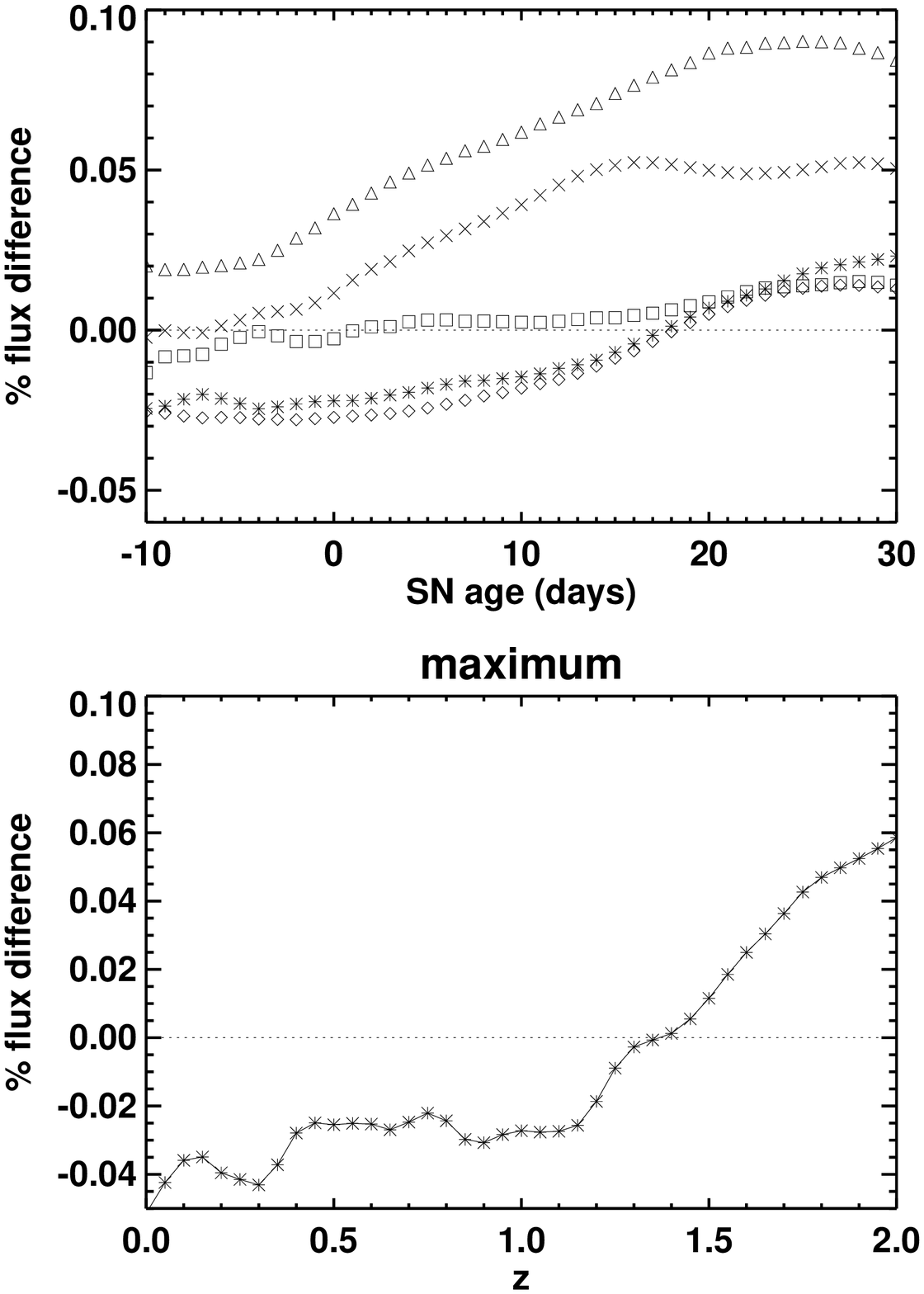}
\end{figure}

\vfill
\eject

\begin{figure}[h]
\vspace*{50mm}
\includegraphics{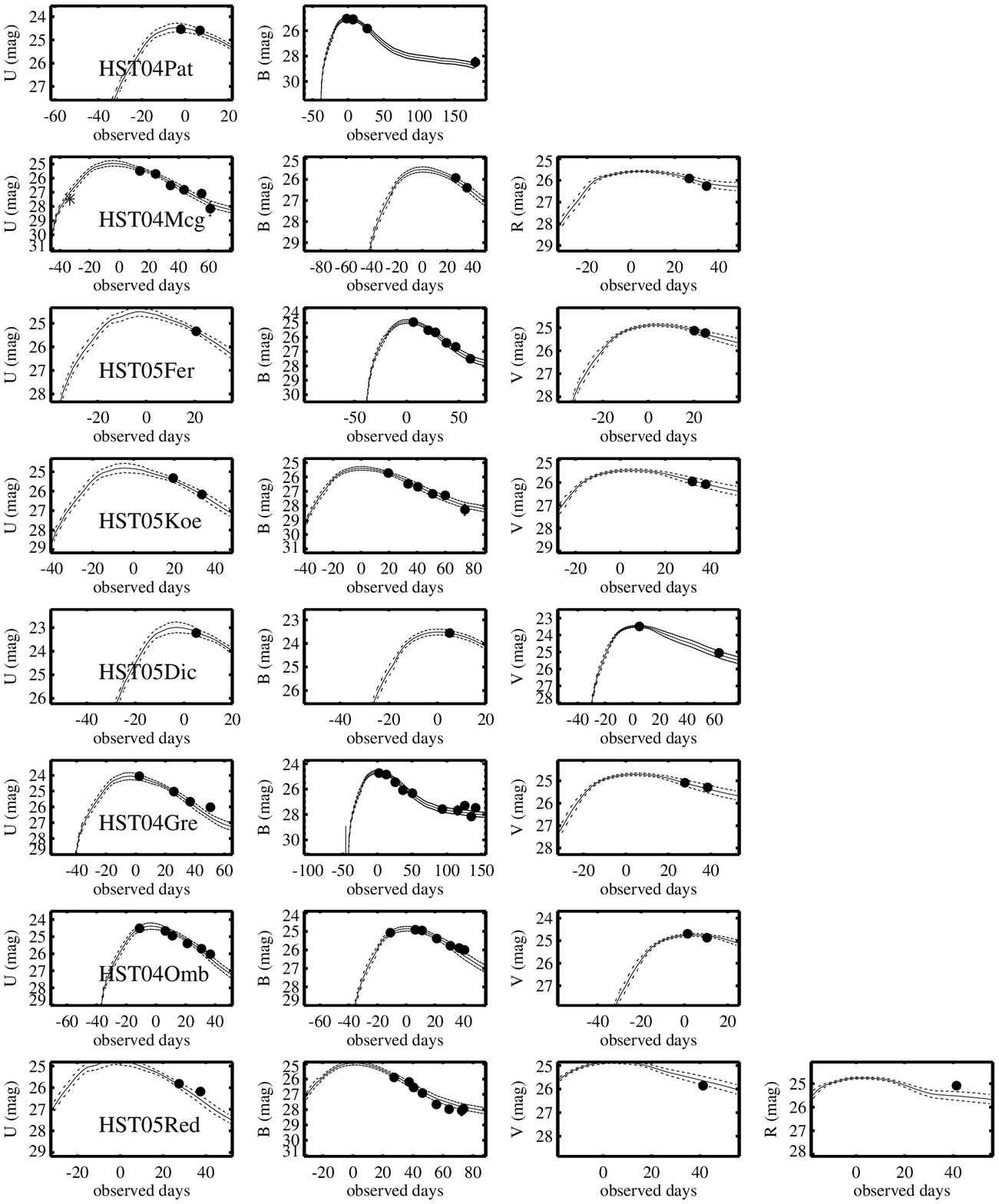}
\end{figure}

\vfill
\eject

\begin{figure}[h]
\vspace*{50mm}
\includegraphics{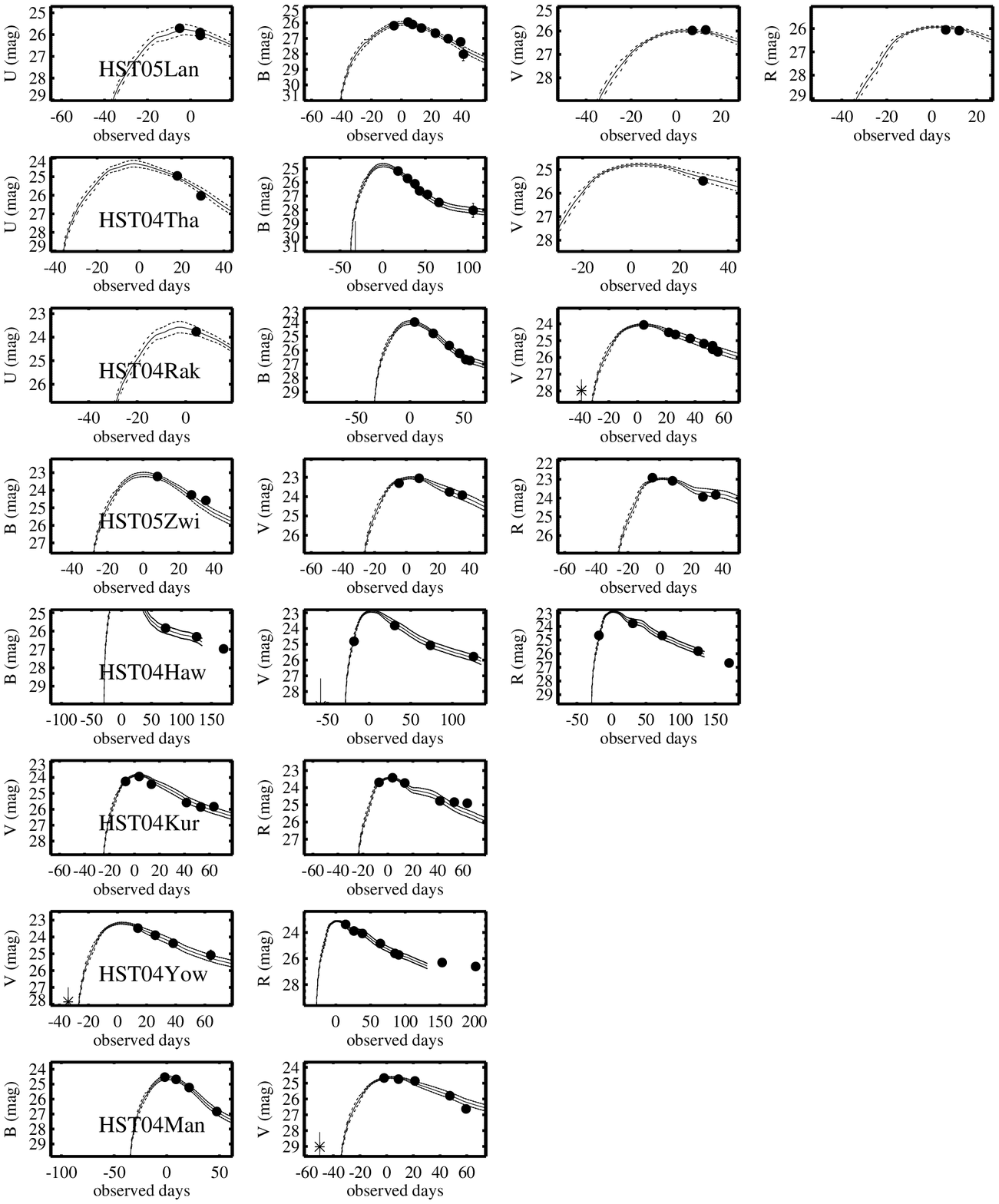}
\end{figure}

\vfill
\eject

\begin{figure}[h]
\vspace*{50mm}
\includegraphics{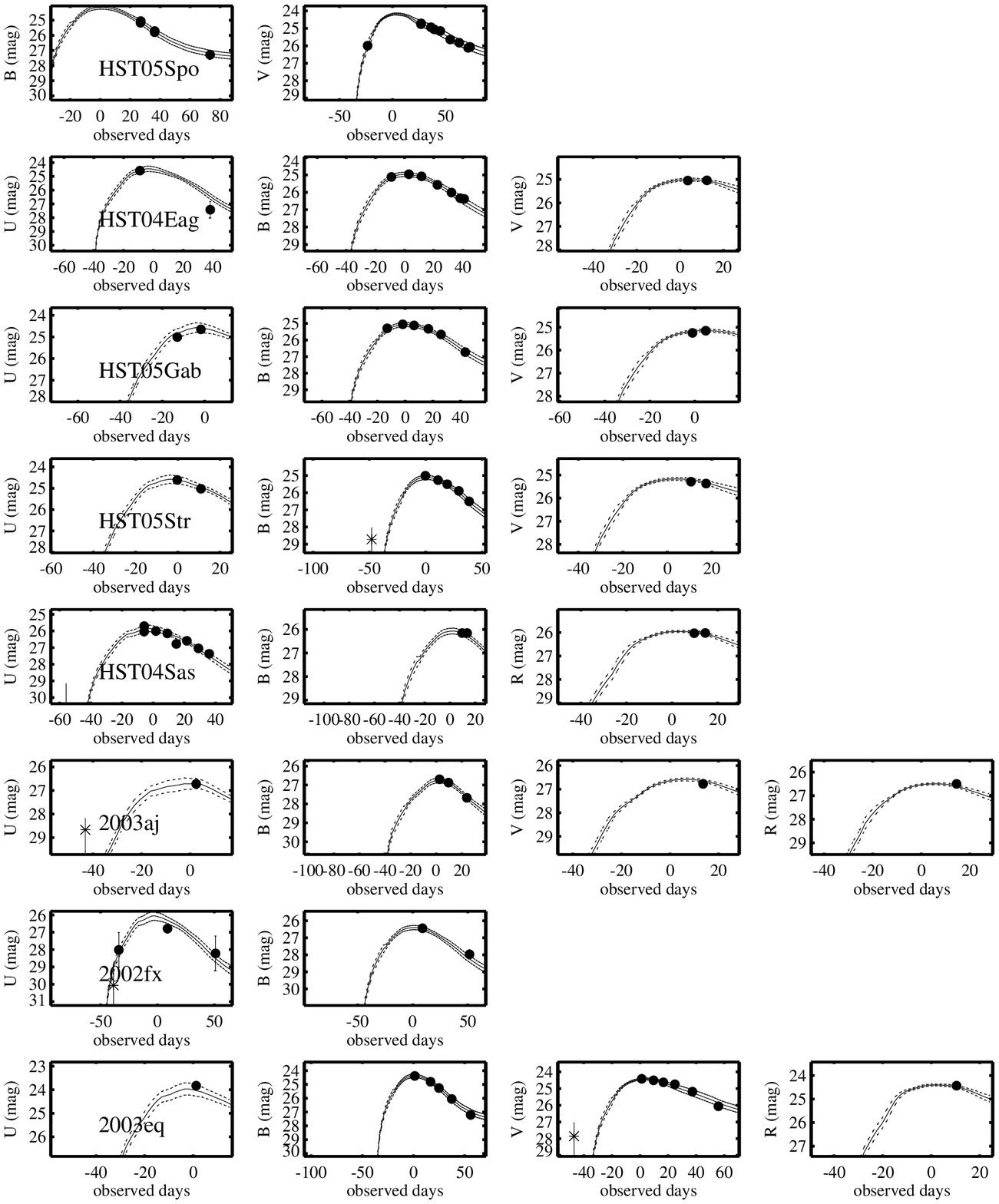}
\end{figure}

\vfill
\eject

\begin{figure}[h]
\vspace*{50mm}
\includegraphics{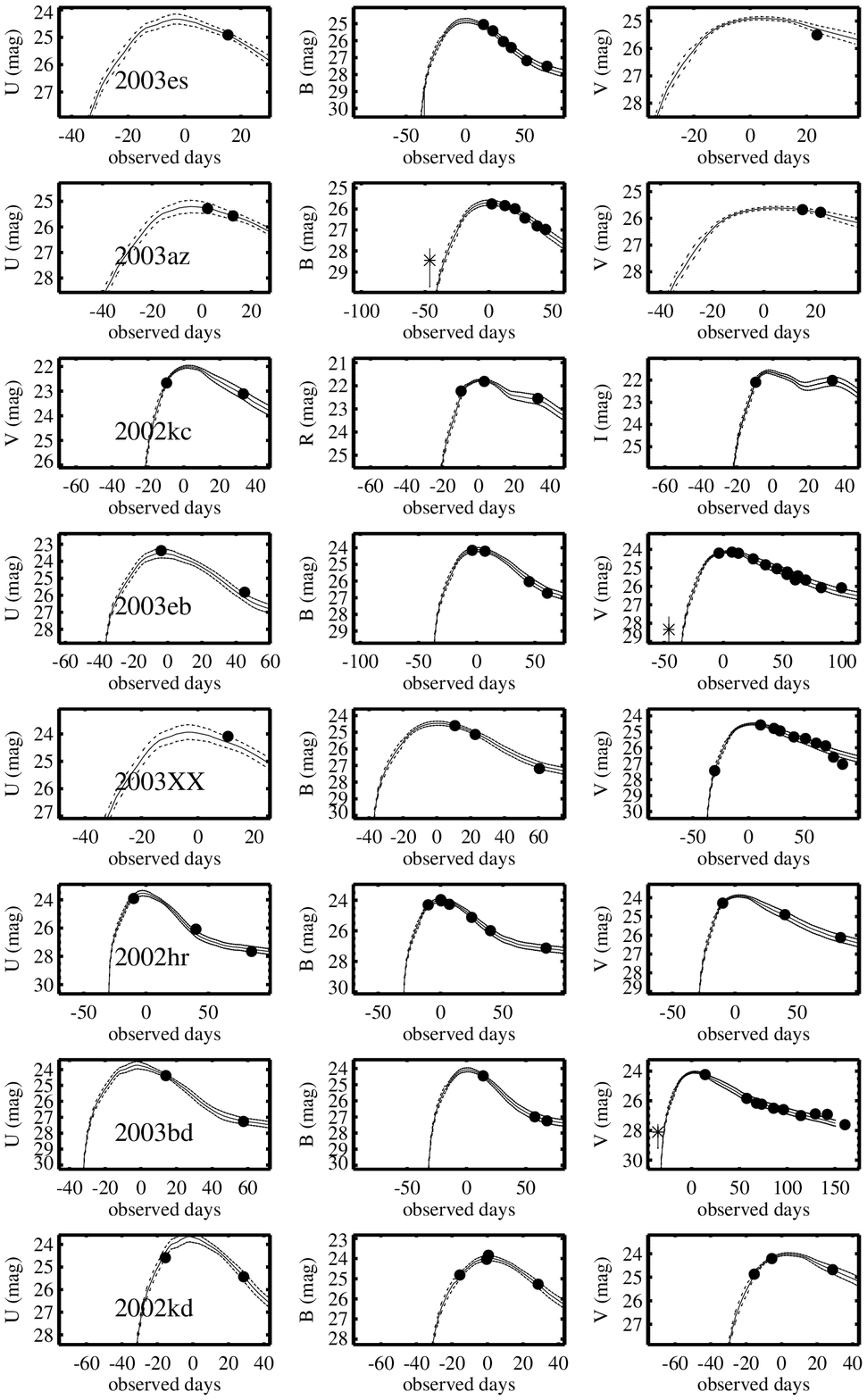}
\end{figure}

\vfill
\eject

\begin{figure}[h]
\vspace*{50mm}
\includegraphics{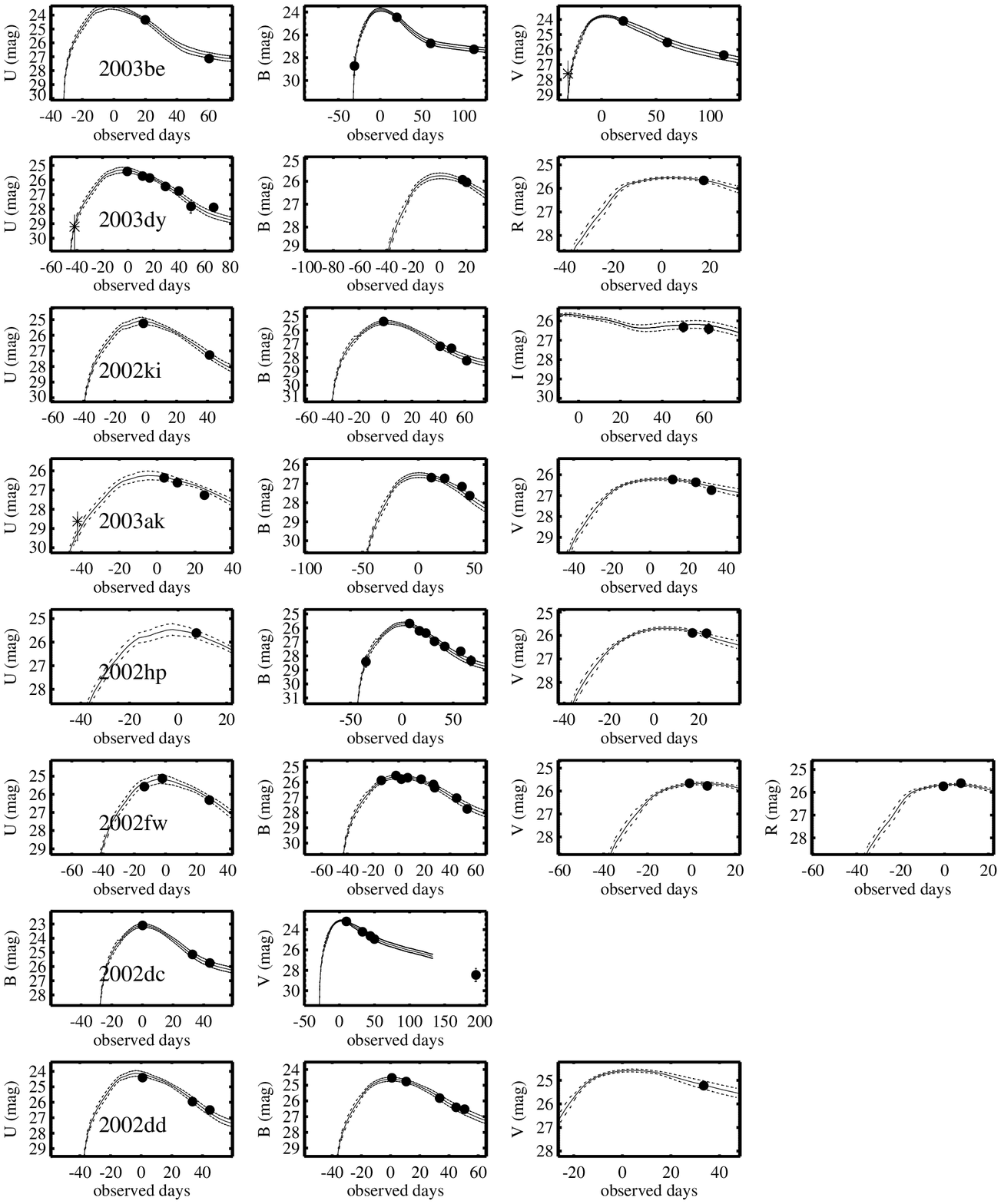}
\end{figure}

\vfill
\eject

\begin{figure}[h]
\vspace*{120mm}
\includegraphics{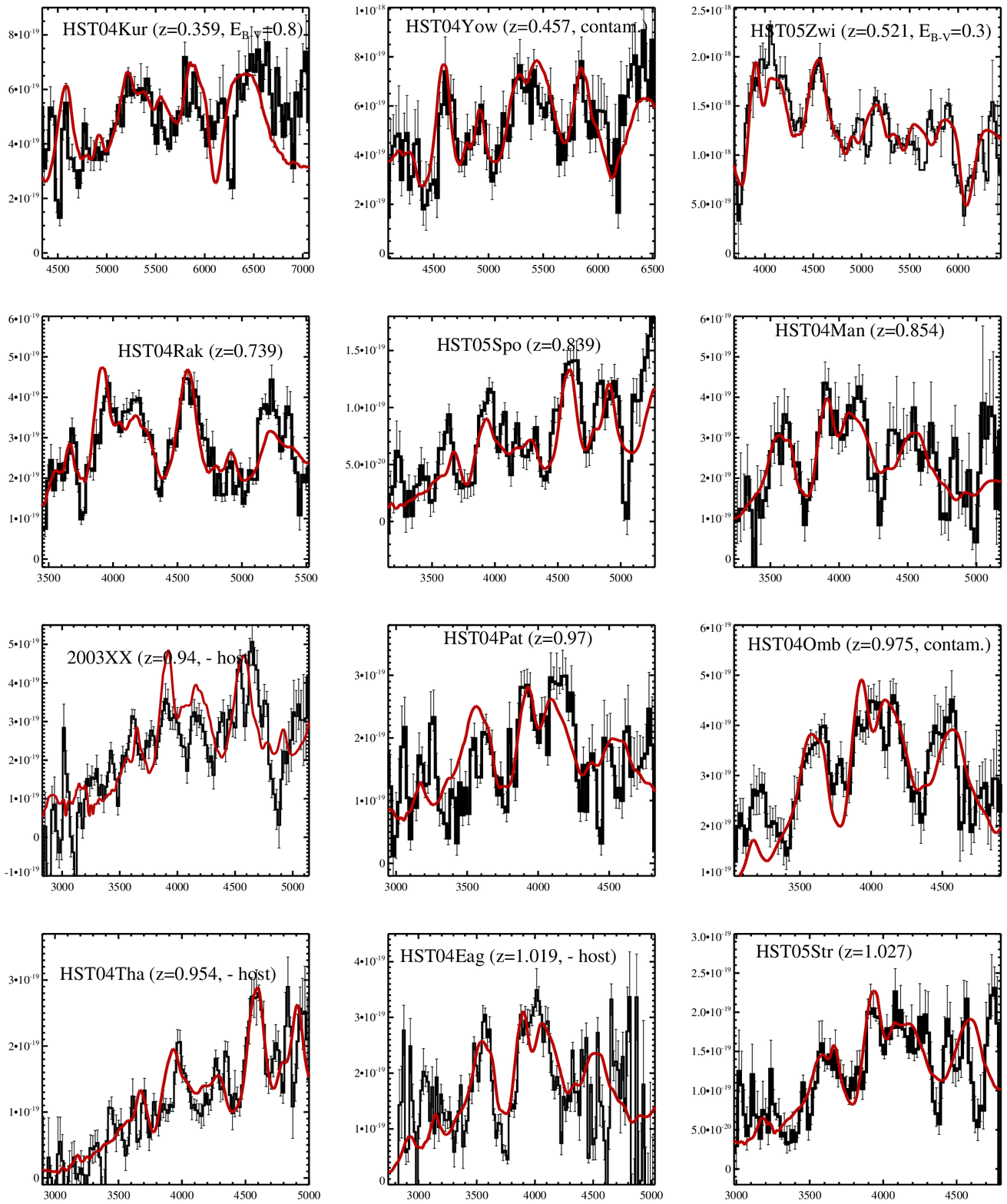}
\end{figure}

\vfill \eject

\begin{figure}[h]
\vspace*{120mm}
\includegraphics{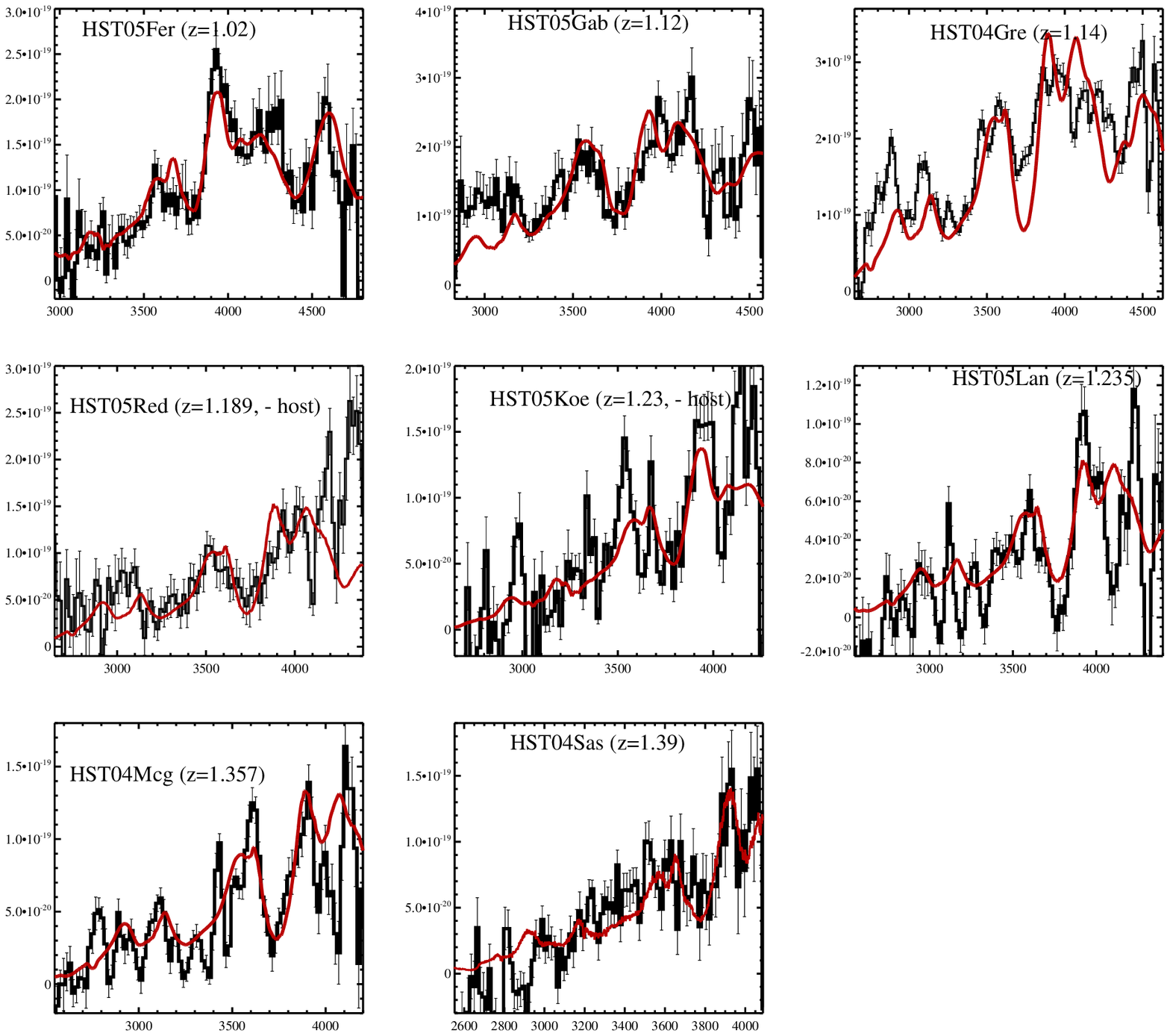}
\end{figure}

\vfill
\eject

\begin{figure}[h]
\vspace*{120mm}
\includegraphics{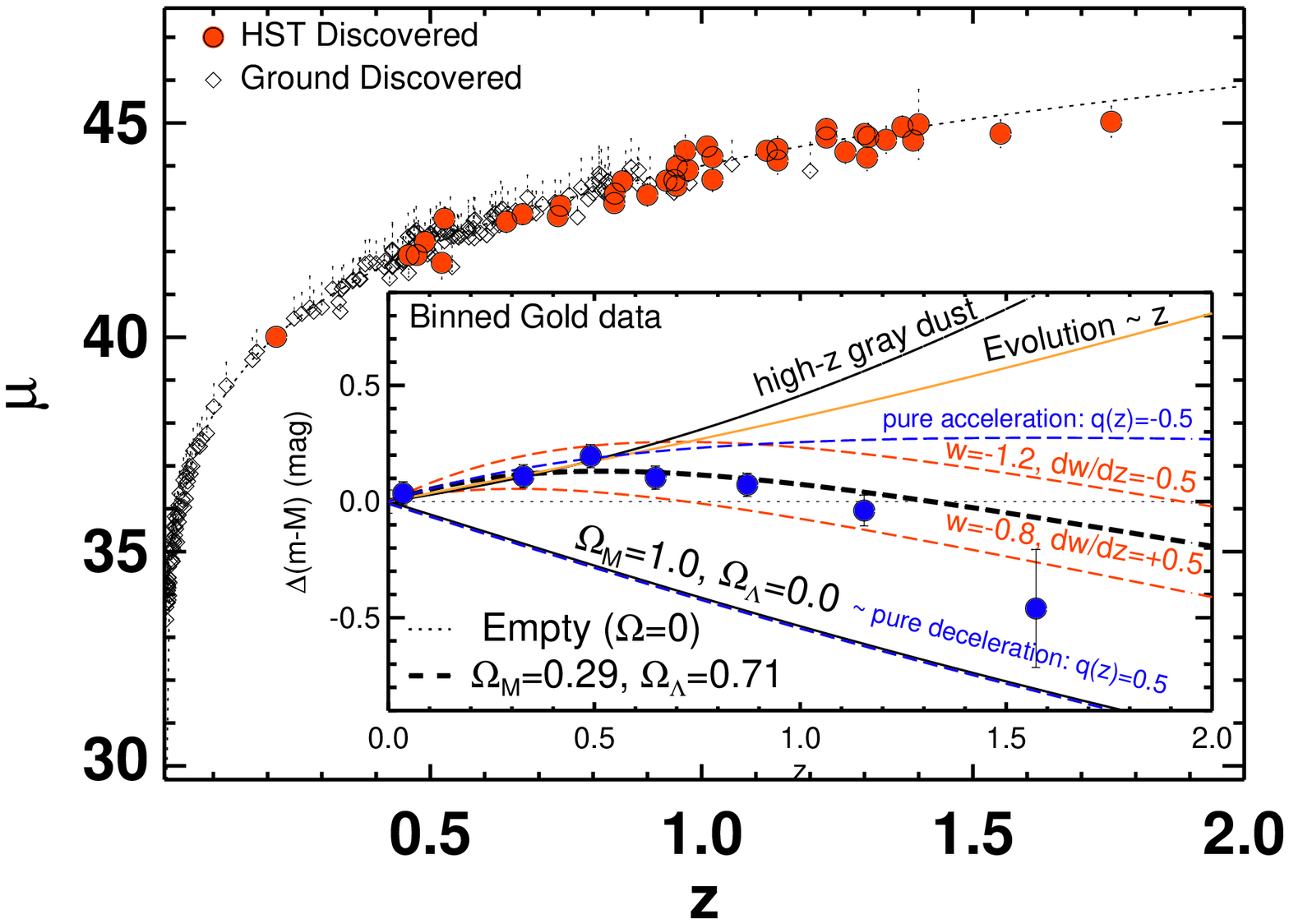}
\end{figure}

\vfill \eject

\begin{figure}[h]
\vspace*{120mm}
\includegraphics{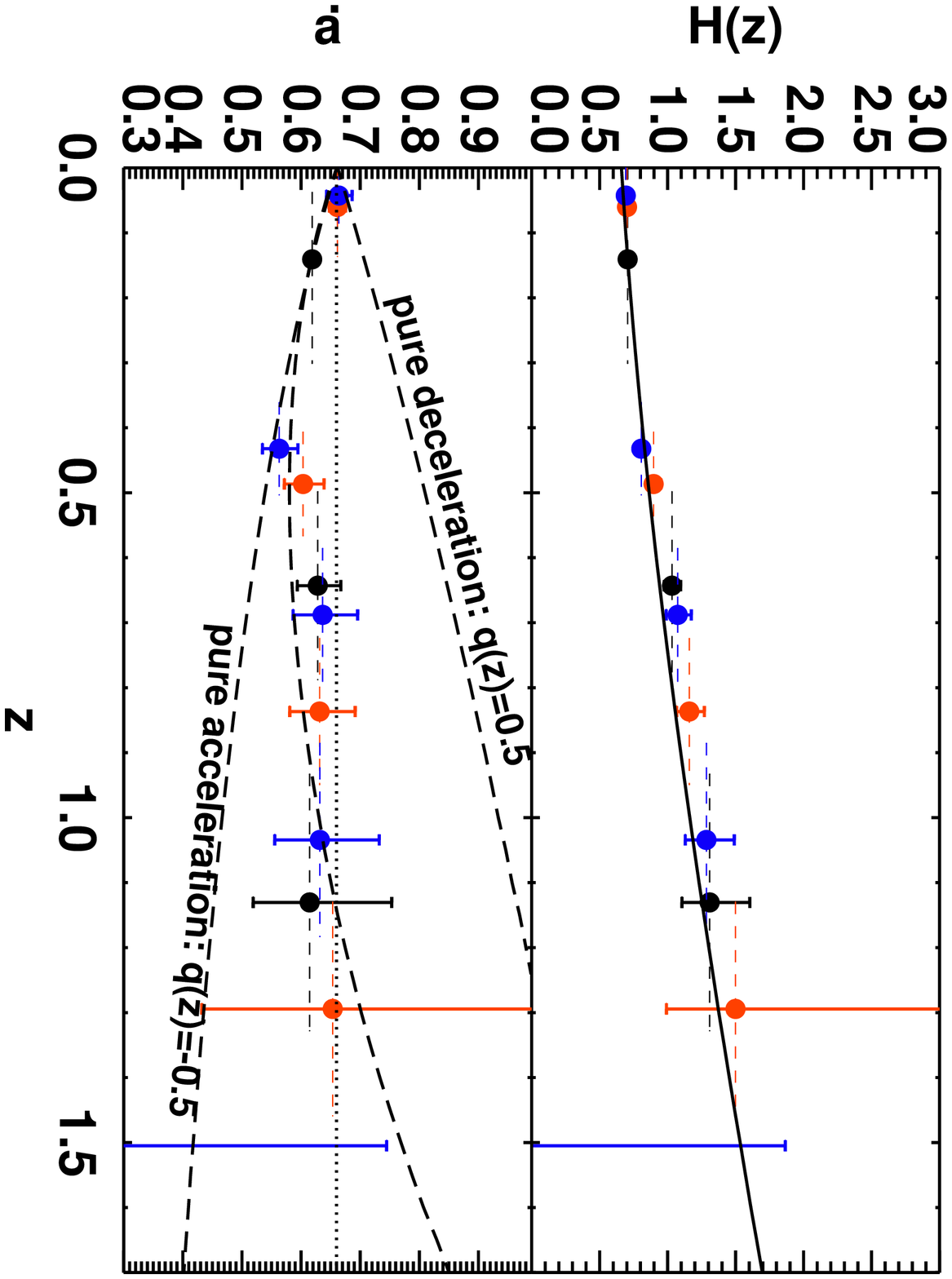}
\end{figure}

\vfill \eject

\begin{figure}[h]
\vspace*{120mm}
\includegraphics{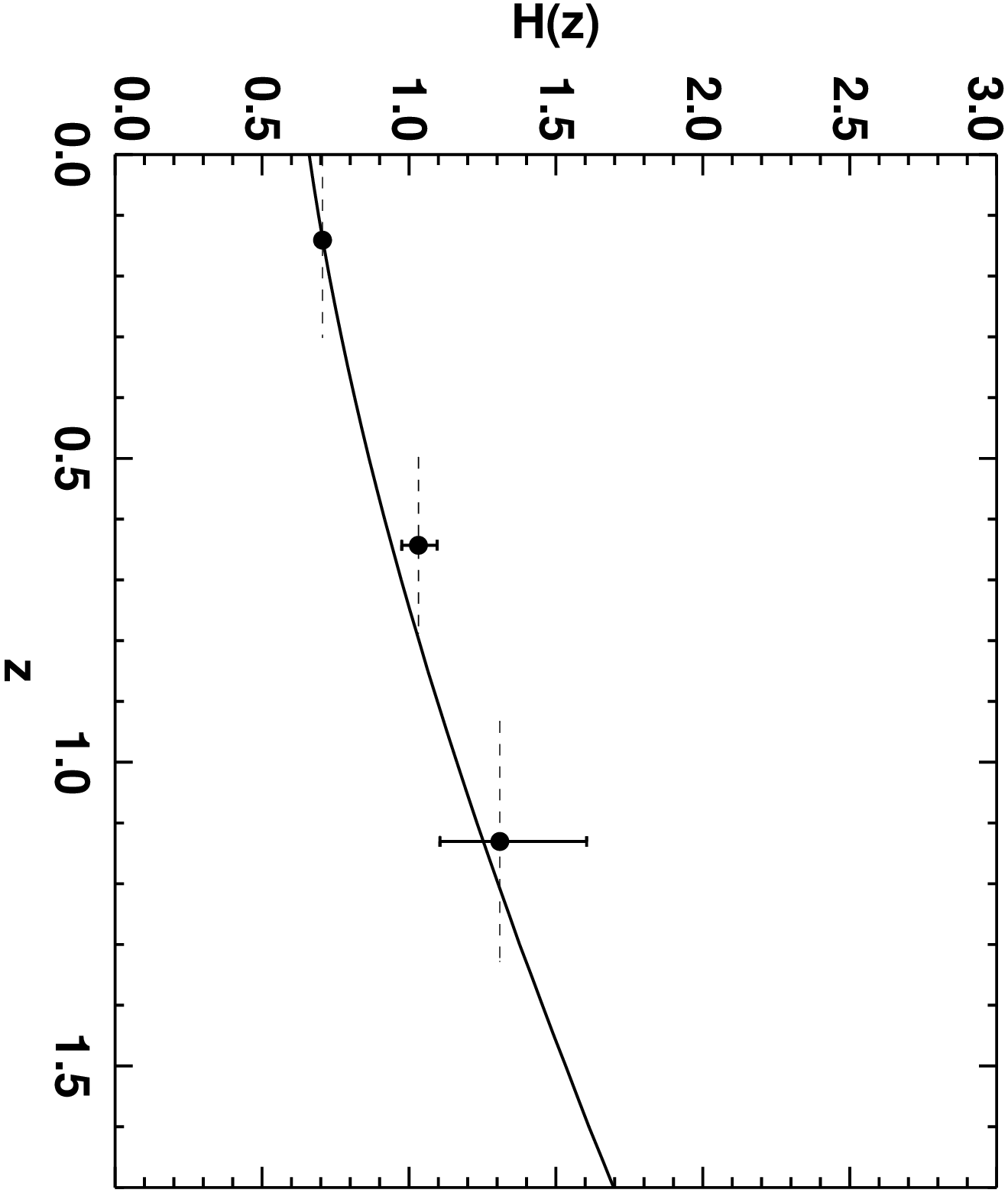}
\end{figure}

\vfill \eject

\begin{figure}[h]
\vspace*{120mm}
\includegraphics{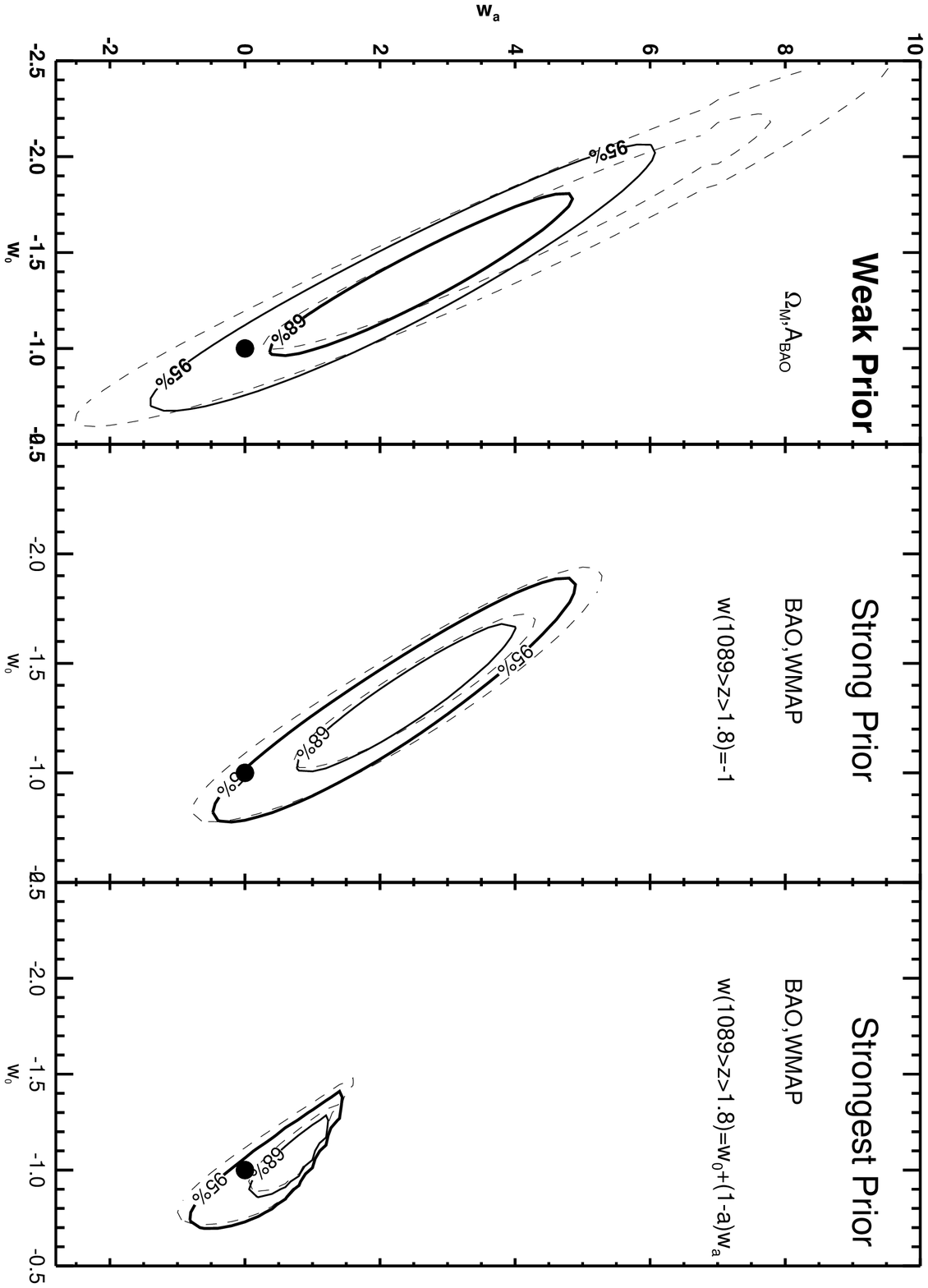}
\end{figure}

\vfill \eject

\begin{figure}[h]
\vspace*{120mm}
\includegraphics{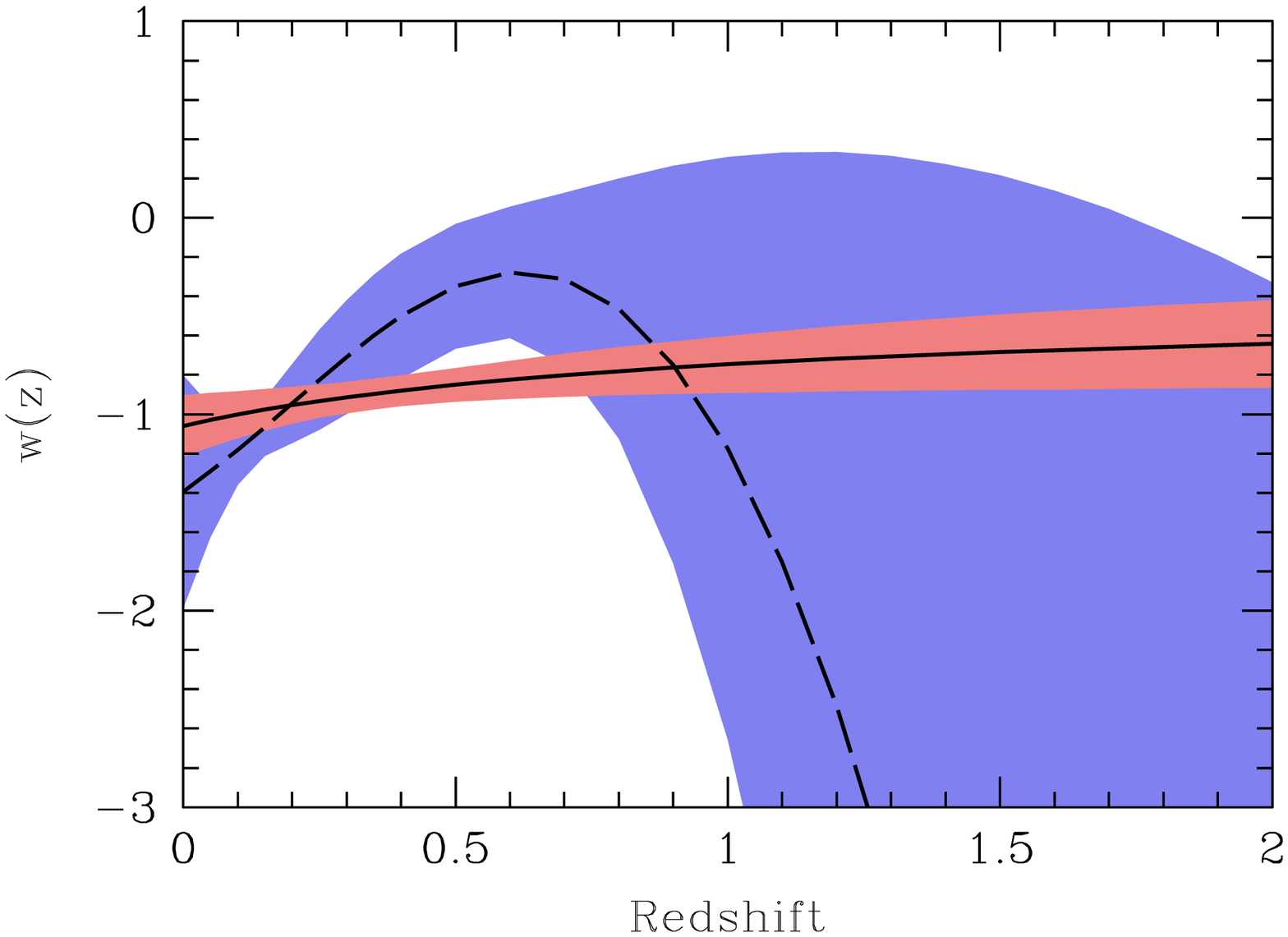}
\end{figure}

\vfill \eject

\begin{figure}[h]
\vspace*{120mm}
\includegraphics{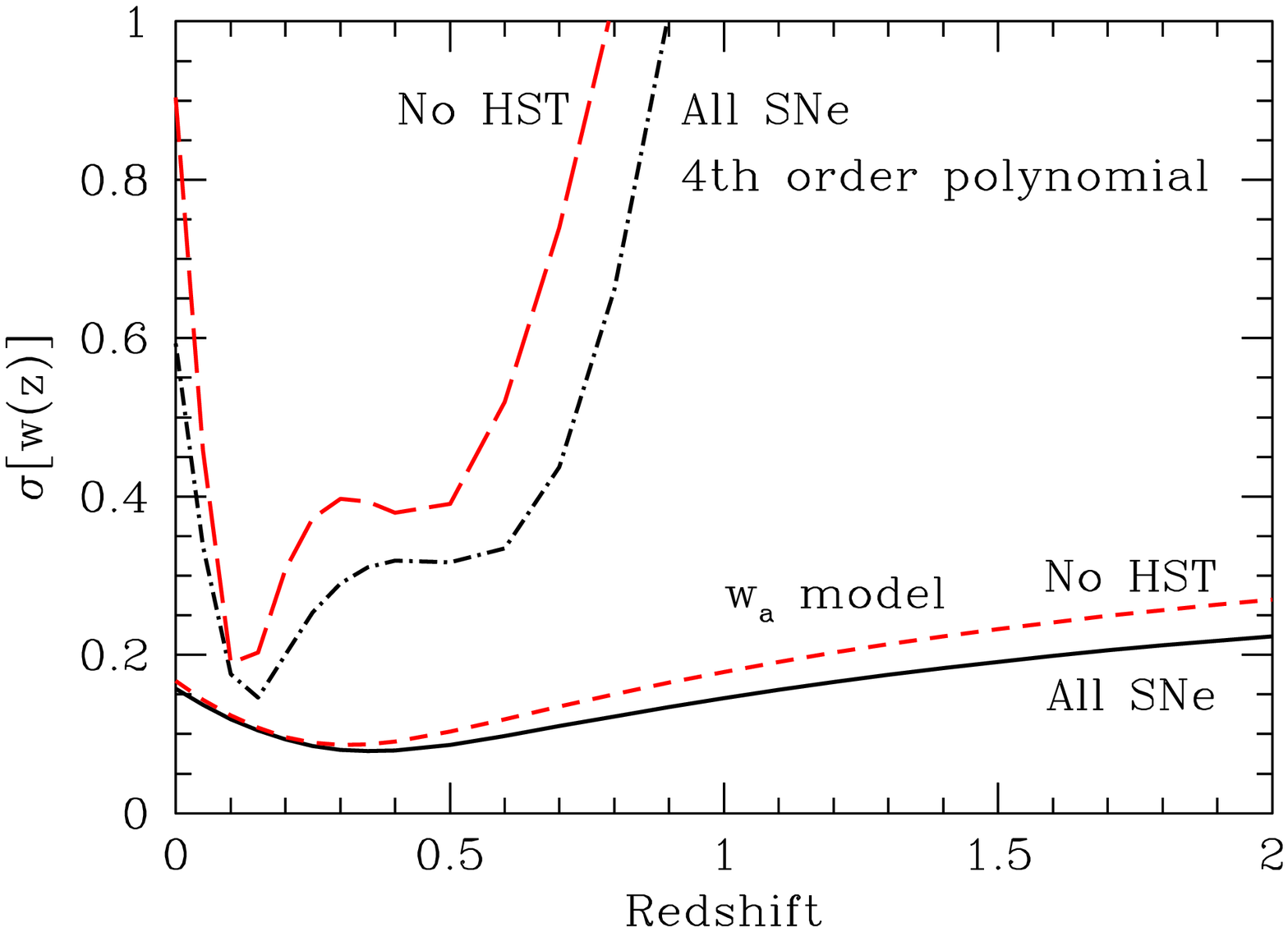}
\end{figure}

\vfill \eject

\begin{figure}[h]
\vspace*{120mm}
\includegraphics{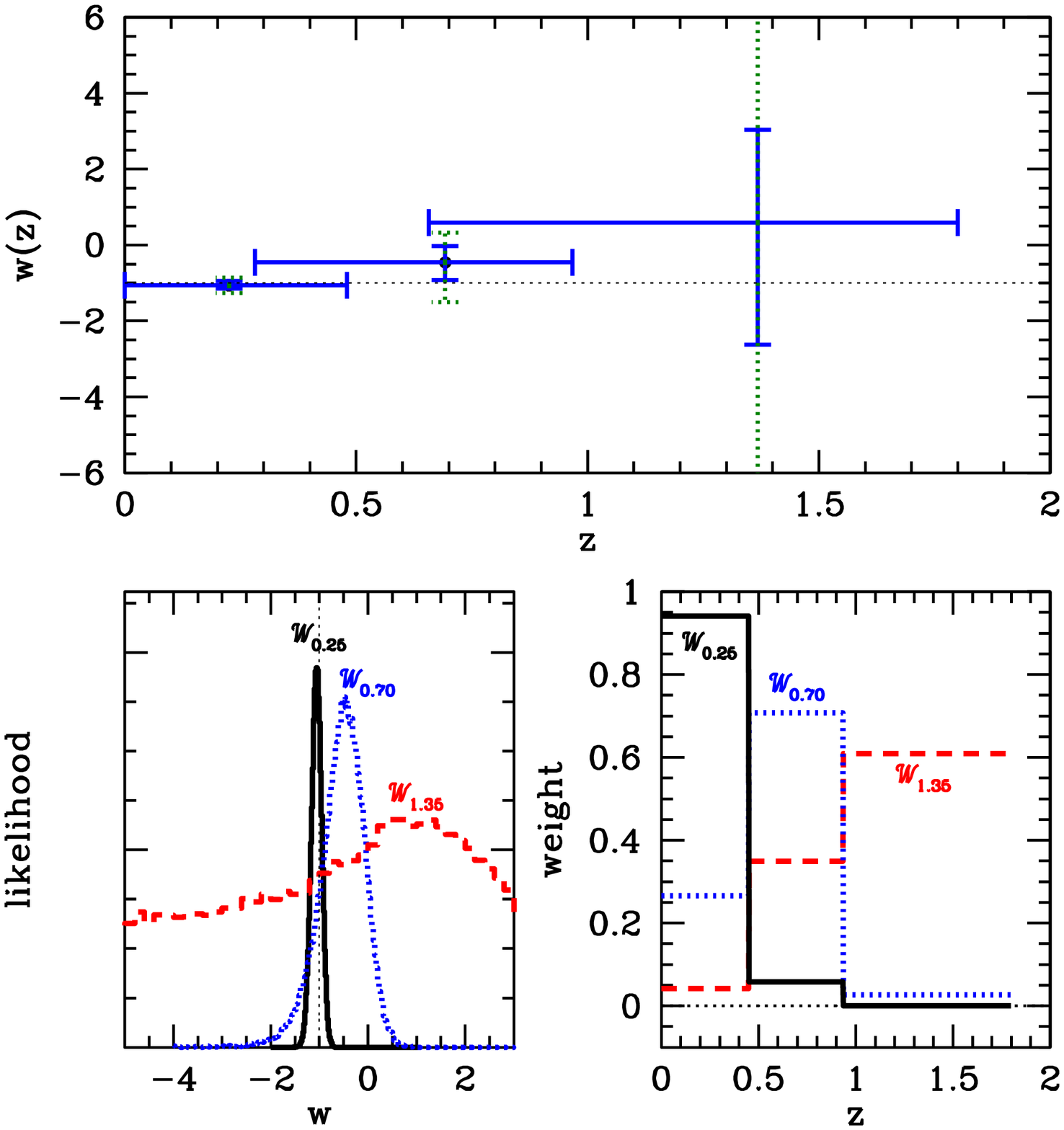}
\end{figure}

\vfill \eject

\begin{figure}[h]
\vspace*{120mm}
\includegraphics{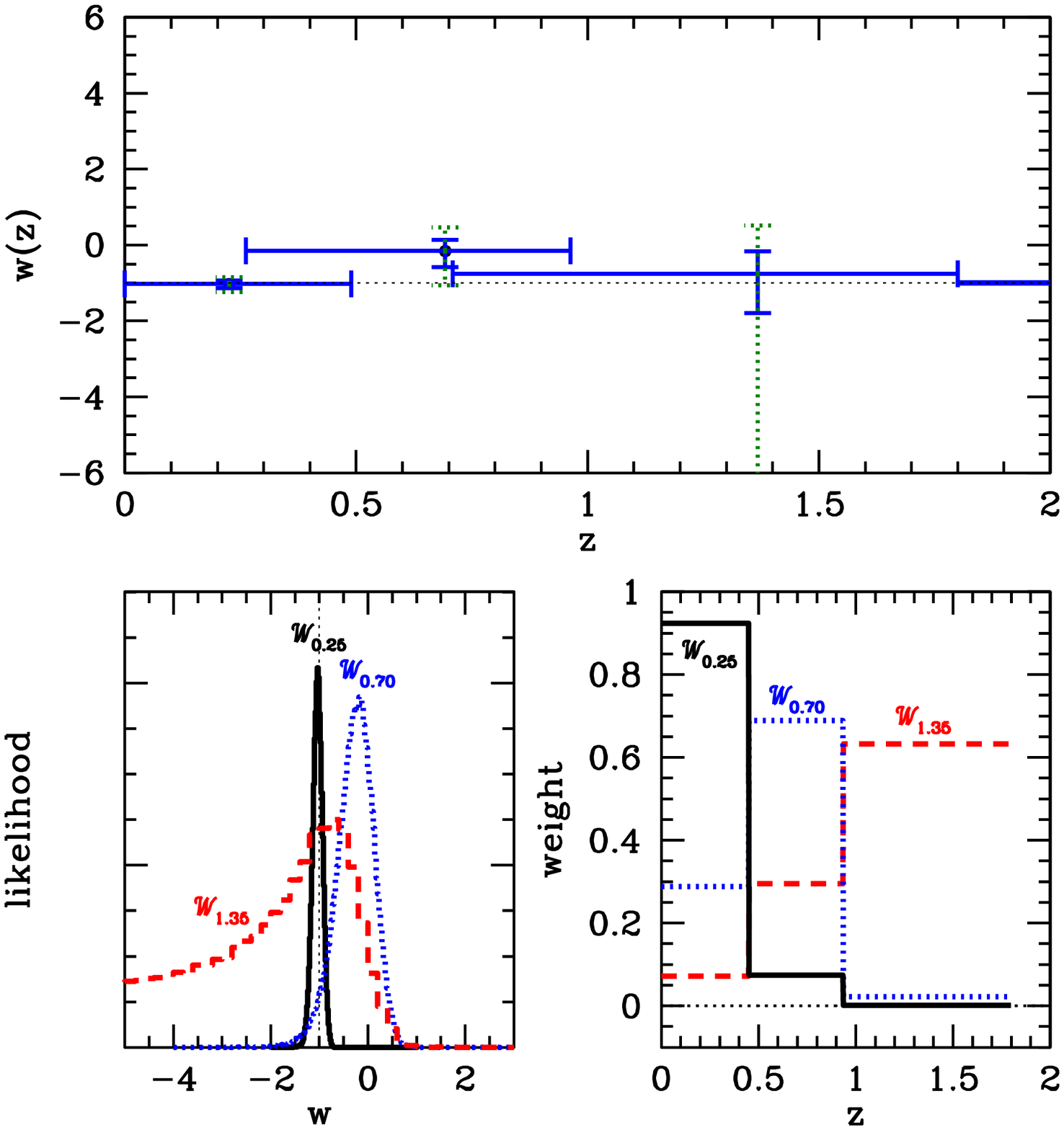}
\end{figure}

\vfill \eject

\begin{figure}[h]
\vspace*{120mm}
\includegraphics{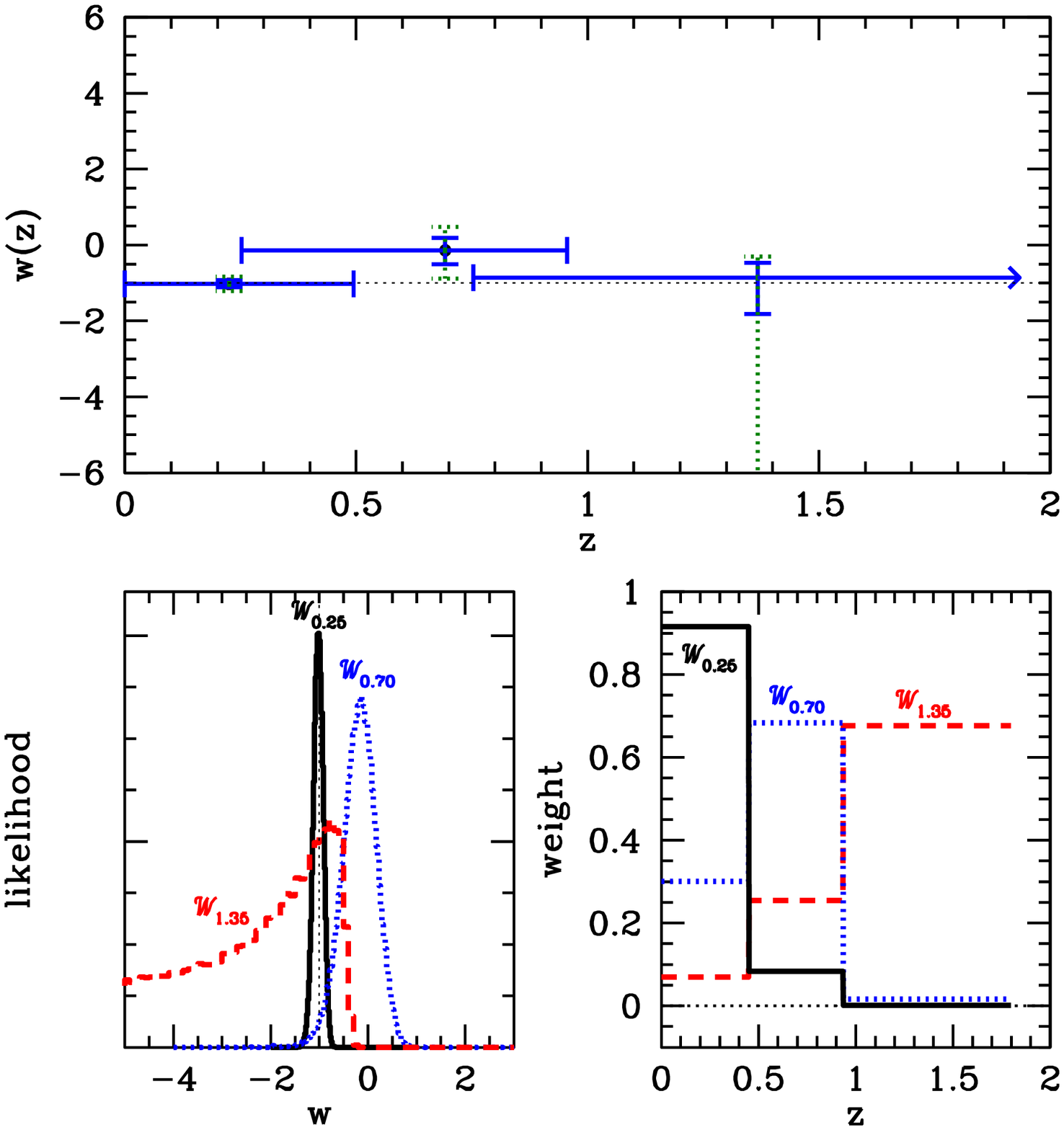}
\end{figure}

\vfill \eject

\begin{figure}[h]
\vspace*{120mm}
\includegraphics{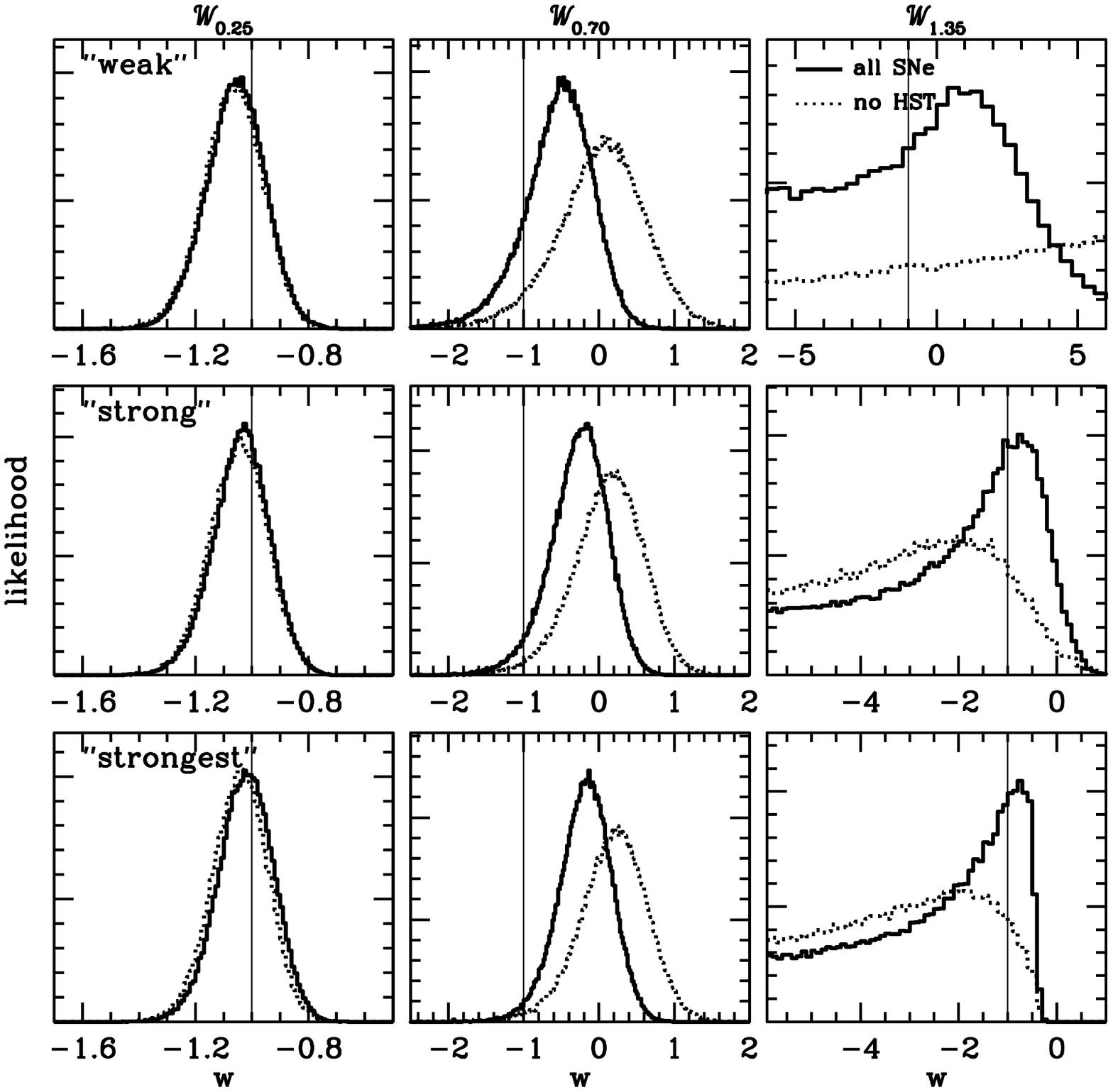}
\end{figure}

\vfill \eject

\begin{figure}[h]
\vspace*{120mm}
\includegraphics{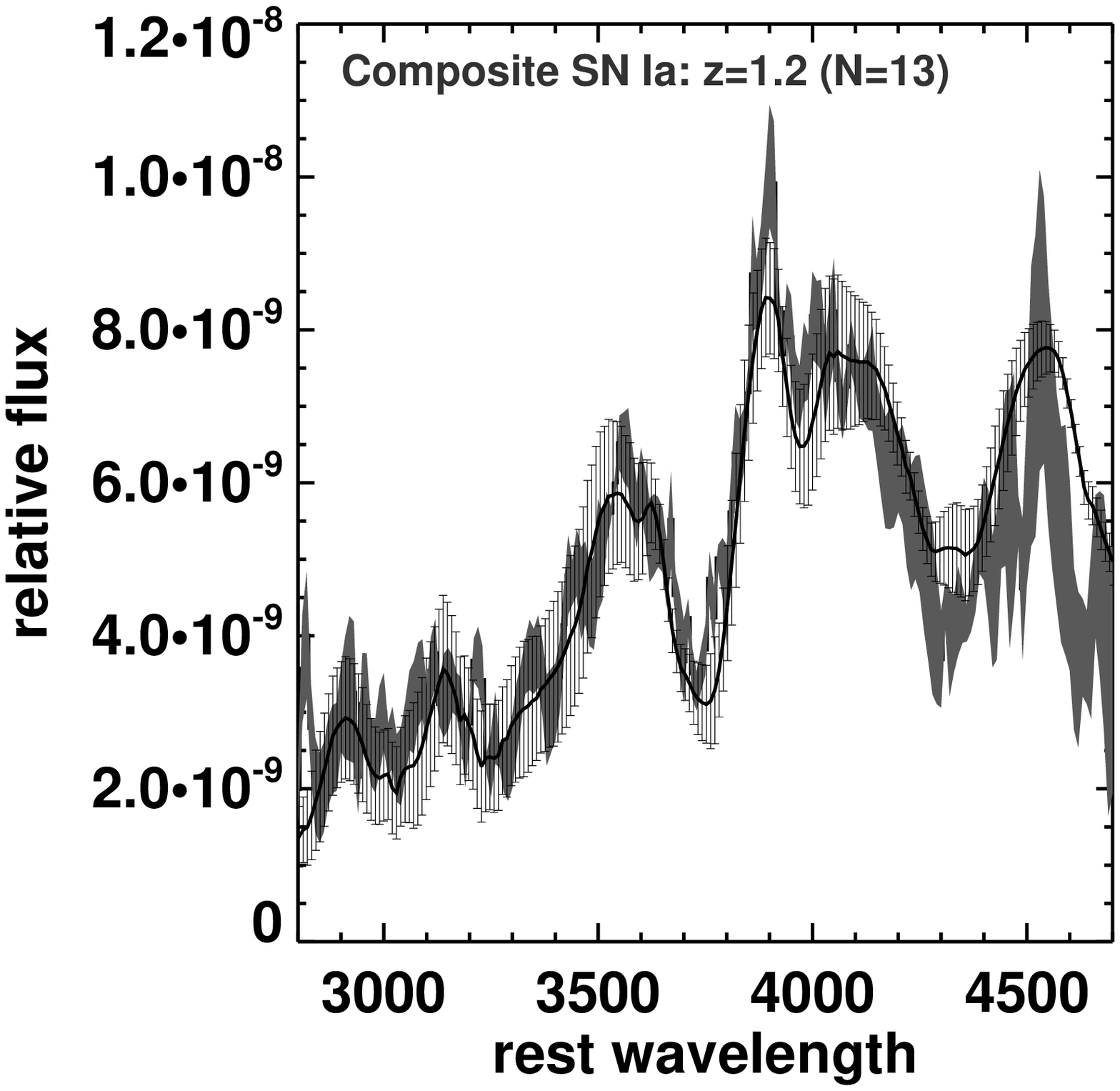}
\end{figure}

\vfill \eject

\begin{figure}[h]
\vspace*{120mm}
\includegraphics{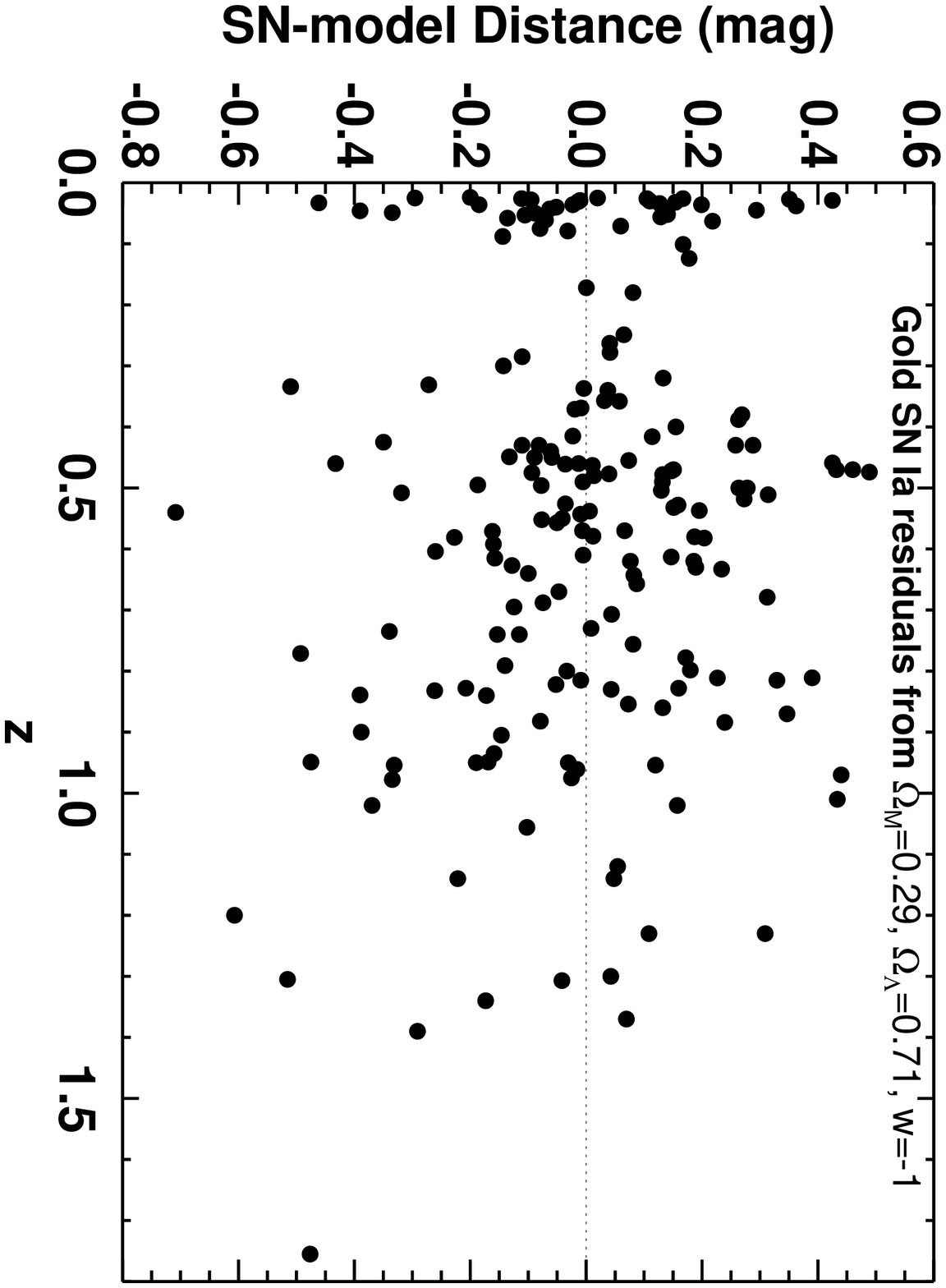}
\end{figure}

\vfill \eject

Figure Captions

Figure 1: Discovery-image sections from ACS $F850LP$ images around
each SN.  Panels on the left and middle show the discovery epoch and
the preceding (template) epoch, respectively. The panels on the right
show the results of the subtraction (discovery epoch minus
template). Arrows indicate position of the SNe.  Image scales and
orientations are given.

Figure 2: Color images of hosts with site of SN indicated. Filters $F850LP$, $F606W$, and $F435W$ correspond to red, green, and blue, respectively.

Figure 3: Corrections to single-model PSF fitting photometry in the $F850LP$ band for SNe Ia due to the red halo.   At red wavelengths, the PSF model and aperture corrections for ACS WFC strongly depend on the monochromatic wavelength of incident light as measured from narrow-band filters.  Thus the use of a single PSF model or aperture correction to measure photometry in a wide passband necessitates the use of a correction derived synthetically by weighting the monochromatic photometric variations by the appropriate SN SED.  In the upper panel we show this correction for $F850LP$ as a function of SN Ia age for different redshifts: 0.75 (asterisks), 1.00 (diamonds), 1.30 (squares), 1.50 (Xs), and 1.70 (triangles).   In the lower panel we show the correction as a function of redshift for a SN Ia at maximum light.

Figure 4: Multi-color light curves of SNe~Ia.  For each SN~Ia,
multi-color photometry transferred to rest-frame passbands is plotted.
The individual, best-fit MLCS2k2 model is shown as a solid line, with
a $\pm 1 \sigma$ model uncertainty, derived from the model covariance
matrix, above and below the best fit.

Figure 5: Identification spectra (in $f_{\lambda}$) of 12 of the new
{\it HST}-discovered high-redshift SNe~Ia all shown in the rest frame.
Classification features are analyzed in \S 3.  The data are compared
to nearby SN~Ia spectra of a similar age as determined by the light
curves (see Table 3).

Figure 6: MLCS2k2 SN~Ia Hubble diagram.  SNe~Ia from ground-based
discoveries in the Gold sample are shown as diamonds, 
{\it HST}-discovered SNe~Ia are shown as filled symbols.  
Overplotted is the best fit for a flat cosmology: $\Omega_M=0.27$,
$\Omega_\Lambda=0.73$.  Inset: Residual Hubble diagram and models after subtracting empty Universe model.  The Gold sample is binned in equal spans of $n\Delta z=6$ where $n$ is the number of SNe in a bin and $\Delta z$ is the redshift range of the bin.

Figure 7:  Uncorrelated estimates of the expansion history.  Following the method of Wang \& Tegmark (2005) we derive 3, 4, or 5 independent measurements of $H(z)$ from the Gold sample using $n\Delta z$ = 40, 20, and 15, respectively.  The bottom panel shows the derived quantity ${\dot a}$ versus redshift.  In this plane a positive or negative sign of the slope of the data indicates deceleration or acceleration of the expansion, respectively.

Figure 8: Same as upper panel of Figure 7 comparing the improvement to the highest-redshift measure of $H(z)$ due only to the newest {\it HST} data, i.e. since R04.

Figure 9: Joint confidence intervals derived from SN samples for a
two-parameter model of the equation-of-state parameter of dark energy,
$w(z) = w_a + w_a z/(1+z)$.  For each panel, constraints from a SN
sample are combined with the indicated prior to yield the indicated
confidence intervals.  The position of a cosmological constant
$(-1,0)$ is shown as a filled symbol.

Figure 10: Best solution and uncertainty for a quartic polynomial fit (blue),
and simple $w_0-w_a$ parameterization using a strong prior (red).   As seen, the simple parameterization highly constrains the behavior of $w(z)$ as compared to the polynomial.  The greater constraint on $w(z)$ implied by the $w_0-w_a$ parameterization derives from implicit and unjustified priors on dark energy: that its evolution is monotonic, linear, and most important at low redshifts.  

Figure 11: As in Figure 10, 
uncertainty for $w(z)$ for $w_0-w_a$ and quartic polynomial 
parameterizations with and without high-redshift {\it HST} data.

Figure 12: Measurement of 3 uncorrelated components of $w(z)$ using the Gold
sample of SNe Ia and the weak prior.  Following the method of Huterer and Cooray (2005) we derived measurements of $w(z)$ in the same 3 redshift bins used in Figure 7 ($n\Delta z$=40).  Using their covariance matrix we derived
new, uncorrelated components of $w(z)$ shown in the upper panel with window functions given in the lower right panel and likelihoods given in the lower left panel.

Figure 13: Same as in Figure 12 but using the strong prior.

Figure 14: Same as in Figure 12 but using the strongest prior.

Figure 15: Difference between Figure 12,13 and 14 with and without the HST-discovered SNe Ia for the weak, strong and strongest prior.
However, without the {\it HST} data, any measurement of the highest-redshift value,
${\mathcal{W}}_{1.35}$, is not very meaningful
because the sample would contain no SNe at $z \sim 1.35$.

Figure 16:  Average spectrum derived from HST ACS grism spectra of thirteen SNe Ia  $z>1$.  The high redshift average and dispersion (mean $z=1.1$) shown in thick gray compares well to the low redshift average (over the 10 days following maximum) and dispersion (day-to-day, for the 10 days post maximum) shown as the heavy line with error bars.  

Figure 17: Distance difference in magnitudes for all Gold SNe ($cz > 7000$ km s$^{-1}$) between
the measured distance and that predicted for the concordance cosmology ($\Omega_M=0.29$, $\Omega_\Lambda=0.71$) with $w=-1$.  As discussed in the text, increased complexity in the description of $w(z)$ is not presently justified by the improvement in the fit.

\vfill
\eject

{\bf References}

\refitem
Afshordi, N., Loh, Y.-S., \& Strauss, M. S. 2004, Phys. Rev. D,
   69, 083524

\refitem
Aguirre, A. N. 1999a, ApJ, 512, L19

\refitem
------. 1999b, ApJ, 525, 583

\refitem
Aguirre, A., Haiman, Z., 2000, ApJ, 532, 28

\refitem
Albrecht, A. \& Bernstein, G., 2006, astro-ph/0608269

\refitem Albrect, A. \& Skordis,  Phys Rev. Lette., 2000, 84, 2076

\refitem
Aldering, G., 2005, astro-ph/0507426, New Astron. Rev. 49, 346

\refitem
Allen, S. W., Schmidt, R. W., Ebeling, H., Fabian, A. C., \&
   van Speybroeck, L. 2004, MNRAS, 353, 457

\refitem
Astier, P., et al. 2006, A\&A, 447, 31

\refitem
Balland, C. et al. 2006, A\&A, 445, 387 

\refitem
Barris, B., et al. 2004, ApJ, 602, 571

\refitem
Ben\'\i tez, N., Riess, A., Nugent, P., Dickinson, M., Chornock, R.,
   Filippenko, A. V. 2002, ApJ, 577, L1

\refitem
Benot, M.C., Bertolami, O., \& Sen, A.A., 2002, PhRvD, 66,4, 3507

\refitem
Blondin, S., et al. 2006, AJ, 131, 1648

\refitem
Bohlin, R. C., \& Gilliland, R. L. 2004, AJ, 127, 3508

\refitem
Bohlin, R. C., Lindler, \& Riess, A. G. 2005, STScI NICMOS ISR, 2005-002

\refitem
Bond, J. R., Efstathiou, G., \& Tegmark, M. 1997, MNRAS, 291, L33

\refitem
Boughn, S., \& Crittenden, R. 2004, Nature, 427, 45

\refitem
Caldwell, R. R., Dav\'e, R., \& Steinhardt, P. J. 1998, Ap\&SS, 261, 303

\refitem
Caldwell, R. R., \& Linder, E. V. 2005, Phys. Rev. Let., 95, 141301

\refitem
Cappellaro, E., et al. 2001, ApJ, 549, L215

\refitem
Carroll, S. M., Duvvuri, V., Trodden, M., \& Turner, M. S. 2004,
  Phys. Rev. D, 70, 043528

\refitem
Casertano, S., \& Riess, A. G. 2007, in preparation

\refitem
Chevalier, \& Polarski, D., 2001, Int. J. Mod Phys, D10, 213 

\refitem
Coil, A. L., et al. 2000, ApJ, 544, L111

\refitem
Cole, S., et al. 2005, MNRAS, 362, 505

\refitem
Conley, A., et al. 2006, ApJ, 644, 1

\refitem
Cooray, A. \& Huterer, D., ApJ, 1999, 513, L95

\refitem
Daly, R. A., \& Djorgovski, S. G. 2004, ApJ, 612, 652

\refitem
Deffayet, C., Dvali, G., \& Gabadadze, G. 2002, Phys. Rev. D, 65044023

\refitem
de Jong, R.. S. et al 2006 astro-ph 0604394

\refitem 
Di Pietro, E., \& Claeskens, J. 2003, MNRAS, 341, 1299 

\refitem
Drell, P. S., Loredo, T. J., \& Wasserman, I. 2000, ApJ, 530, 593

\refitem
Einstein, A. 1917, SPAW, 142

\refitem
Eisenstein, D. J., et al. 2005, ApJ, 633, 560

\refitem
Filippenko, A. V. 1997, ARA\&A, 35, 309 

\refitem
Filippenko, A. V. 2004, in Carnegie Observatories Astrophysics Series,
  Vol. 2: Measuring and Modeling the Universe, ed. W. L. Freedman (Cambridge:
  Cambridge Univ. Press), 270

\refitem
Filippenko, A. V. 2005, in White Dwarfs: Cosmological and 
  Galactic Probes, ed. E. M. Sion, S. Vennes, \& H. L. Shipman 
  (Dordrecht: Springer), 97

\refitem
Fosalba, P., et al. 2003, ApJ, 597, L89

\refitem
Freedman, W., et al. 2001, ApJ, 553, 47

\refitem
Freese, K. 2005, New Astronomy Reviews, 49, 103

\refitem
Giavalisco et al 2004, ApJ 600, 93

\refitem
Goobar, A., Bergstrom, L., \& M\"{o}rtsell, E. 2002, A\&A, 384, 1

\refitem
Guy, J., Astier, P., Nobili, S., Regnault, N., \& Pain, R.
   2005, A\&A, 443, 781

\refitem
Hamuy, M., et al. 2003, Nature, 424, 651

\refitem
Holz, D. E. 1998, ApJ, 506, L1

\refitem
Hook, I.M., et al. 2005, AJ, 130, 2788

\refitem
Howell, D. A., et al. 2005, ApJ, 634, 1190

\refitem
Huterer, D., \& Cooray, A. 2005, Phys. Rev. D, 71, 023506

\refitem
Huterer, D., \& Turner, M., 2001, Phys Rev D., 64, 123527

\refitem
Jha, S. 2002, Ph.D. thesis, Harvard University

\refitem
Jha, S., Riess, A. G., \& Kirshner, R. P. 2006, submitted

\refitem
J\"{o}nsson, J., Dahl\'{e}n, T., Goobar, A., Gunnarsson, C.,
  M\"{o}rtsell, E., \& Lee, K. 2006, ApJ, 639, 991

\refitem
Knop, R., et al. 2003, ApJ, 598, 102

\refitem
Knox, L., Christensen, N., \& Skordis, C. 2001, ApJ, 563, L95

\refitem
Kolb, E. W., et al. 2006, Dark Energy Task Force White Paper

\refitem
Krisciunas, K., et al. 2005, AJ, 130, 2453

\refitem
Kulkarni, V. P., Fall, S. M., Lauroesch, J. T., York, D. G.,
   Welty, D. E., Khare, P., \& Truran, J. W. 2005, ApJ, 618, 68

\refitem
Landolt, A. U.,  1992, AJ, 104, 340

\refitem
Leibundgut, B. 2001, ARAA, 39, 67

\refitem
Liddle, A. R. 2004, MNRAS, 351, 49

\refitem
Linder, E. V. 2003, Phys. Rev. Lett., 90, 91301

\refitem
Miknaitis, G. A. et al. 2007, in prep

\refitem
Nolta, M. R., et al. 2004, ApJ, 608, 10

\refitem
Nugent, P., Kim, A., \& Perlmutter, S. 2002, PASP, 114, 803

\refitem
\"{O}stman, L., \& M\"{o}rtsell, E. 2005, JCAP, 2, 005

\refitem
Peebles, P. J., \& Ratra, B. 2003, Rev. Mod. Phys., 75, 559

\refitem
Perlmutter, S. 1999, ApJ, 517, 565

\refitem
Petric, A., Telis, A., Paerels, F., Helfand, D. J., 2006, ApJ in press

\refitem
Phillips, M. M., et al. 1999, AJ, 118, 1766

\refitem
Pirzkal, N., et al. 2005, ApJ, 622, 319

\refitem 
Quinn,J.L., Garnavich, P.M., Li, W., Panagia, N., Riess, A., Schmidt, B.P., \& Della Valle, M., 2006, ApJ, submitted

\refitem 
Rana, N. C. 1979, Ap\&SS, 66, 173

\refitem 
Rana, N. C. 1980, Ap\&SS, 71, 123

\refitem
Riess, A. G., \& Livio, M. 2006, astro-ph/0601319

\refitem
Riess, A. G., Press, W. H., \& Kirshner, R. P. 1995, ApJ, 438, L17

\refitem
Riess, A. G., Press, W. H., \& Kirshner, R. P. 1996, ApJ, 473, 588

\refitem
Riess, A. G., et al. 1998, AJ, 116, 1009 

\refitem
------. 1999, AJ, 118, 2668

\refitem
------. 2001, ApJ, 560, 49

\refitem
------. 2004a, ApJ, 600, L163

\refitem
------. 2004b, ApJ, 607, 665 (R04)

\refitem
------. 2005, 627, 579

\refitem
Schmidt, B. P., et al. 1994, ApJ, 434, L19

\refitem
Scranton, R., et al. 2005, ApJ, 633, 589

\refitem
Shapiro, C. \& Turner, M. S. astro-ph/0512586

\refitem
Sirianni, M., et al. 2005, PASP, 117, 1049

\refitem
Spergel, D. N., et al. 2006, astro-ph/0603449, in press

\refitem
Strolger, L. et al 2004 ApJ 613, 200

\refitem
Sullivan, M., et al. 2003, MNRAS, 340, 1057

\refitem
Sullivan, M., 2005, private communication

\refitem
Szydlowski, M., Kurek, A. and Krawiec, A.  2006 astro-ph/0604327

\refitem
Tegmark, M., et al. 2004, ApJ, 606, 702

\refitem
Tonry, J. T., et al. 2003, ApJ, 594, 1

\refitem
Tripp, R., \& Branch, D. 1999, ApJ, 525, 209

\refitem
Wang, L., Strovink, M., Conley, A., Goldhaber, G., Kowalski, M.,
  Perlmutter, S., \& Siegrist, J. 2006, ApJ, 641, 50

\refitem
Wang, Y. 2005, New Astronomy Reviews, 49, 97

\refitem
Wang, Y., \& Tegmark, M. 2005, Phys. Rev. D, 71, 103513

\refitem
Wang, Y., \& Mukherjee, P. 2006, ApJ (in press), astro-ph/0604051

\refitem 
Wetterich, C., 1995, A\&A, 301, 321

\refitem
Wirth, G. D. et al 2004 AJ 127, 3121

\refitem
Wood-Vasey, W. M., et al. 2007, ApJ, submitted

\refitem
Wright, E. L. 2002, BAAS, 161.17 (astro-ph/0201196)

\end{document}